\documentclass[fleqn,usenatbib]{mnras}

\usepackage{newtxtext,newtxmath}  
\usepackage[T1]{fontenc}

\DeclareRobustCommand{\VAN}[3]{#2}
\let\VANthebibliography\thebibliography
\def\thebibliography{\DeclareRobustCommand{\VAN}[3]{##3}\VANthebibliography}

\usepackage{graphicx}	% Including figure files
\usepackage{amsmath}	% Advanced maths commands
\usepackage{soul} % Temporatily added by Uma
\usepackage{afterpage} % Temporatily added by Uma
\usepackage{xcolor} %
\usepackage{booktabs}  % tables on book style

\def\eqp#1{(\ref{eq:#1})}
\def\eql#1{\label{eq:#1}}
\newcommand{\be}{\begin{equation}}
\newcommand{\ee}{\end{equation}}
\newcommand{\ba}{\begin{eqnarray}}
\newcommand{\ea}{\end{eqnarray}}
\newcommand{\bas}{\begin{eqnarray*}}
\newcommand{\eas}{\end{eqnarray*}}
\newcommand\rmH{\mathrm{H}}
\newcommand\SO{\mathrm{SO}}
\newcommand\cm{\mathrm{cm}}
\newcommand\au{\mathrm{au}}
\newcommand\adsorb{\mathrm{adsorb}}
\newcommand\warmdust{\mathrm{warm\ dust}}
\newcommand\nH{n_\rmH}
\newcommand\NH{N_\rmH}
\newcommand\rms{\mathrm{s}}

\newcommand\p{\partial}
\newcommand\calP{\mathcal{P}}
\newcommand\calE{\mathcal{E}}
\newcommand\calQ{\mathcal{Q}}
\newcommand\ppt{\frac{\p}{\p t}}
\newcommand\ppr{\frac{\p}{\p r}}
\newcommand\enter{\mathrm{enter}}
\newcommand\etaenter{\eta_\enter}
\newcommand\renter{r_\enter}

\newcommand\rc{r_\mathrm{c}}
\newcommand\rmK{\mathrm{K}}

\newcommand\Mzdot{\dot{M}_0}
\newcommand\bfq{\mathbf{q}}

\newcommand\bfF{\mathbf{F}}
\newcommand\bfS{\mathbf{S}}
\newcommand\bfH{\mathbf{H}}
\newcommand\half{\frac{1}{2}}

\newcommand\Sigmat{\widetilde{\Sigma}}

\newcommand\uK{u_\rmK}
\newcommand\GammaK{\Gamma_\rmK}
\newcommand\Mdot{\dot{M}}
\newcommand\GMRfac{\left(\frac{GM}{R}\right)^{1/2}}
\newcommand\GMrfac{\left(\frac{GM}{r}\right)^{1/2}}
\newcommand\rmi{\mathrm{i}}

\newcommand\et{\vec{e}_\theta}
\newcommand\Hrmax{H(\rmax)}
\newcommand\rmmp{\mathrm{mp}}

\newcommand\ppz{\frac{\p}{\p z}}
\newcommand\Hpcl{H_\mathrm{pcl}}
\newcommand\Tpcl{T_\mathrm{pcl}}
\newcommand\tauP{\tau_\mathrm{P}}

\newcommand\rcb{r_\mathrm{CB}}

\newcommand\Msolar{M_\odot}
\newcommand\peryear{\mathrm{yr}^{-1}}

\newcommand\Rinit{R_\mathrm{\,i}}
\newcommand\Hshock{H_\mathrm{shock}}
\newcommand\rhomp{\rho_\mathrm{mp}}

\newcommand\Rgas{R_\mathrm{gas}}
\newcommand\Htilde{\widetilde{H}}
\DeclareMathOperator{\erf}{erf}
\newcommand\qbar{\overline{\bfq}}
\newcommand\Htwo{\mathrm{H}_2}
\newcommand\nHtwo{n_{\Htwo}}
\newcommand\mHtwo{m_{\Htwo}}
\newcommand\rrshock{r_\mathrm{shock}}

\newcommand\rmin{r_\mathrm{min}}
\newcommand\rmax{r_\mathrm{max}}
\newcommand\nut{\nu_\mathrm{t}}

\newcommand\MrB{(M_r)_\mathrm{ballistic}}
\newcommand\urpost{|u_r|_\mathrm{post}}

\title[Protostellar disks with infall]{Protostellar disks subject to infall: a one-dimensional inviscid model \\and comparison with ALMA observations}     

\author[K. Shariff, U. Gorti, \& J.D. Melon Fuksman]{
Karim Shariff,$^{1}$\thanks{E-mail: Karim.Shariff@nasa.gov}
Uma Gorti$^{1,2}$
Julio David Melon Fuksman$^{3}$
\\
$^{1}$NASA Ames Research Center, Moffett Field, CA 94035, USA\\
$^{2}$SETI Institute, Mountain View, CA 94043, USA\\
$^{3}$Max Planck Institute for Astronomy, K\"onigstuhl 17, D-69117 Heidelberg, Germany
}

\date{Accepted XXX. Received YYY; in original form ZZZ}

\pubyear{2022}

\begin{document}
\label{firstpage}
\pagerange{\pageref{firstpage}--\pageref{lastpage}}
\maketitle
    
\begin{abstract}
A new one-dimensional, inviscid, and vertically integrated disk model with prescribed infall is presented.  The flow is computed using a second-order shock-capturing scheme.  Included are vertical infall, radial infall at the outer radial boundary, radiative cooling, stellar irradiation, and heat addition at the disk-surface shock.  Simulation parameters are chosen to target the L1527 IRS disk which has been observed using ALMA (Atacama Large Millimeter Array).  The results give an outer envelope of radial infall and $u_\phi \propto 1/r$ which encounters a radial shock at $\rrshock \sim 1.5\ \times$ the centrifugal radius ($\rc$) across which the radial velocity is greatly reduced and the gas temperature rises from a pre-shock value of $\approx 25$ K to $\approx 180$ K over a spatially thin region calculated using a separate shock structure code.  At $\rc$, the azimuthal velocity $u_\phi$ transitions from being $\propto 1/r$ to being nearly Keplerian.  These results qualitatively agree with recent ALMA observations which indicate a radial shock where SO is sublimated as well as a transition from a $u_\phi \sim 1/r$ region to a Keplerian inner disk.  However, in one set of observations, the observed position-velocity map of cyclic-C$_3$H$_2$, together with a certain ballistic maximum velocity relation, has suggested that the radial shock coincides with a ballistic centrifugal barrier, which places the shock at $\rrshock = 0.5 \rc$, i.e, inward of $\rc$, rather than outward as given by our simulations.  It is argued that radial velocity plots from previous magnetic rotating-collapse simulations also indicate that the radial shock is located outward of $\rc$.  The discrepancy with observations is analyzed and discussed, but remains unresolved.
\end{abstract}


%%%%%%%%%%%%%%%%% BODY OF PAPER %%%%%%%%%%%%%%%%%%

\section{Introduction}
\label{sec:intro}
Protoplanetary disk formation, which lasts $\sim 10^5$ yr for stars of $\sim$ one solar mass, is characterized by infall of material from a parent cloud core and is rich in phenomena.  Bipolar jets and outflows launched likely due to rotating magnetic field lines near the star or from the disk itself are common \citep[e.g.,][]{Pudritz_and_Ray_2019}.  Though the available sample is limited due to their short duration, episodic accretion outbursts known as FU Orionis and EX Lupi events are also thought to be common \citep{Audard_etal_2014}.  Finally, ring-and-gap structures \citep{Sheehan_etal_2020}
in protostellar disks (aged $\sim 0.1$--1 Myr) suggest that planet formation might begin earlier than thought.

Disks form as a result of the gravitational collapse of rotating dense cores in molecular clouds.  Due to conservation of angular momentum, infalling particles are deflected away from the rotation axis, and each particle lands on the equatorial plane at a radius that increases with its initial specific angular momentum.  According to a ballistic model \cite[][hereafter UCM]{Ulrich_1976, Cassen_and_Moosman_1981}, infalling particles are on zero energy orbits, i.e., parabolas.  Those orbits having a small inclination angle, $\theta_0$ (measured from the pole), have a smaller angular momentum (in addition to having a smaller cylindrical radius to begin with) and land closer to the star in the equatorial plane.  On the other hand, orbits with $\theta_0$ closer to the equatorial plane have a larger angular momentum and land further away from the star and with a smaller vertical velocity and greater radial velocity.  

The \textit{centrifugal radius}, $\rc$, is defined to be the furthest radius where infalling particles impact the midplane with a non-zero vertical speed $u_z$ (in absence of a disk).  With increasing epoch, $t_0$, since the initiation of collapse, material arrives at the disk from greater distances and with greater angular momentum.  As a result, $\rc$ grows as $\propto t_0^3$ (Equation \ref{eq:rc2}).

\begin{figure*}
\vskip 0.5truecm
\centering
\includegraphics[width=5.0truein]{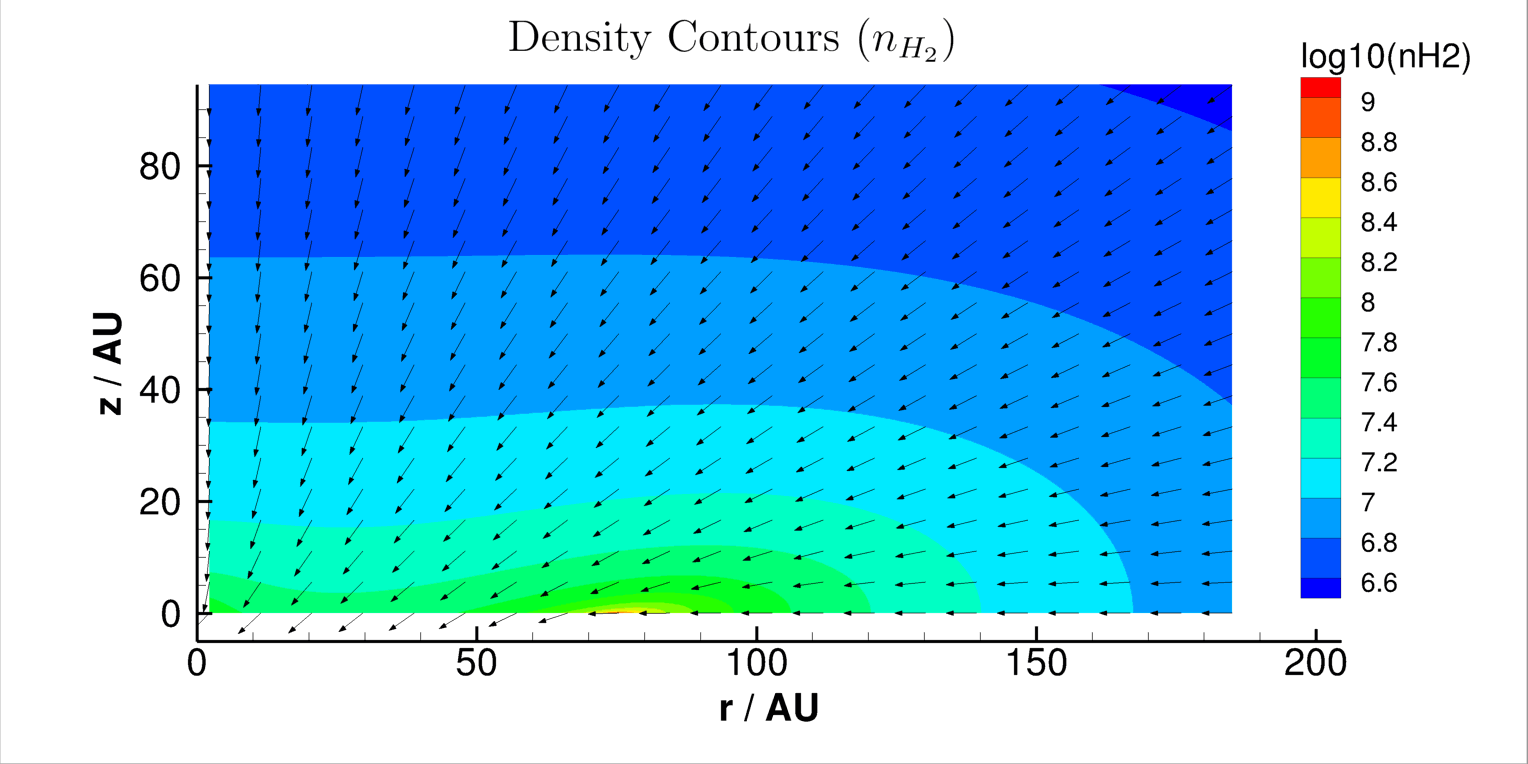}
\centering
\includegraphics[width=5.0truein]{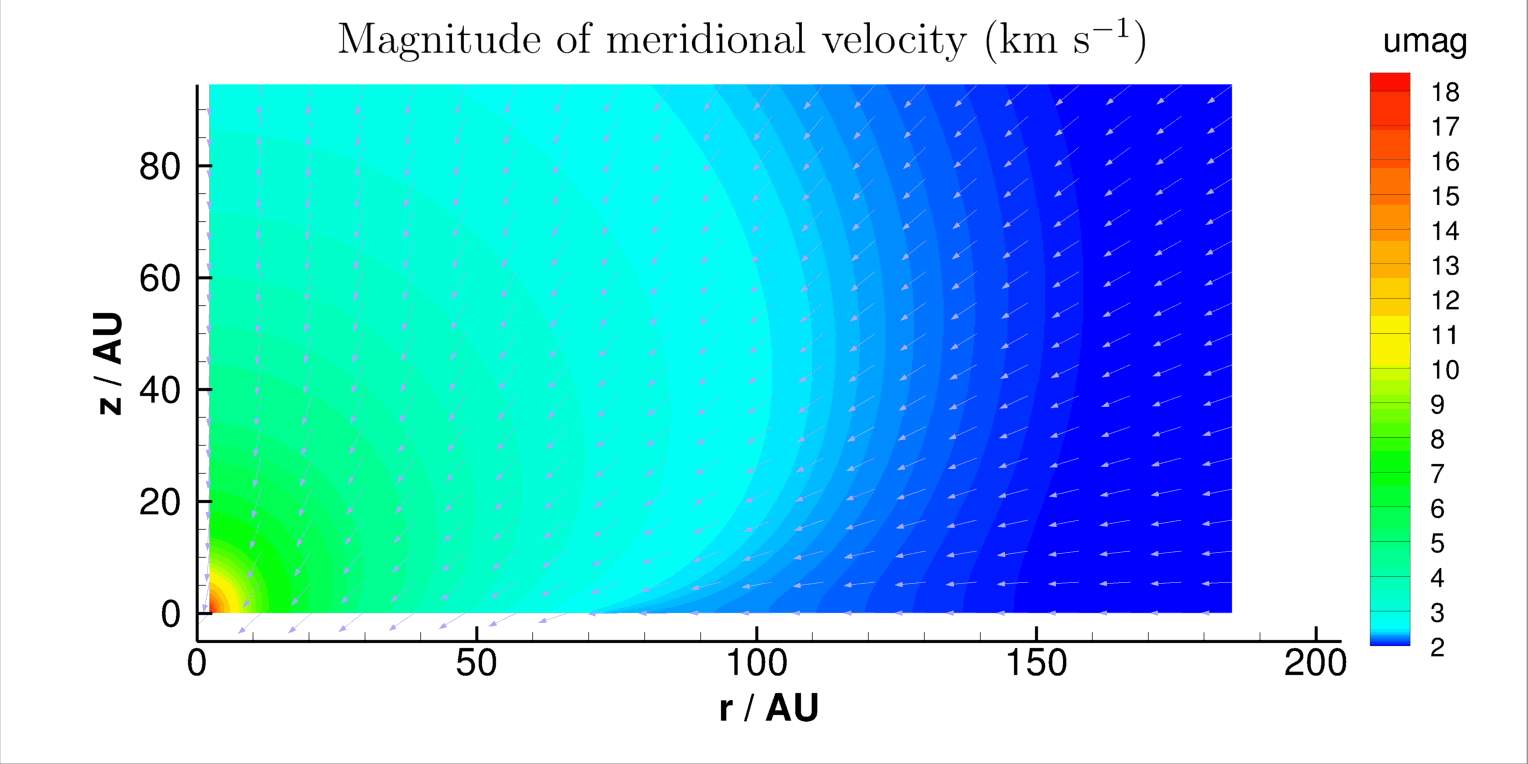}
\caption{Ulrich-Cassen-Moosman (UCM) infall with arrows showing the direction of the meridional velocity superimposed on color contours of (a) $\log_{10}(\nHtwo)$ which form a pseudo-disk, and (b) Magnitude of meridional velocity.  The same parameters as the simulation were used giving a centrifugal radius $\rc = 74$ au.  The vertically integrated model uses the infall flow evaluated at the midplane $z=0$.  Note that at $r = \rc = 74$ au, the vertical component becomes zero at the midplane.}
\label{fig:CM}\end{figure*}

The above ideas are illustrated in Figure~\ref{fig:CM} which was constructed using the analytical UCM model using parameters of our simulation, which are described later in \S\ref{sec:results}.  The centrifugal radius for the figure is $\rc = 74$ au.

Particles that land near $\rc$ have orbits with inclination angle $\theta_0 \to \pi/2$, i.e., grazing the midplane.  The value of $\rc$ is given by
\be
\rc = {j_z^2}/{(GM)}, \eql{rc_def}
\ee 
where $j_z = u_\phi r$ is the conserved specific angular momentum of orbits with $\theta_0 \to \pi/2$.

%%%%%%%%%%%%%%%%%%%%%%%%%%%%
%%%%%%%%%%%%%%%%%%%%

Substituting $j_z = u_\phi r$ into $\eqp{rc_def}$ one sees that $u_\phi$ is Keplerian at $r = \rc$.  We will see that this is an observationally useful result \citep{Ohashi_etal_2014, Aso_etal_2017}: if \textit{in a region of conserved $j_z$} one identifies the radius where $u_\phi$ is Keplerian, then that radius equals the centrifugal radius.  \footnote{We thank Dr. N. Ohashi for pointing this out.}
Note from Figure~\ref{fig:CM} that $u_r$ is still $< 0$ at $r = \rc$, i.e., particles continue to move inward.

Recent observations by the Atacama Large Millimeter Array (ALMA) are able to penetrate the cocoon of dust surrounding disks with infall and reveal their structure with unprecedented resolution.  Table~\ref{tab:alma}, based upon a similar table in Sakai (\citeyear{Sakai_Slides}), lists sources that are believed to have a similar kinematic structure.
\begin{table*}
\begin{center}
\begin{tabular}{c c c c c c}
\toprule
 Source                     & In                &Reference          & Class    &  $M/M_\odot$& $\rrshock$ (au)\\
 \midrule
 IRAS 04368+2557  & L1527        &\cite{Sakai_etal_2014_Nature}  & 0/I& 0.18     & 100 (SO ring)\\
 "         &  "   &\cite{Ohashi_etal_2014}  & "& 0.30     & 120 (jump in modeled $u_r$; Model 2)\\
 "         &  "  & \cite{Sakai_etal_2017_MNRAS}                       & "   & 0.18     & 100 (SO ring)\\
  "        &  "  & \cite{Aso_etal_2017}& "                                    & 0.45$^\dag$ & -\\
 IRAS 15398-3359   & Lupus 1 MC                 &\cite{Okoda_etal_2018}               & 0 & 0.007  & 40 (PV diagram of CCH )\\
 IRAS 16293-2422A & $\rho $ Ophiuchi cloud &\cite{Oya_etal_2016}                    & 0 & 0.5--1.0      &40--60 (CH$_3$OH \& HCOOCH$_3$ ring)\\
 IRAS 16293-2422B & "  &\cite{Oya_etal_2018}                   & 0 & 0.2--0.8       &30--50 (Similar to above)\\
 IRAS 04365+2535  &TMC-1A     & \cite{Aso_etal_2015}                 & I  & 0.68       & -\\
 "                              &"                 & \cite{Sakai_etal_2016_TMC1}    & "  & 0.53       & $\sim$ 50 (radius of SO peak displaced from continuum)\\
 IRAS 18148-0440   & L483          &\cite{Oya_etal_2017}        & 0 & 0.10--0.20        & $100^{+100}_{-70}$ (size of SO peak centered on continuum and \\
                                 &                   &                                          &   &                         & SiO emission displaced from continuum peak)\\
%HH212 & Orion B GMC& \cite{Lee_etal_2017}.      & 0  & 0.25 & $44 \pm 9$ ($(1/2) \rc$, $\rc$ determined from angular momentum)\\
HH212.    & Orion B GMC       & \cite{Codella_etal_2018} & "& $\sim 0.2$      & $\sim 60$ (CH$_3$CHO \& HDO rings above and below midplane)\\
 L1489 IRS                 &  Taurus MC              &\cite{Yen_etal_2014}                    & I  &  1.6         &250--390 (SO ring)\\
 "                                 &  "                           &\cite{Sai_etal_2020}                     & "  &  $1.64\pm0.12$             & - \\
 %IRAS 19347+0727 & B335 &\cite{Imai_etal_2019}& 0 & 0.02--0.06 & $< 5$\\
 %
\midrule
\bottomrule
\end{tabular}\end{center}
\caption{Forming disks observed with ALMA.  The table was inspired by a similar one in the presentation of Sakai (\citeyear{Sakai_Slides}).  $\rrshock$ is the radius of a putative radial shock and the parenthetical remark indicates how the presence and location of the shock was determined.
The value of $M$ is as reported in each cited reference.  ($\dag$) This is the stellar mass we use for the simulation; see \S\ref{sec:preamble} for the justification. MC: Molecular Cloud.  GMC: Giant molecular cloud.}
\label{tab:alma}\end{table*}
The best resolved and the one we target for simulation is the class 0/I protostar IRAS 04368+2557 in L1527 IRS which was observed with improving resolution over successive ALMA cycles \citep{Sakai_etal_2014_Nature, Sakai_etal_2014_ApJ, Ohashi_etal_2014,  Oya_etal_2015, Aso_etal_2017,Sakai_etal_2017_MNRAS}.  It has a cold infalling and rotating envelope (IRE) with an angular momentum preserving rotational velocity ($u_\phi \propto 1/r$), in which carbon chain molecules (CCH, c-C$_3$H$_2$) and CS are present.  Where the IRE transitions to a rotationally supported disk with a Keplerian $u_\phi$, there is a ring of SO emission due to a putative radial shock that sublimates/desorbs SO from icy grains.  As we will show later, our simulations confirm the shock but not its radial location with respect to the ballistic \textit{centrifugal barrier} radius ($\rcb$) as inferred by Sakai etal.
We now discuss what is meant by a centrifugal barrier.

Many observations listed in Table~\ref{tab:alma} identify the radial shock with the ballistic centrifugal barrier, which conflicts with our simulation results as will be seen later.  The radius, $\rcb$, of the centrifugal barrier is defined to be \citep[e.g.,][Methods Section]{Sakai_etal_2014_Nature} the radius of closest approach to the star (the periastron) of a ballistic parabolic orbit lying in the midplane.  In other words, it is the innermost radius a ballistic particle traveling in the midplane can reach while conserving angular momentum $j_z$ and mechanical energy $E$ (both per unit mass).  Since $E = 0$ on a parabolic orbit, we have that
\be
  E = 0 = \frac{1}{2} \left(u_r^2 + u_\phi^2\right) - GM/r = \frac{1}{2}\left(u_r^2 + j_z^2/r^2\right) - GM/r.  \eql{Em}
\ee
Setting $u_r = 0$ in \eqp{Em} gives
\be
   \rcb = {j_z^2}/{(2GM)}. \eql{rcb}
\ee
Comparing \eqp{rc_def} and \eqp{rcb} one sees that
\be
   \rcb = \rc/2. \eql{half}
\ee

At $r = \rcb$ all the kinetic energy is in the azimuthal component and the potential energy is a local minimum in the orbit.  Therefore, $u_\phi$ is a local maximum at $r = \rcb$ \citep[][hereafter Sa14a]{Sakai_etal_2014_Nature}.

It should be noted at the outset that the ballistic approximation is valid only where: (i) Pressure gradient is negligible compared to radial advective acceleration:
\be
   \frac{\p p}{\p r} \ll \frac{\p}{\p r} \rho u_r^2,
\ee
which implies that the radial Mach number $M_r^2 \gg 1$.  
Therefore, the ballistic approximation breaks down inward of the radial shock since $M_r < 1$ there.
(ii) Ballistic trajectories do not cross or collide.

Note in passing: The ballistic model discussed above considers only those particles that are infalling along the midplane.  However, Figure~\ref{fig:CM} shows that particles can also enter the disk via vertical infall at radii $\renter < \rc$.  The ballistic motion of such particles after they enter the disk is considered in Appendix~\ref{sec:extended_ballistic}, where it is shown that their trajectories are ellipses and each $\renter$ has a different periastron/centrifugal barrier inward of $\rc/2$.  We caution the reader that the limitations stated in the previous paragraph apply equally to the analysis in Appendix~\ref{sec:extended_ballistic}.

We end the introduction by discussing the role of magnetic fields \citep[e.g., reviews by][]{Tsukamoto_2016, Zhao_etal_2020} which are not included in the present work.  Early simulations assuming ideal magnetohydrodynamics (MHD) and a magnetic field initially aligned with the rotation axis, showed that formation of a rotationally supported disk is suppressed by magnetic braking, i.e., angular momentum removal by magnetic torque.  Since this is at odds with the observed existence of rotationally supported disks with radii $\sim 100$ au, several scenarios have been explored that can mitigate magnetic braking.  These include misalignment of the magnetic field and rotation axis, turbulence, and non-ideal MHD.  A recent study \citep{Zhao_etal_2021} reinforced their earlier work which showed that non-ideal MHD effects are a robust mechanism for averting magnetic braking.   Finally, magnetic fields launch outflows which impinge upon the infall, cutting it off at some locations on the disk surface.  This was pointed out by \cite{Velusamy_and_Langer_1998} and can be observed in simulations \citep[e.g.,][]{Zhao_etal_2016}.  We briefly consider this effect in \S\ref{sec:radial_infall_case} by turning off vertical infall and allowing only radial infall.

The next section describes our vertically integrated model; a detailed comparison with previous such models is provided in \S\ref{sec:review}.  An implicit assumption of vertically integrated models is that infalling fluid quantities are instantaneously homogenized vertically with their disk counterparts.  The main differences of our model compared to previous ones are that we include radial infall and do not assume Keplerian $u_\phi$.  However, unlike most previous models, ours has no turbulent viscosity.  Our philosophy, inspired by \citet{Stahler_etal_1994}, is that one should first understand the basic state structure of the flow and then analyze and simulate turbulence instability mechanisms with the goal of constraining the value of the turbulent viscosity.

%%%%%%%%%%%%%%%%%%%%%%
\section{Model description}
\label{sec:modeldesc}
\subsection{Overview}

\begin{figure*}
\centering
\vskip 0.20truecm
\includegraphics[width=4.5truein]{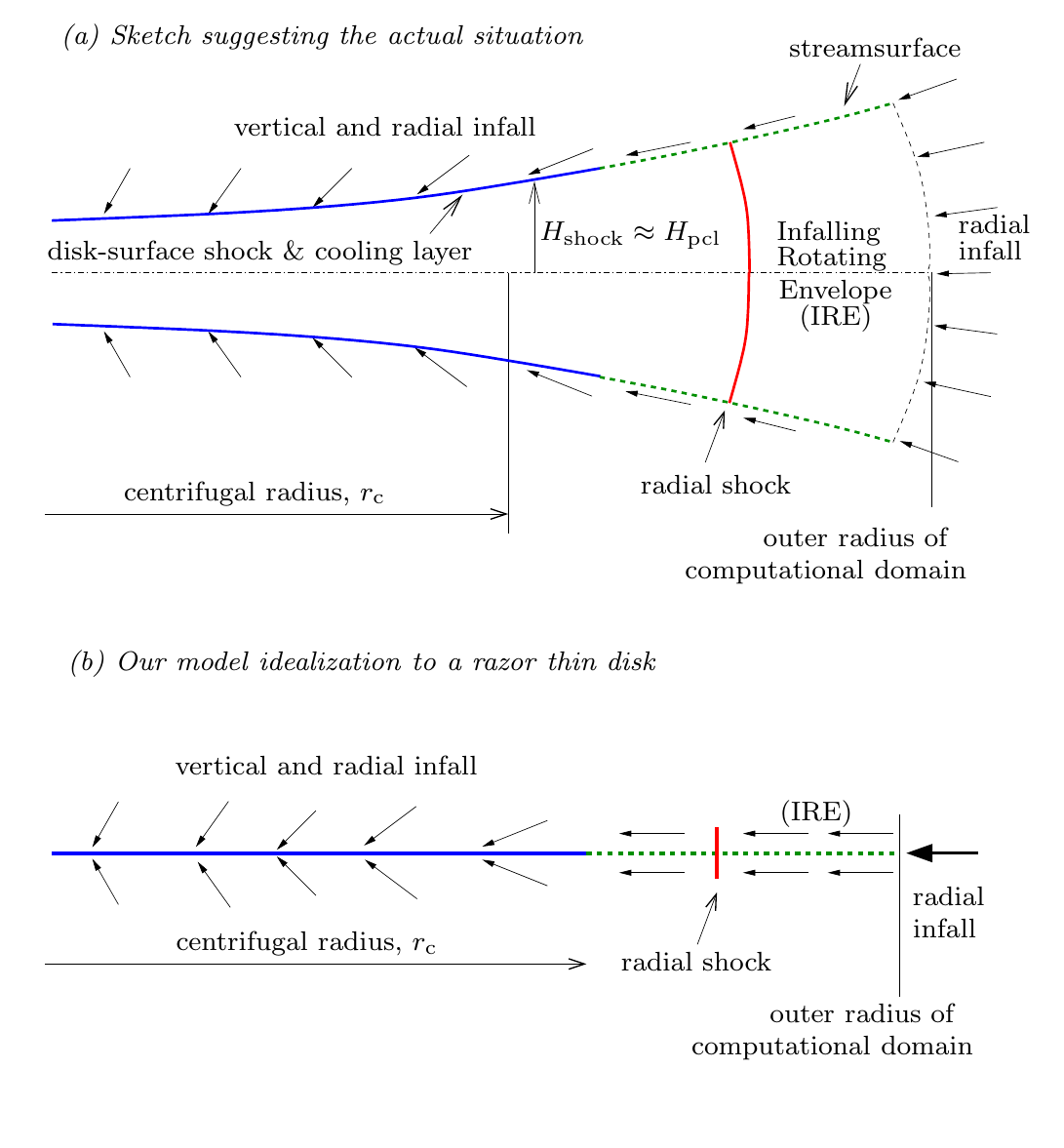}
\caption{Model schematic.  (a) Sketch suggesting the actual situation.
Moving rightward along the disk-surface shock (blue), the infall becomes more tangent and less normal to the shock.  The shock ends where the shock-normal velocity vanishes.  Note that the centrifugal radius $\rc$, which is defined in the midplane, is slightly inward of where the shock terminates.  This is due to the finite height, $\Hshock$, of the shock.
Both the radial and disk surface shocks are followed by a thin cooling layer whose properties were studied by \citet{Neufeld_and_Hollenbach_1994}.
We assume that the post cooling-layer (pcl) height, $\Hpcl$ of the disk surface shock $\approx \Hshock$.
 (b) The present model where the disk is razor thin and the infall is evaluated at the midplane.  
The radial shock and its relative location is an \textit{outcome} of the simulation and not imposed by the model.  IRE:  Infalling Rotating Envelope.}
\label{fig:schematic}
\end{figure*}
Figure \ref{fig:schematic} is a schematic of the model and is best studied along with Figure \ref{fig:CM}.  Figure \ref{fig:schematic}a is suggestive of the actual situation.  There is a disk-surface shock (colored blue) at height $\Hshock(r)$ which reduces the velocity component normal to it.  This shock is followed in the normal direction by a thin cooling layer whose properties have been studied in detail \citep[][hereafter NH94]{Neufeld_and_Hollenbach_1994}.  The "post cooling-layer" (pcl) height is denoted $\Hpcl(r)$ and will be assumed to equal $\Hshock(r)$ owing to the thinness of the cooling layer.  The shock surface terminates where the normal component of the infall velocity goes to zero.  Due to finite disk height, this point is slightly outward of the centrifugal radius where $u_z \to 0$ at the midplane.  At this point, the shock abuts a streamsurface (green dotted line) along which the flow is tangent by definition.  Strictly speaking, the shock ends where the normal velocity becomes subsonic rather than zero; this distinction is ignored in what follows.  
Our model, shown in panel (b), idealizes the above picture to a razor-thin disk.  The UCM infall flow (Appendices \ref{sec:infall_field} \& \ref{sec:infall_inward_of_rc}) evaluated at the disk midplane, provides boundary conditions to the vertically integrated disk equations given in the next subsection.  The infall into the disk has two parts: (i) For $r<\rc$, there is both a vertical and radial component, infall being more vertical closer to the star and more radial, less vertical, and more dense as $r \to \rc$.  (ii) At $r = r_\mathrm{max}$, the outer radial boundary of the simulation, radial infall  is applied as a boundary condition (Appendix \ref{sec:radial_infall}).
  
%%%%%%%%%%%%%%%%%%%%%%%%%%
\subsection{Mass and momentum equations}\label{sec:mass_and_mom}

Assuming symmetry about the midplane, the vertically integrated density and pressure are
\be
   \Sigma(r, t) \equiv  2 \int_{0}^{Z(r,t)} \rho\, dz, \hskip 0.5truecm \calP(r, t) \equiv 2\int_{0}^{Z(r, t)} p\, dz,
\ee
where $Z(r, t)$ is the upper disk surface defined as
\be
Z(r, t) = \begin{cases}
H_\mathrm{shock}^+(r, t),   \hskip 0.25truecm r \leq \rc;\\
S(r, t), \hskip 0.25 truecm r > \rc.
\end{cases}
\ee
Here $H_\mathrm{shock}^+ > 0$ denotes a height just above the disk-surface shock where infall quantities are specified.  For $r > \rc$, the surface $Z = S(r, t)$ is the streamsurface that joins the shock.

Performing a vertical integration of the conservation equations for mass, radial and angular momentum, we obtain
\begin{align}
\ppt\left(\Sigma r\right) + \ppr\left(r\Sigma u_r\right) + 2 r\left(\rho u_z\right)_1 &= 0, \eql{mass}\\
\ppt\left(\Sigma u_r r\right) + \ppr\left[r(\calP + \Sigma u_r^2)\right] - \Sigma u_\phi^2 + 2r\left(\rho u_r u_z\right)_1 &= \calP + r\Sigma g_r,\eql{rmom}
\\
\ppt\left(\Sigma u_\phi r^2\right) + \ppr\left(r^2\Sigma u_\phi u_r\right) + 2r\left(\rho u_\phi r u_z\right)_1 &= 0 ,\eql{amom}
\end{align}
where each equation has been multiplied by $r$ to allow a finite-volume numerical treatment after averaging over a computational cell $[r_{i-1/2}, r_{i+1/2}]$ (see \S\ref{sec:numerics}).  The terms with a `1' subscript are pre-shock quantities that arise from $z$-integration of $\p/\p z$ terms from $z = 0$ to $Z(r)$.  Appendix \ref{sec:vert_int} discusses a subtlety in the vertical integration that can be ignored for the present assumption of a razor thin disk, but should be kept in mind when finite disk thickness is accounted for. 

Note that in its usual form, the right-hand-side of the radial momentum equation (multiplied by $r$) has a $-r \p\calP/\p r$ term.  We have written this as
\be
   -r \frac{\p\calP}{\p r} = -\left[\frac{\p}{\p r}\left(r\calP\right) - \calP\right]
\ee
and moved the $-\p\left(r\calP\right)/\p r$ to the left hand side, leaving a $\calP$ on the right hand side of \eqp{rmom}.  This was done in order that radial fluxes have the same form as for Cartesian coordinates which permits direct use of shock-capturing methods developed for Cartesian coordinates.

A key assumption of the vertical integration procedure is that the incoming mass and momenta (and below, internal energy) are instantaneously mixed vertically into the disk.  Also, note that turbulent viscosity has not been included in the above equations.

%%%%%%%%%%%%%%%%%%%
\subsection{Energy equation}

The equation of state for internal energy (per unit volume) is taken to be $e = p/(\gamma - 1)$, where $\gamma \equiv c_p/c_v$.  For typical temperatures in our simulation, which range from $T = 10$ and 100 K, $\gamma$ varies between $5/3$ and $7/5$ as the rotational mode of H$_2$ is excited.  For simplicity we take $\gamma = 7/5$.

For axisymmetric flow, the internal energy satisfies
\begin{multline}
   \frac{\p e}{\p t} + \ppz\left(u_z e\right) + \frac{1}{r} \frac{\p}{\p r}\left(r u_r e\right) = 
   -p\left[\frac{\p u_z}{\p z} + \frac{1}{r}\ppr\left(r u_r\right)\right] \\
   - q_\mathrm{cool} + q_\mathrm{heat}. \eql{e}
\end{multline}
Since vertical compressional heating $-p \p u_z/\p z$ involves the product of a Heaviside and $\delta$ function at the disk-surface shock and there is a cooling layer following the shock, we take a slightly different approach to vertically integrating \eqp{e} than was done for the mass and momentum equations.
The cooling layer is thin: using the NH94 code we find that its column density varies between $0.002$ and $0.005$ gm cm$^{-2}$ for the simulation parameters.  This allows one to perform the integration up to $\Hpcl$, the post cooling-layer (subscript pcl) height with negligible loss of consistency with the shock height used for integrating the mass and momentum equations.  This avoids having to deal with the pressure compression term at the shock.
Making the assumption that the pressure compression term $-p \p u_z/\p z$ can be neglected \textit{within} the disk compared to the radial pressure compression, vertical integration of \eqp{e} gives
\begin{multline}
\frac{\p}{\p t}\left(\calE r\right) + (2 \rho u_z)_1 c_v \Tpcl r + \frac{\p}{\p r}\left(r u_r \calE\right) = - \calP\frac{\p}{\p r}\left(r u_r\right) \\
- r\calQ_\mathrm{cool} + r\calQ_\mathrm{heat}, \eql{e_vint}
\end{multline}
where
\be
   \calE \equiv 2 \int_0^{\Hpcl} e \, dz.
\ee
For the second term on the left-hand-side of \eqp{e_vint} we have used the fact that the mass flux $\rho u_z$ is preserved from the pre-shock state to the end of the post-shock cooling layer.  

The shock is effectively a jump and immediately across it, the gas temperature rises to a large value.
For instance, for our simulation parameters we have at $r/\rc = 0.2$ a pre-shock velocity of $(u_z)_1 = 4.64$ km s$^{-1}$ and a density of $n_\mathrm{H} = 6.6 \times 10^7$ (see Figure~\ref{fig:infall_quants}).  For a pre-shock temperature of $T = 20$ K, the NH94 code then gives $T = 1066$ K immediately behind the shock.  The dominant emission in the cooling layer following the shock is from rotational transitions of H$_2$O and CO (see Figure 9, top left panel in NH94).  However, these details are not important for specifying the temperature at the end of the cooling layer.  Given that at the end of the cooling layer, gas and grain temperatures have equilibrated, we set $\Tpcl$ equal to an estimate for the minimum grain temperature provided by NH4 (their Equation 43) which we slightly modify by adding a term for the flux from the surrounding cloud:
\be
\sigma_\mathrm{SB} T_\mathrm{pcl}^4 = \sigma_\mathrm{SB} T_0^4 + \frac{1}{2} (\rho u_z^3)_1. \eql{Tpcl}
\ee
Equation (\ref{eq:Tpcl}) assumes that all of the vertical kinetic energy flux is annulled across the shock and radiated.
The above neglects pre-heating of the gas by stellar and shock irradiation \citep{Chick_and_Cassen_1997}.

For radiative cooling $\calQ_\mathrm{cool}$ and stellar heating $\calQ_\mathrm{heat}$, the formulation due to \cite{Nakamoto_and_Nakagawa_1994} is used:
\be
   \calQ_\mathrm{heat} - \calQ_\mathrm{cool}(r) = \frac{4\tauP}{1 + 2\tauP}\left[\frac{F_\mathrm{irr}}{2} + \sigma_\mathrm{SB} \left(T_0^4 - T^4\right)\right],
   \eql{rad}
\ee
where 
\be
   \tauP \equiv \kappa_\mathrm{P} \Sigma,
\ee
is the vertical optical depth.
The Planck mean opacity, $\kappa_\mathrm{P}$, was obtained using the subroutines of \cite{Semenov_etal_2003}.  The opacities, given in cm$^2$ per gm of gas, are functions of gas density and temperature and for these we use the characteristic values
\be
   \overline{\rho} = \Sigma/H, \hskip 0.5truecm \overline{T} = \calE / (\Sigma c_v), \eql{char_vals}
\ee
with $H = c_\mathrm{s} / \Omega$, $c_\mathrm{s}$ being the sound-speed.
$F_\mathrm{irr}$ is the stellar flux and is given by
\be
   F_\mathrm{irr}(r)  = \frac{L_*}{4\pi R^2} \cos\gamma_\mathrm{irr}.
\ee
The factor of two dividing $F_\mathrm{irr}(r)$ in \eqp{rad} accounts for the fact that half of the flux enters the disk while the other half is radiated to space.
%
%%%%%% Figure 3
\begin{figure}
\centering
\includegraphics[width=3.7truein]{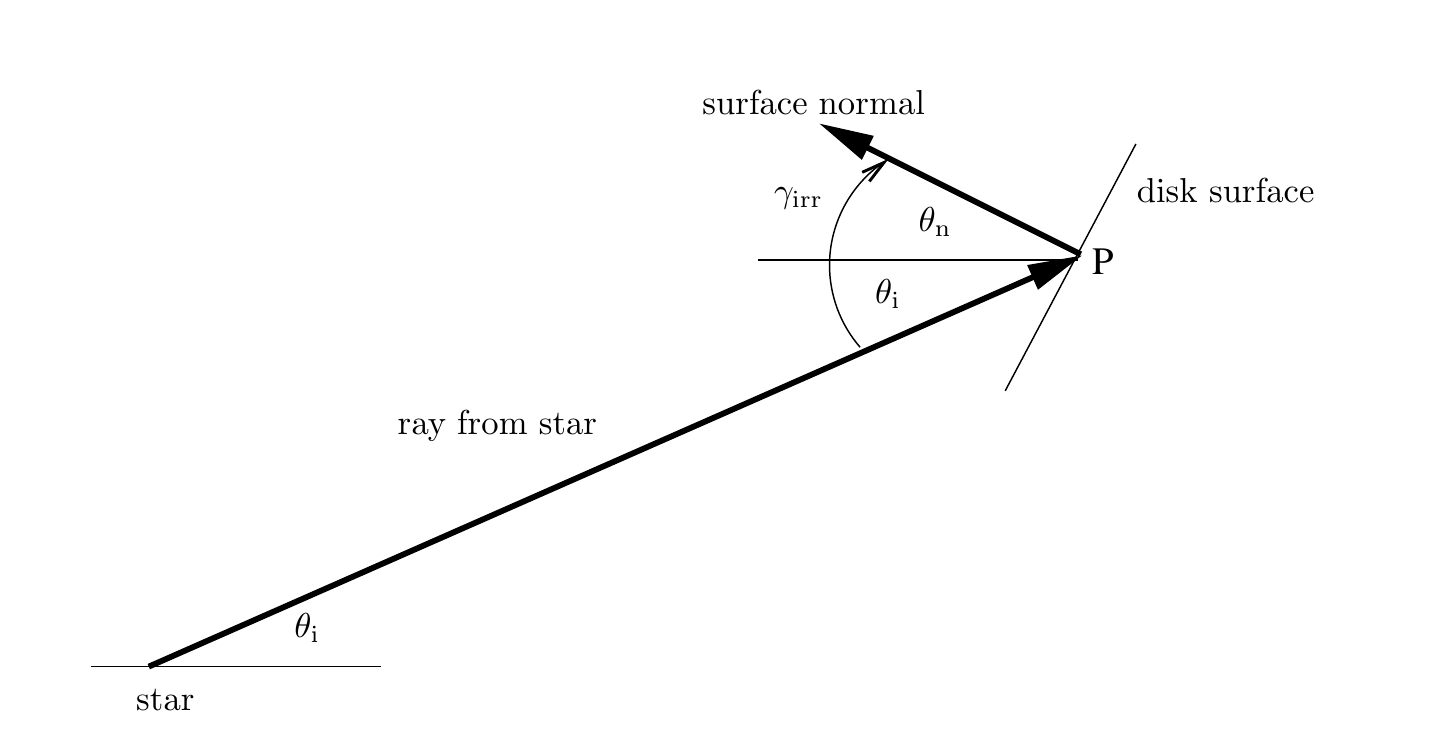}
\caption{The angle $\gamma_\mathrm{irr}$, needed for calculation of stellar irradiation, is measured between a ray incident on a point $P$ on the scale height surface, $(r, H(r))$, and the surface normal at $P$.  We have that $\gamma_\mathrm{irr} = \theta_\mathrm{i} + \theta_\mathrm{n}$.} 
\label{fig:irrad}\end{figure}
%%%%%

The quantity $\gamma_\mathrm{irr}$ is the angle between a ray incident on a point $P = (r, z = H(r))$ on the scale height surface and the normal to the surface.  Consulting Figure \ref{fig:irrad} we have that $\gamma_\mathrm{irr} = \theta_\mathrm{i} + \theta_\mathrm{n}$, where
\be 
\theta_\mathrm{i} = \tan^{-1} z/r \mathrm{\ and\ }\theta_\mathrm{n} = \tan^{-1} dr/dz,
\ee
are the inclination angles, relative to the midplane, of the incident ray and the surface normal, respectively.
This treatment assumes that the radiation in each ray is absorbed at point $P$ where the ray meets the surface $(r, H(r))$.  In reality, the absorption will take place over a distributed region.  
Stellar heating is applied to only those parts of the surface that face the star.
It was found that when the numerical scale height surface was used, large oscillations resulted because star facing portions of the surface receive stellar irradiation, while non-star-facing portions receiving none.  This increases the amplitude of oscillations.  To avoid this, a quartic polynomial is fit to the scale height surface for the purposes of calculating $F_\mathrm{irr}$.  The accuracy of the fit is assessed \textit{a posteriori}; see the dashed line in Figure \ref{fig:concluded}b.
We use $L_* = 2 L_\odot$ which is the bolometric luminosity for L1527 IRS \citep{Aso_etal_2017}.

%%%%%%%%%%%%%%%%%%%%%%%%%%%%%%%%%%%%
\subsection{Numerical treatment}\label{sec:numerics}

The transport equations \eqp{mass}--\eqp{amom} and \eqp{e_vint} have the form
\be
   \p_t \bfq + \p_r \bfF = \bfS(\bfq), \eql{euler}
\ee
where
\be
   \bfq = r \left(\Sigma, \Sigma u_r, \Sigma u_\phi r, \calE\right)^T
\ee
is a column vector of evolved quantities.  The vector $\bfF(\bfq)$ contains the radial advective fluxes of $\bfq$, and $\bfS(\bfq)$ contains the rest of the terms.

The computational domain $[r_\mathrm{min}, r_\mathrm{max}]$ is divided into $N$ equal cells of width $\Delta r$.  Let  a bar denote a cell average.  Then averaging \eqp{euler} over the $i$th cell gives:
\be
   \frac{d\,\qbar_i}{dt} + \frac{\bfH_{i + \half} - \bfH_{i - \half}}{\Delta r} = \overline{\bfS}_i
\ee
where $\bfH_{i + \half}$ is a numerical flux evaluated at a cell face.  The numerical fluxes are evaluated using the second-order \cite{Kurganov_and_Tadmor_2000} scheme, specifically their equation (4.4) .  
The third-order TVD (total variation diminishing) Runge-Kutta scheme is used for time advancement.
At $r_\mathrm{min}$, a one-way ``diode'' boundary condition is applied: if there is inflow into the domain ($u_r > 0$) at the end of a sub-step, then we set $u_r = 0$.  At $r_\mathrm{max}$, radial infall is specified as described in Appendix \ref{sec:radial_infall}.

%%%%%%%%%%%%%%%%%%%%%%%%%%%%%%%%%%%%%
\section{Results of model simulations}\label{sec:results}

\subsection{Preamble} \label{sec:preamble}

Simulation parameters (Table~\ref{tab:params}) were chosen to approximate L1527 IRS.  According to \cite{Aso_etal_2017}, $u_\phi = 2.31$ km s$^{-1}$ at $r  = 74$ au (corrected for beam resolution) where the transition from a $j_z$-preserving $u_\phi$ to a Keplerian $u_\phi$ takes place.  This implies that $M = 0.45 \Msolar$, which is fixed throughout the simulation.  We choose the evolutionary epoch to be $t_0 = 10^5$ yr.  
Then, to obtain $M = 0.45 \Msolar$, \eqp{Mzdot} gives a cloud temperature of $T = 20.1175$ K.  
From the discussion in the introduction, we know if Keplerian $u_\phi$ occurs at a certain radius in a $j_z$-preserving region, then that radius equals $\rc$.
Hence we want $\rc = 74$ au.  This is achieved from \eqp{rc2} by setting the cloud rotation to be $\Omega_0 = 1.5015 \times 10^{-13}$ s$^{-1}$.

\begin{table}
\centering
\begin{tabular}{l c c c c}
\toprule
 Parameter & Value\\
\midrule
Stellar mass, $M$                          &$0.45 M_\odot$\\
Centrifugal radius, $\rc$               &$74$ AU\\
Cloud rotation rate, $\Omega_0$ & $1.5015 \times 10^{-13}$ s$^{-1}$\\
Cloud temperature, $T_0$            & $20.1175$ K\\
Evolutionary epoch, $t_0$            & $10^5$ yr\\
Cloud infall rate, $\Mzdot$                      & $4.5 \times 10^{-6} M_\odot$ yr$^{-1}$\\
Stellar luminosity, $L_*$               & $2 L_\odot$\\
No. of radial grid cells, $n_r$        &2000\\
Computational domain, $[\rmin/\rc, \rmax/\rc]$ & $[0.2, 2.5]$\\
\bottomrule
\end{tabular}
%\end{center}
\caption{Run parameters chosen to target IRAS 04368+2557 in L1527 IRS using observed values reported by \citet{Aso_etal_2017}.}
\label{tab:params}
\end{table}
%%%%%%%%%%%%%%%%%%%%%%%%
%
The simulation is initialized with a very small surface density and zero velocities.  Infall is then turned on. Time elapsed from the start of the simulation is denoted as $\tau$.  To avoid very large imbalances between heating and cooling at the start, the pressure dilatation and radiation terms in the energy equation are ramped up for 3,000 yr to their full values.

How should one interpret the simulation time $\tau$?  Since an infall field appropriate to a star of $M = 0.45 M_\odot$ is turned on at $\tau = 0$ with a non-existent disk, $\tau$ cannot be interpreted as time since the start of collapse.  If the simulation achieved a stationary state as $\tau \to \infty$, one might claim that the stationary state represented the true state of the disk at epoch $t_0$.  Unfortunately, this is not the case here: the disk mass increases with $\tau$.  The claim cannot be made even if the simulation arrived at a stationary state.  This is because the true state of a disk at $t_0$ is very likely dependent on its past history, and our simulation does not trace the same history.

The best we can do is make a comparison at a $\tau$ value when the simulated disk mass is comparable to the observed disk mass.  \citet[][pg. 5]{Nakatani_etal_2020} estimate that $M_\mathrm{disk} = 0.26 M_\odot$ for L1527 IRS.  In our simulation this is achieved at $\tau \approx 40,000$ yr (Figure~\ref{fig:concluded}f).  Fortunately, simulation velocities and temperature are insensitive to $\tau$ and disk mass. 
 
%%%%%%%%%%%%%%%%%%%%%%%%%%%%%%%%%%%%% Figure 4
\begin{figure*}
\vskip 0.5truecm
\centering
\includegraphics[width=3.3truein]{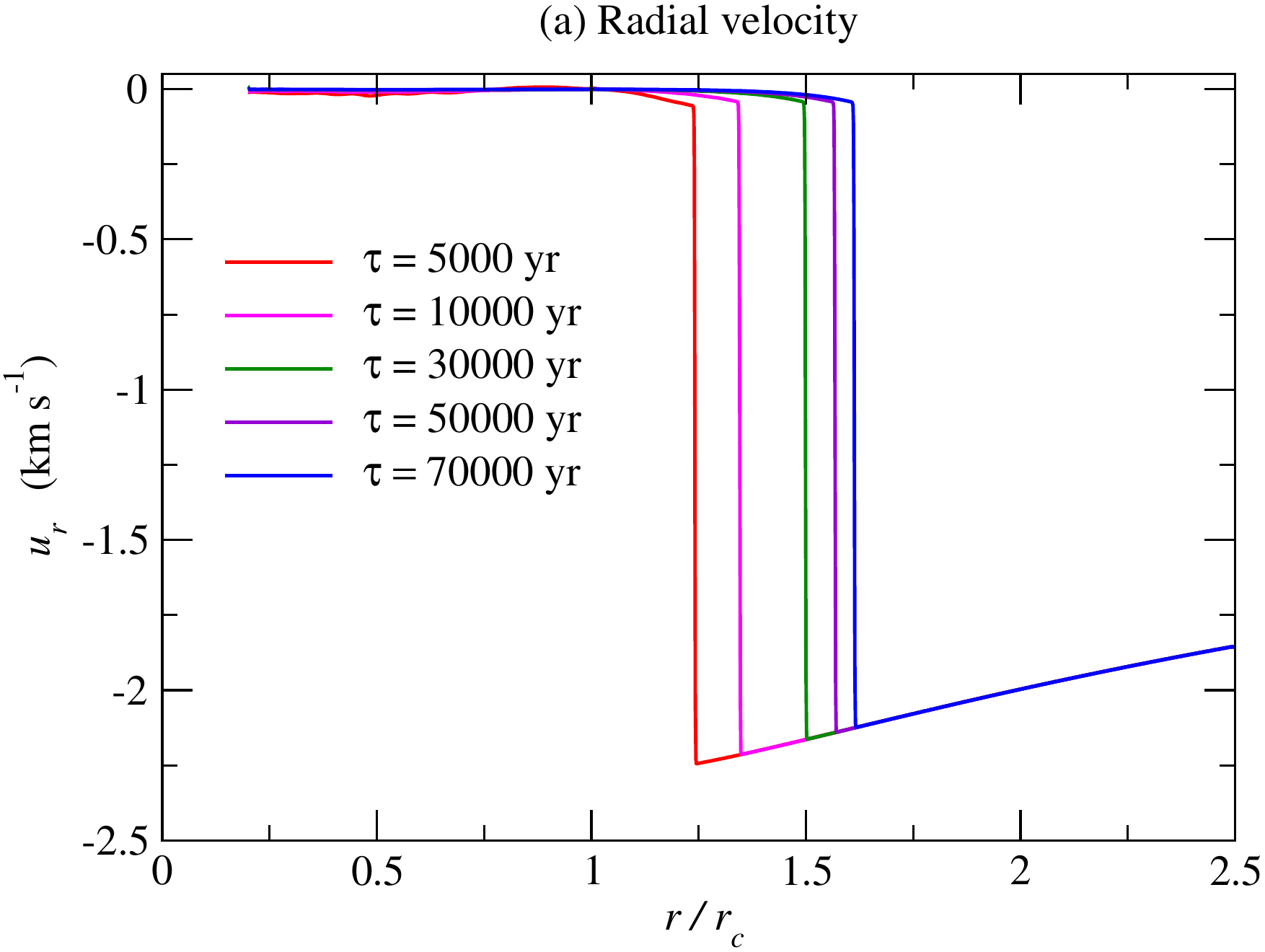}\hfill
\includegraphics[width=3.3truein]{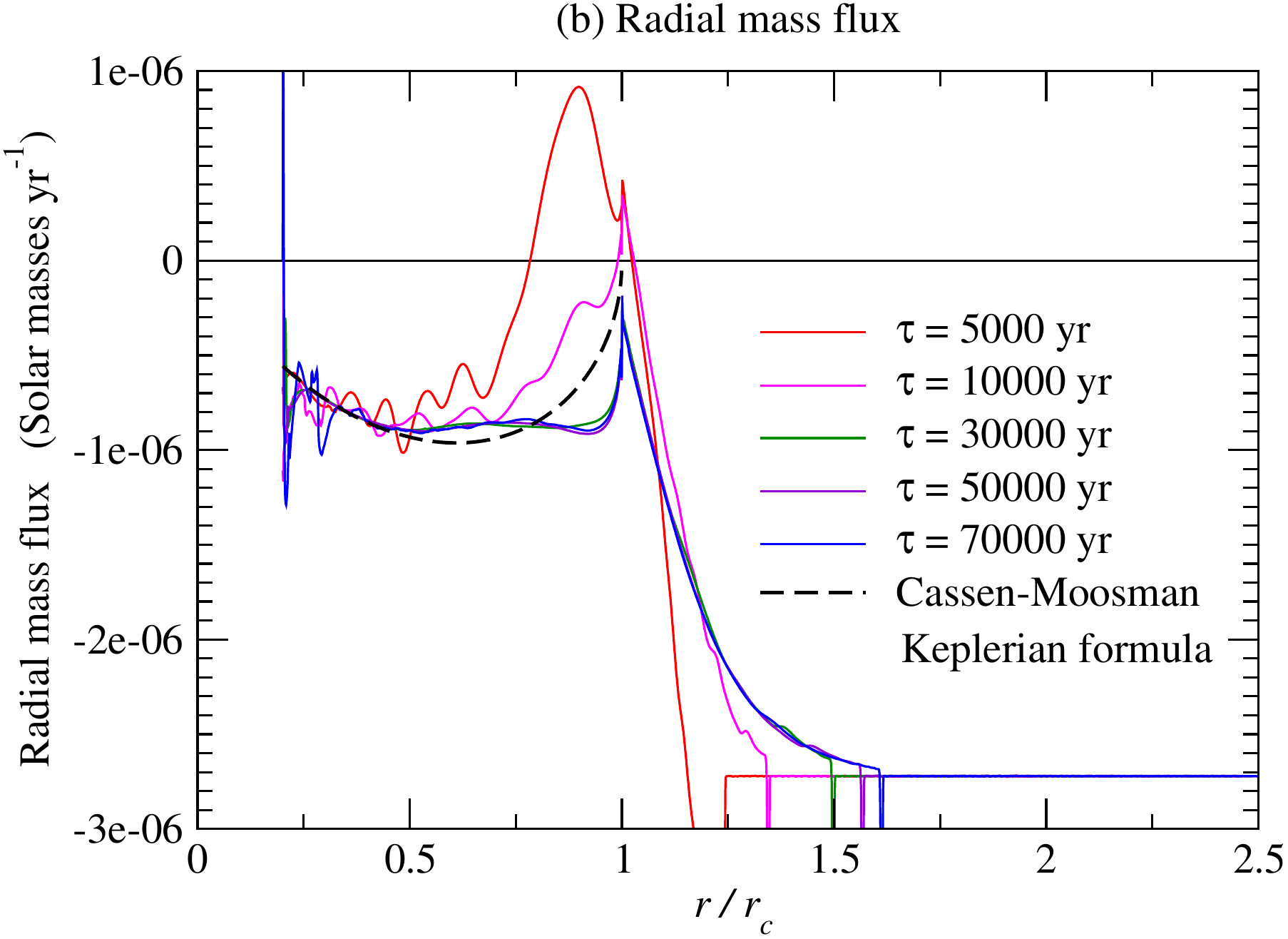}
\vskip 0.50truecm
\centering
\includegraphics[width=3.3truein]{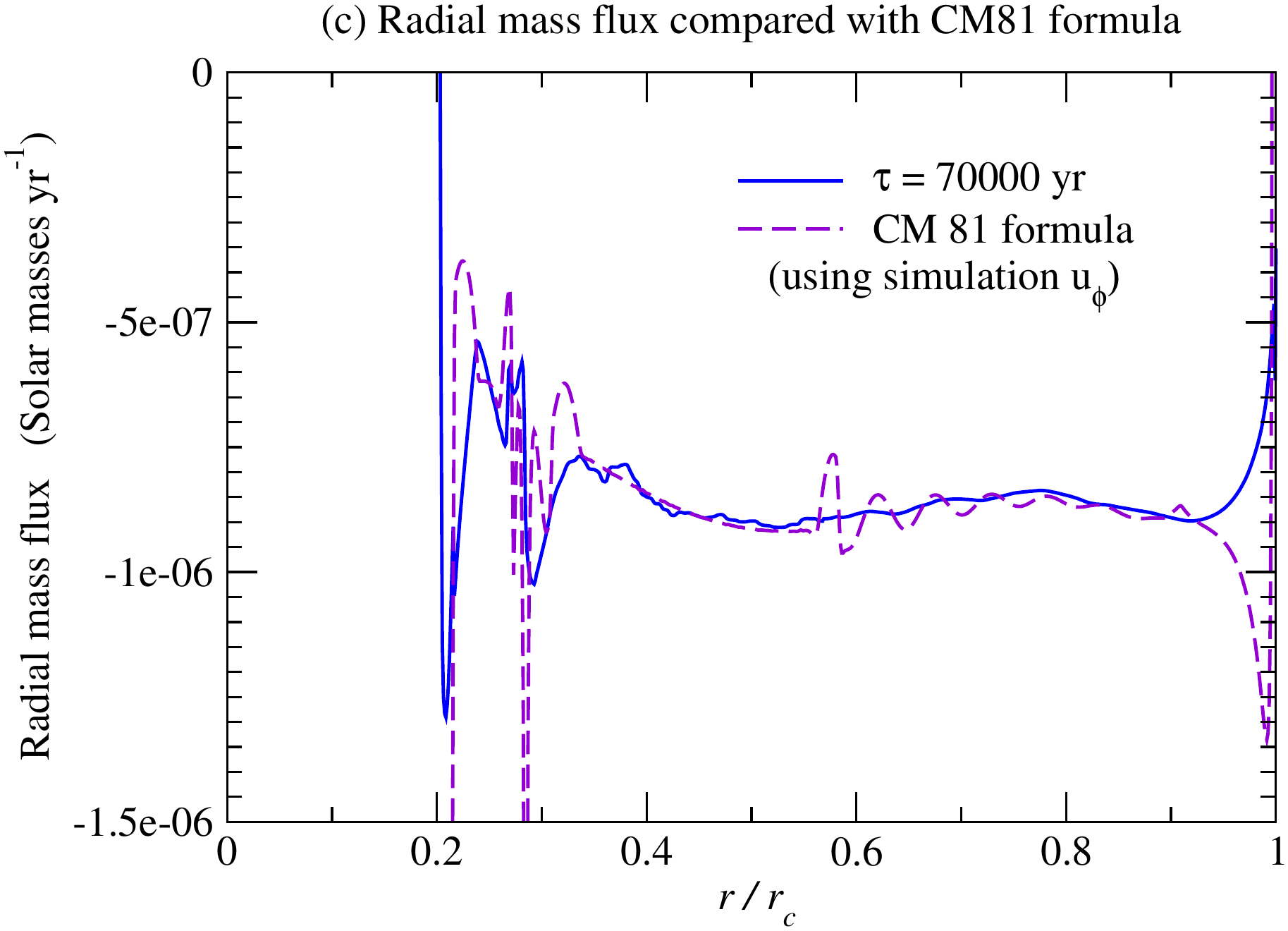}\hfill
\includegraphics[width=3.3truein]{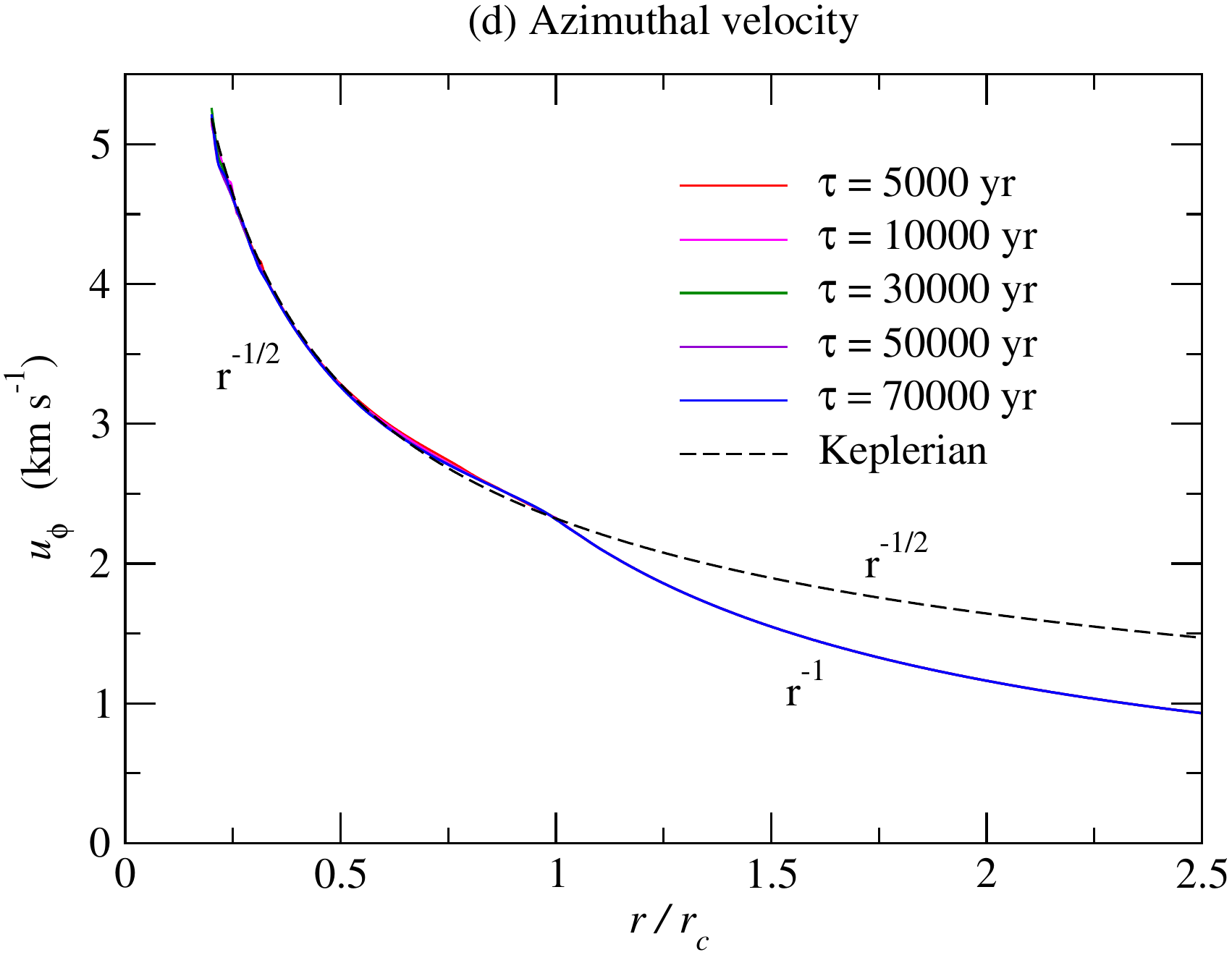}\hfill
\vskip 0.50truecm
\centering
\includegraphics[width=3.3truein]{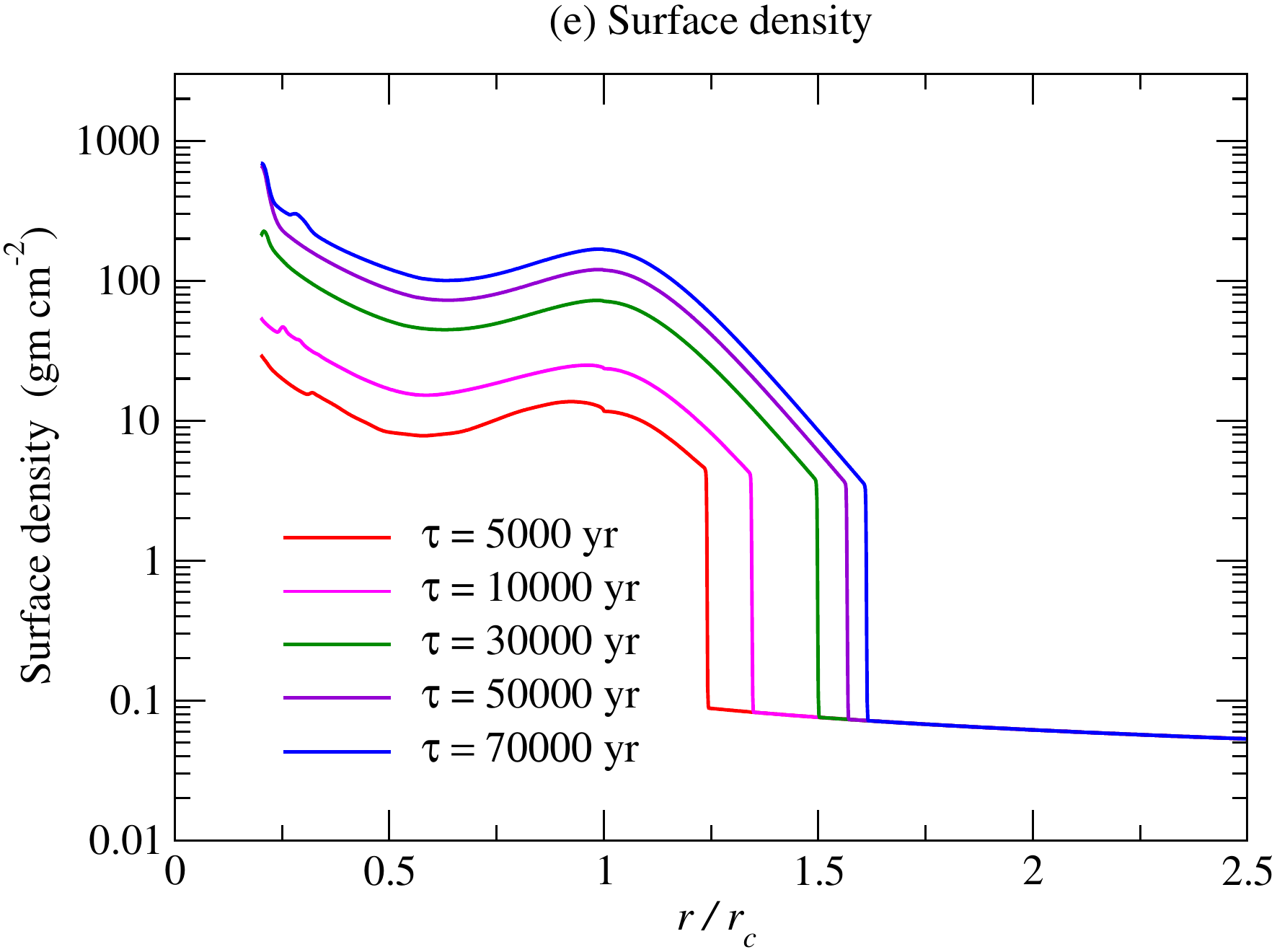}\hfill
\includegraphics[width=3.3truein]{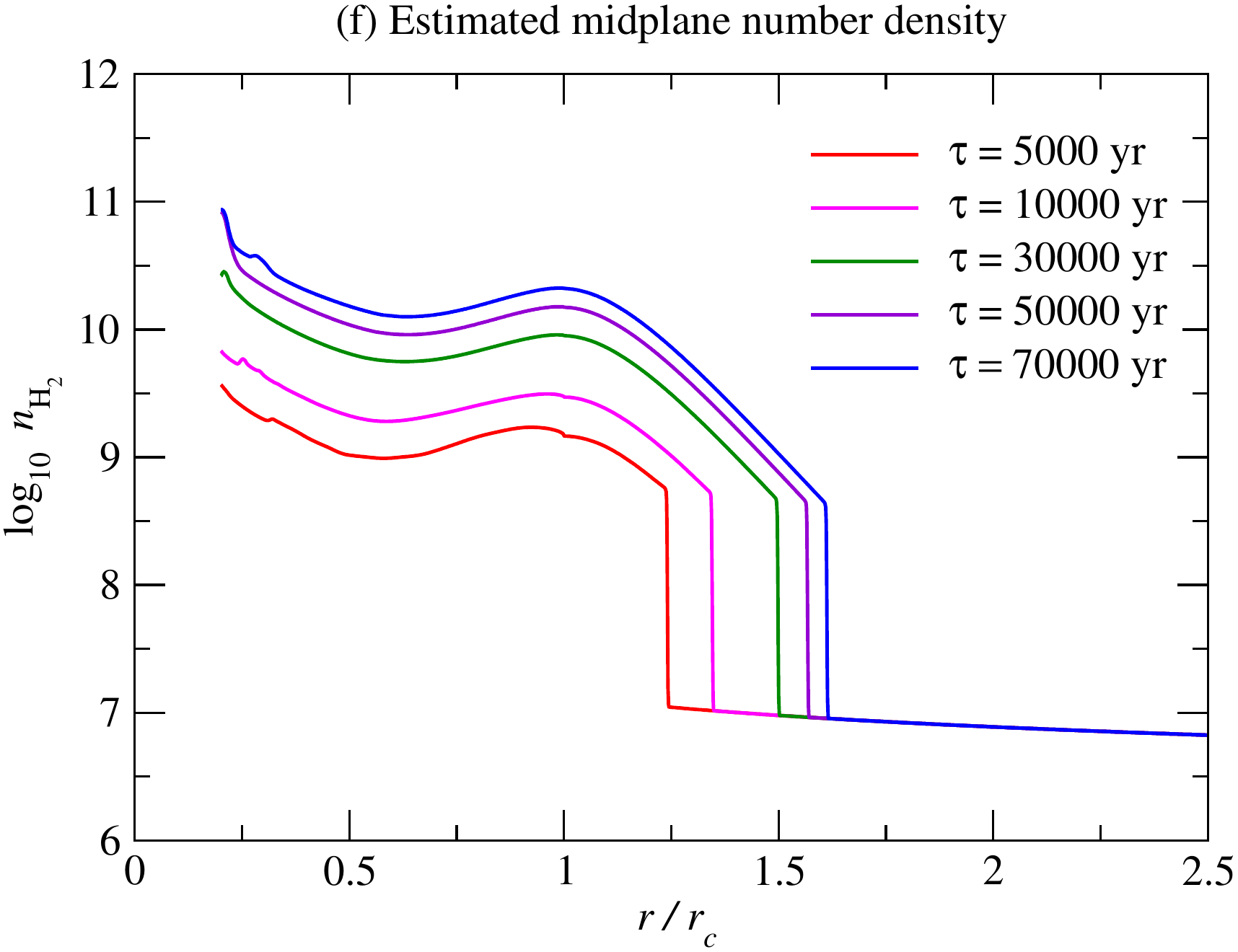}
\caption{First of two figures presenting the results for the simulation targeting L1527 IRS.  In panel (b), the Cassen-Moosman Keplerian formula for mass flux accounts for drag due to sub-Keplerian infall and is given in equation \eqp{Mdot_CM}.  The formula \eqp{Mdot_CM_general} used for panel (c) is valid for arbitrary $u_\phi$.}
\label{fig:case}
\end{figure*}
%%%%%%%%%%%%%%%%%%%%%%%%%%%%%

\subsection{First set of plots}

Figure \ref{fig:case} shows the first set of radial profiles of various quantities at different times.

The radial velocity $u_r$ (Figure~\ref{fig:case}a) most clearly shows a radial shock at $r/\rc \sim 1.5$, i.e, outward of the centrifugal radius, which propagates outward very slowly with decreasing speed.  The shock reduces $|u_r|$ to a small value and $|u_r|$ further decreases as the centrifugal radius is approached.  
It was verified that the region $r > \rrshock$ (the IRE) is well described by zero mechanical energy ballistic flow.  As stated in the introduction, the ballistic approximation breaks down inward of the radial shock since $u_r$ is subsonic there.

The mechanism of shock initiation ($\tau < 1000$ yr) may be qualitatively described as follows.  Consider a shock-free state where
$u_r$ slows smoothly with decreasing $r$ due to angular velocity  build-up, and eventually becomes subsonic.  In the region where $u_r$ is subsonic, pressure waves can travel upstream up to the point where $u_r$ is sonic.  A shock forms as waves pile-up at the sonic point.  As the strength of the shock grows, it propagates into more supersonic flow until the Rankine-Hugoniot jump conditions (including radiative losses) for a steady shock are satisfied.  To our knowledge, no simple analysis can predict where the shock will finally sit.

A better quantity to assess the radial flow in the post-shock, low $|u_r|$ region is the radial mass flux $\Mdot(r) \equiv 2\pi r\Sigma u_r$.
Figure \ref{fig:case}b shows that $\Mdot \approx -2.7 \times 10^{-6} \Msolar \peryear$ in the IRE beyond the radial shock, and is conserved across the shock as it should be.  From the shock inward it decreases to a small valued cusp at the centrifugal radius.  In this region therefore, the surface density must build up with time, which it does and leads to a maximum in $\Sigma$ at $r/\rc = 1$.  The fact that $\Mdot(r)$ is small at $r = \rc$ means that the region $\rc > 1$ feeds a relatively small amount of mass to the region $\rc < 1$, which accumulates mass via vertical infall.

For $0.3 \rc < r < 0.9 \rc$, $\Mdot(r)$ is relatively uniform at $\approx -0.8 \times 10^{-5} \Msolar \peryear$.  The dashed line shows the result of the inviscid formula \eqp{Mdot_CM} following \citet[][hereafter CM81]{Cassen_and_Moosman_1981} which accounts for the drag exerted by the sub-Keplerian infall on a Keplerian disk.  Its agreement with the simulation is only fair.  Much better agreement results (see Figure~\ref{fig:case}c) if we substitute the actual disk angular momentum, $\Gamma(r) \equiv j_z = u_\phi r$, into \eqp{Mdot_CM_general}, which is valid for general but steady $\Gamma(r)$.  The remaining differences with the simulation are attributed to unsteadiness of $\Gamma(r)$ and to the fact that the ratio in \eqp{Mdot_CM_general} becomes indeterminate as $r/\rc \to 1$. 

Figure~\ref{fig:case}d shows that $u_\phi \propto r^{-1}$ (angular-momentum preserving) for $r > \rc$ and is nearly Keplerian for $r < \rc$, however, as noted above there is sufficient deviation from Keplerian $u_\phi$ to account for a qualitative change in the disk $\Mdot$.

Figure \ref{fig:case}e shows the corresponding build-up of surface density.  Its radial profile has a minimum at $r/\rc \approx 0.6$, then a maximum at $\rc \approx 1$ before decreasing exponentially for $\rc < r  < \rrshock$.

Motivated by its presentation in a model fit to observations \citep{Ohashi_etal_2014} to be discussed later, Figure \ref{fig:case}f shows the midplane number density $\nHtwo$ estimated using
\be
   \nHtwo(r) = (2 \pi)^{-1/2} (\Sigma(r) / H(r)) / \mHtwo,
\ee
which assumes a Gaussian (hydrostatically balanced) density profile for $z \in [-\infty, \infty]$.  Here $\mHtwo$ is the molecular mass of H$_2$.  A more accurate result for the region $\rc < 1$ which accounts for the fact that the Gaussian is truncated by the surface shock is given later and shows that the estimate represents an under-prediction, at least for $\rc < 1$.

At the inner boundary, where a one-way diode boundary condition is applied ($u_r$ is set to zero if it is incoming), $u_r$ alternates between periods of low outflow and zero flow.  This would obviously not occur in reality.  Since the outflow is subsonic, there is in reality one incoming characteristic wave (with speed $u_r + c_\rms > 0$) carrying information from the region $r < r_\mathrm{min}$ not in the computational domain.  It is unclear how lack of this information in the simulation affects the disk within the computational domain.  Since $\Mdot(r)$ has a relatively long uniform region away from the computational boundary, one might conjecture that this should continue were it not for the computational boundary. If it did continue, the rate of loss ($\approx  0.09 \times 10^{-5} \Msolar \peryear$) is small compared rate of growth of disk mass ($\approx 0.6 \times 10^{-5} \Msolar \peryear$) and would diminish the rate of disk growth only slightly.
Recall that the CM81 inviscid result that Keplerian $u_\phi$ produces a zero $\Mdot(r)$ at $r = 0$.  If the conjecture is true that $\Mdot(r)$ should remain uniform outside the computational domain, then this can only be due to a small deviation from Keplerian angular momentum.  In reality, turbulent viscosity and outflows must also be taken into account.

\subsection{Final set of simulation plots}

\begin{figure*}
\vskip 0.5truecm
\centering
\includegraphics[width=3.3truein]{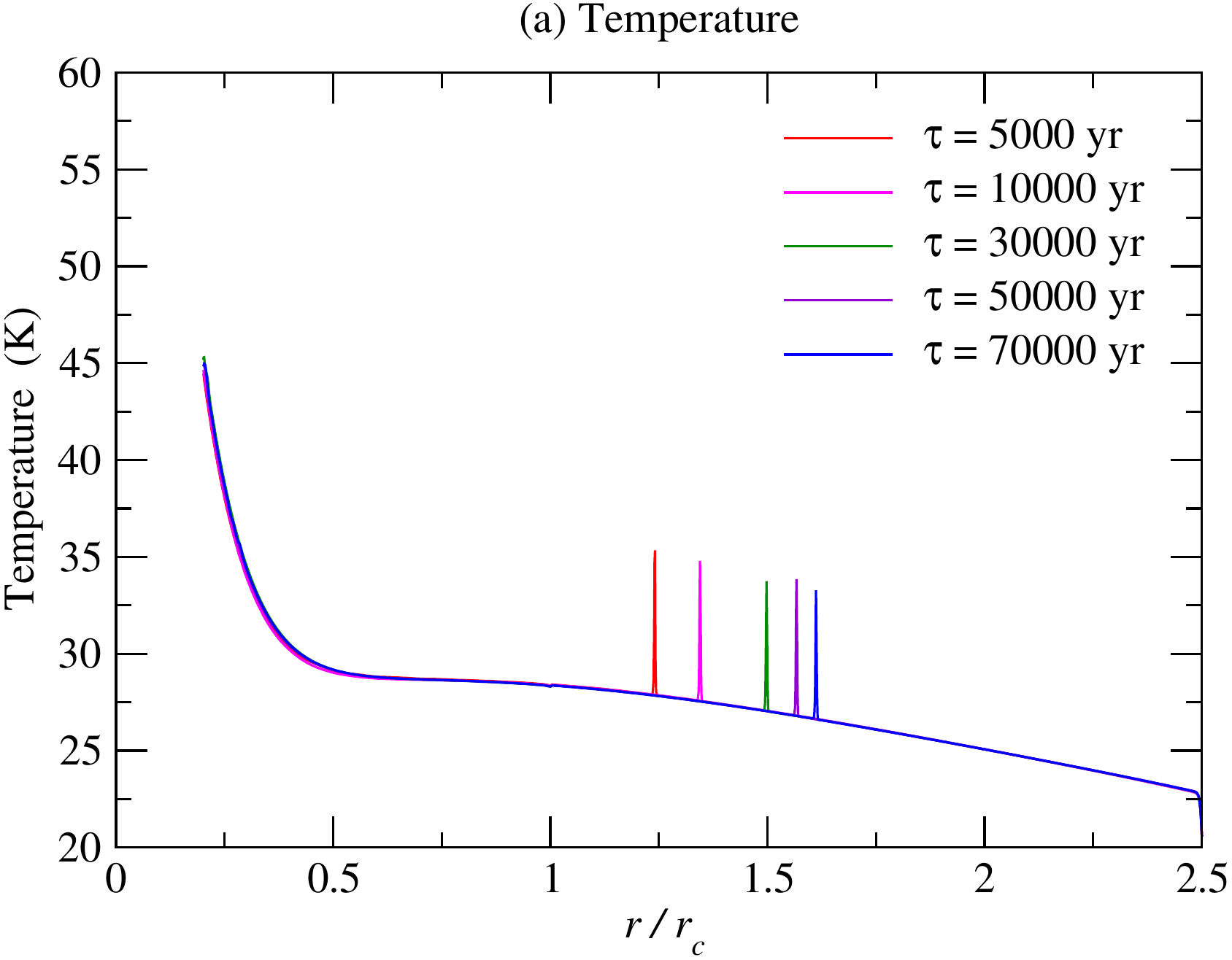}\hfill
\includegraphics[width=3.3truein]{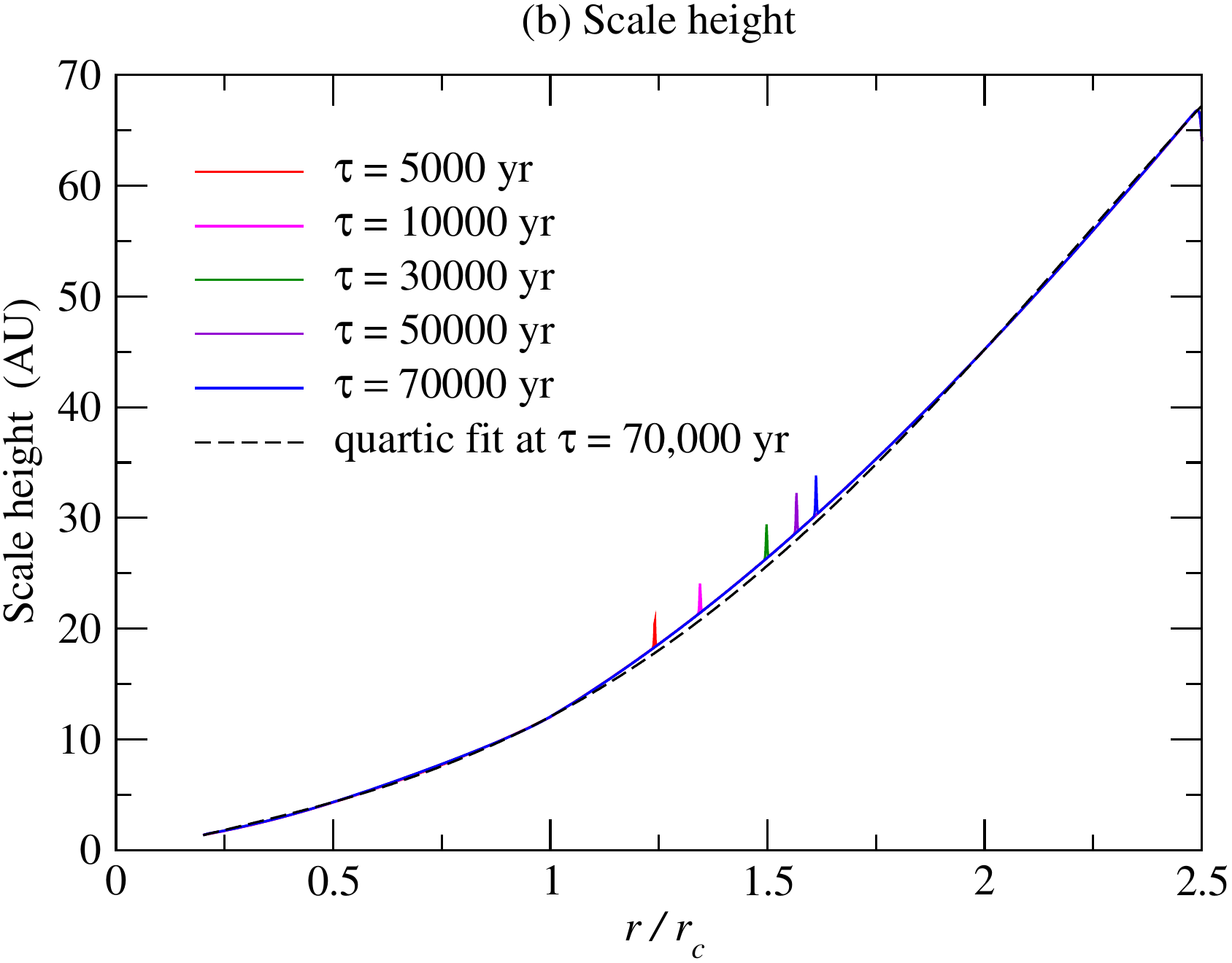}
\centering
\includegraphics[width=3.3truein]{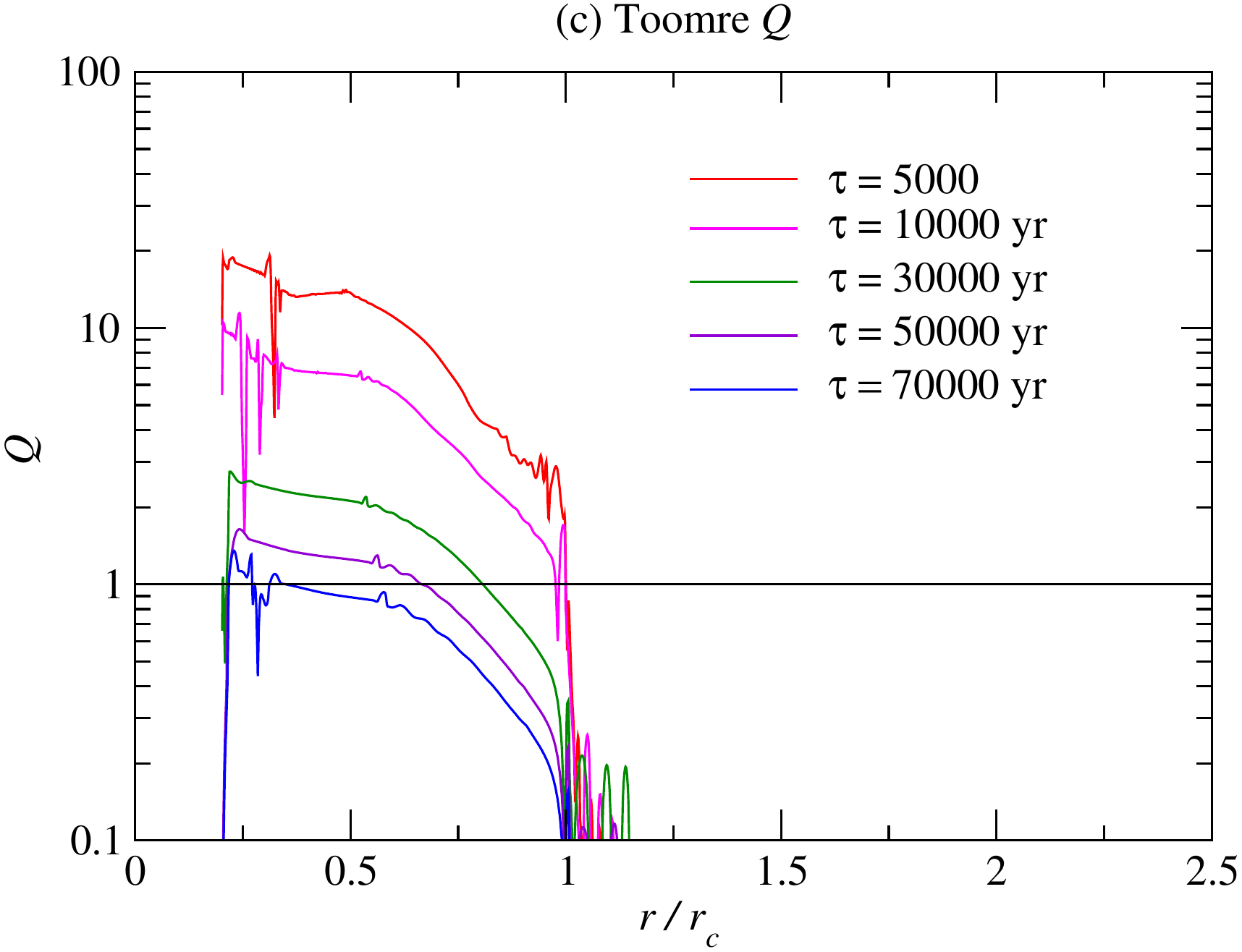}\hfill
\includegraphics[width=3.3truein]{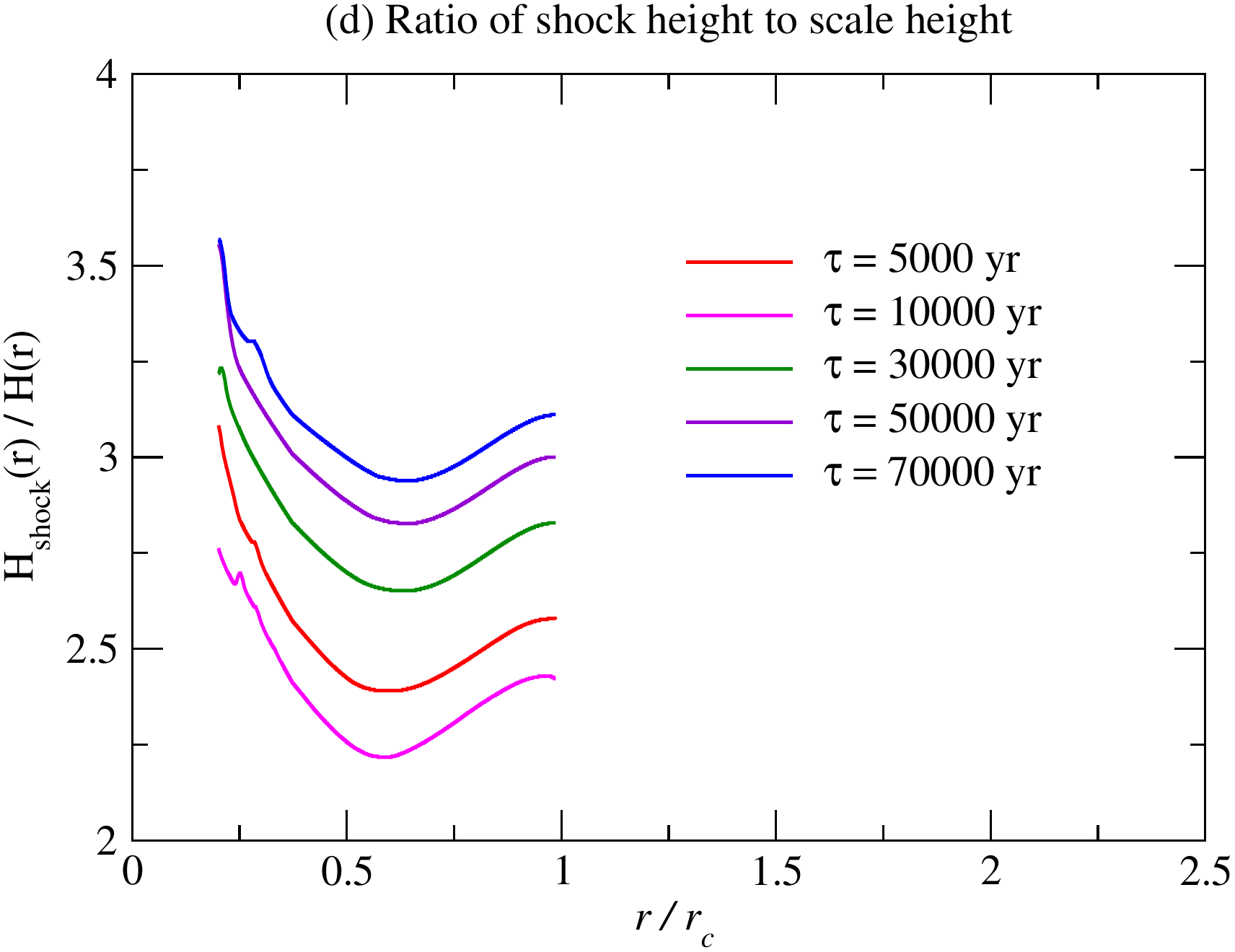}
\centering
\includegraphics[width=3.3truein]{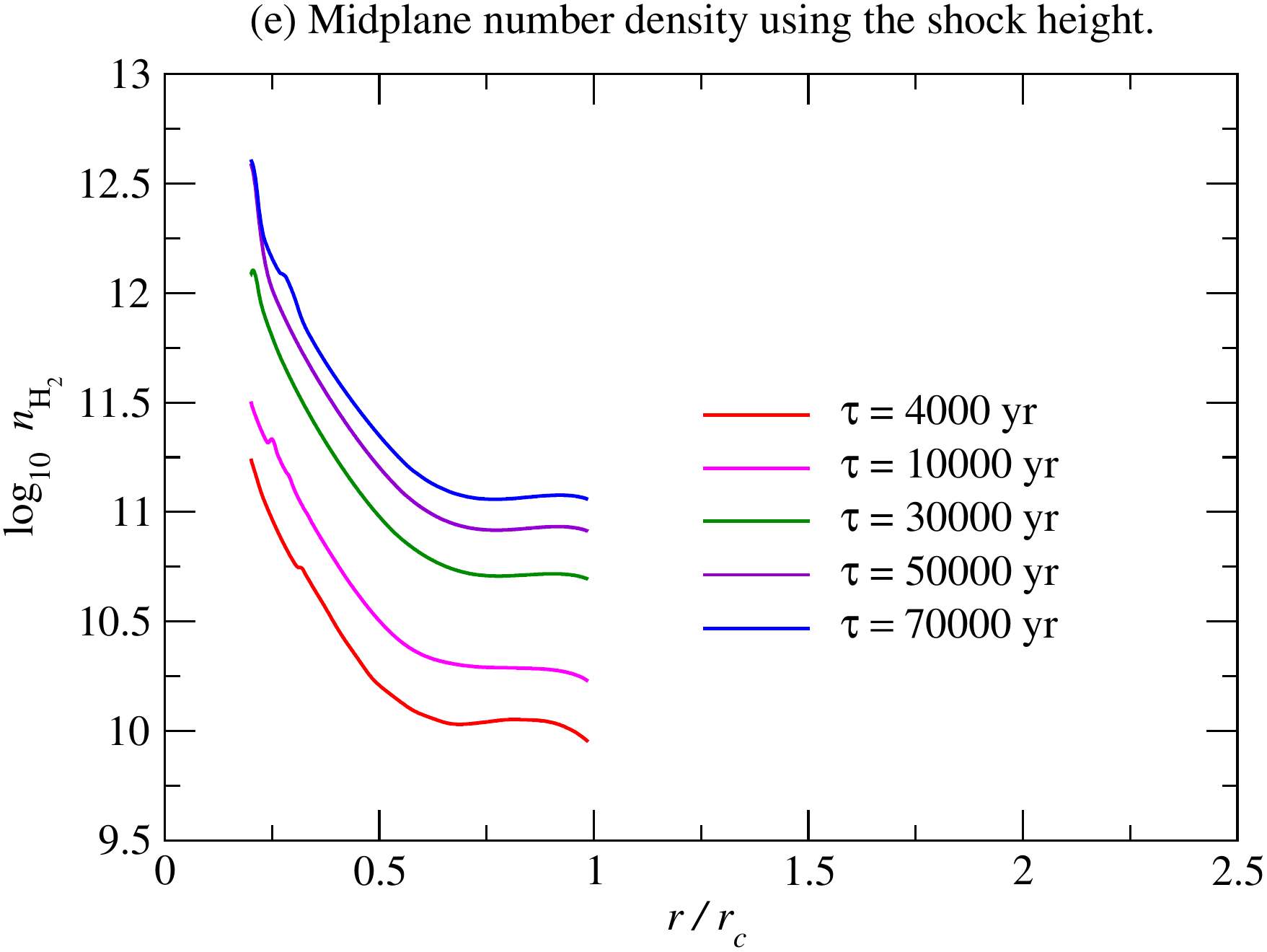}\hfill
\includegraphics[width=3.3truein]{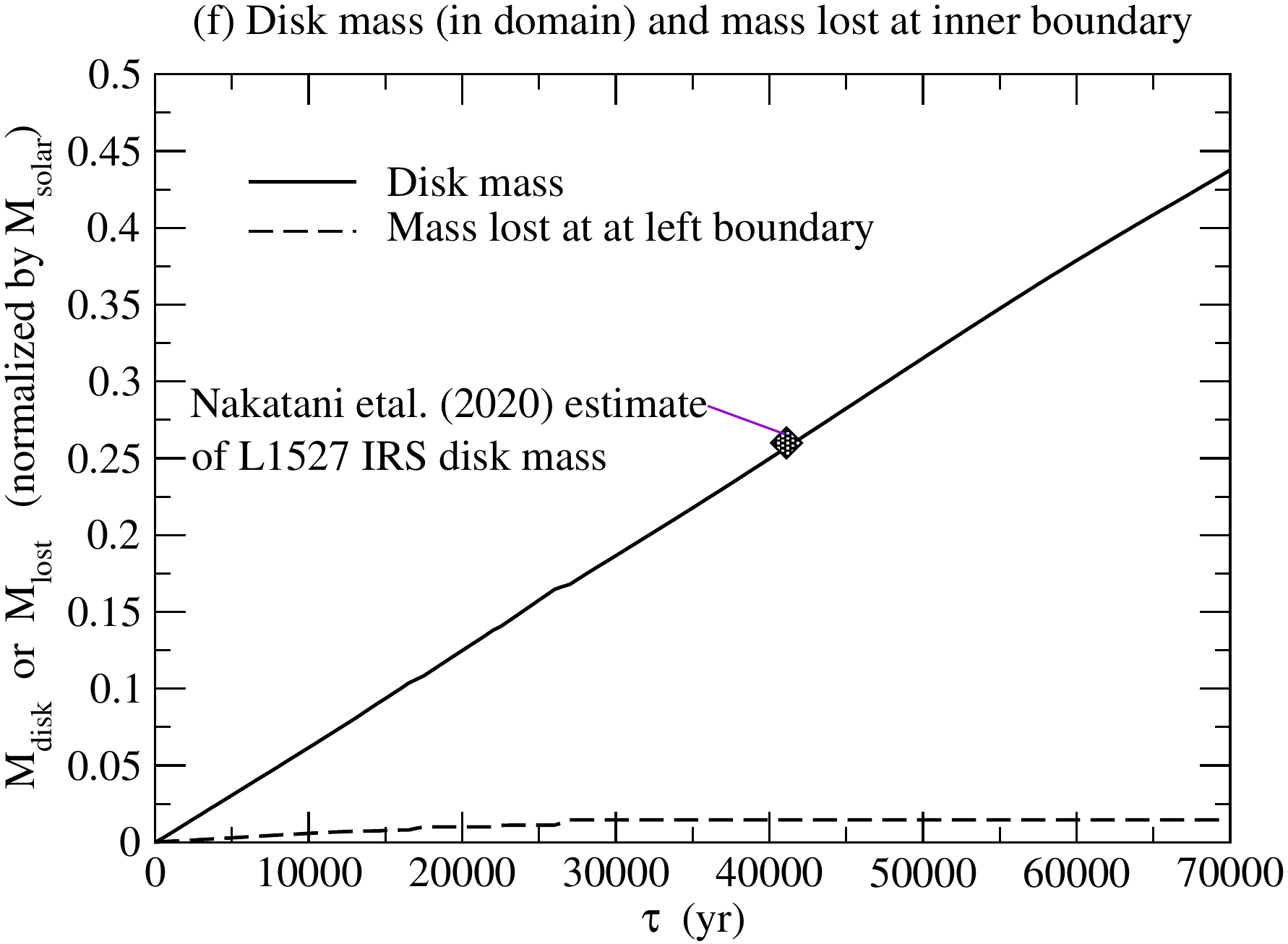}
\caption{Second set of plots for the simulation targeting L1527.}
\label{fig:concluded}
\end{figure*}

Figure \ref{fig:concluded} presents the final set of diagnostics starting with the temperature in panel (a).  The temperature has a sharp spike to $\sim 35$ K due to the radial shock, with a corresponding change in scale height $H(r)$ shown in panel (b).
This temperature rise is insufficient to account for SO desorption in the ALMA observations and will be discussed in more detail in \S\ref{sec:SO} by appealing to non-LTE (non local thermodynamic equilibrium) processes in the shock which are not captured in the simulation.

%%%%%%%%%%%%%%%%%%%%%%%%%%%%%%%% Figure 5
%%%%%%%%%%%%%%%%%%%%%%%%%%%%%%%%%%%%

Figure~\ref{fig:concluded}c shows the Toomre parameter
\be
   Q \equiv \frac{c_s \kappa}{\pi G \Sigma},
\ee
where
\be
   \kappa= \left(4 \Omega^2 + r \frac{d\Omega^2}{dr}\right)^{1/2}
\ee
is the epicyclic frequency for arbitrary rotation profiles.  $Q < 1$ implies gravitational instability (GI) to axisymmetric modes (assuming a purely rotational flow) and spiral features occur for $Q \lesssim 1.7$.   The simulation has $Q < 1$ at all $\tau$ in the region $1 < \rc < \rrshock$ where $u_r$ is small and $u_\phi \propto r^{-1}$.  Axisymmetric GI is therefore possible in this region.  For $\tau > 10000$ yr a growing portion of the region $r < \rc$ also develops $Q < 1$.  In this region $u_r$ is even smaller which renders the Toomre criterion valid.
The sharp dip in $Q$ near the inner boundary is an artifact of the boundary condition; this was confirmed by running a simulation with $\rmin/\rc =0.15$ instead of  $\rmin/\rc = 0.20$.

Figure \ref{fig:concluded}d shows the disk-surface shock height $\Hshock(r)$ calculated as described in Appendix \ref{sec:height}.  This calculation extends radially up to end of the shock where $u_z$ at the midplane becomes subsonic.  The figure shows that the ratio $\Hshock(r)/H(r)$ is always $> 1$ and increases with $\tau$.  Since the pre-shock density and vertical speed are constant with $\tau$, the density ratio across the disk-surface shock is also constant.  On the other hand, the disk density increases with $\tau$.  Hence, the height of the disk-surface shock must increase with $\tau$.

Figure \ref{fig:concluded}e shows that the corresponding midplane number density, obtained from \eqp{rhomp}, is larger than the estimate plotted in Figure~\ref{fig:case}b based on the assumption of a Gaussian density profile not truncated by a shock.  
%%%%%%%%%%%%%%%%%%%%%%%%%%%%%%%%
\begin{figure}
\centering
\includegraphics[width=3.5truein]{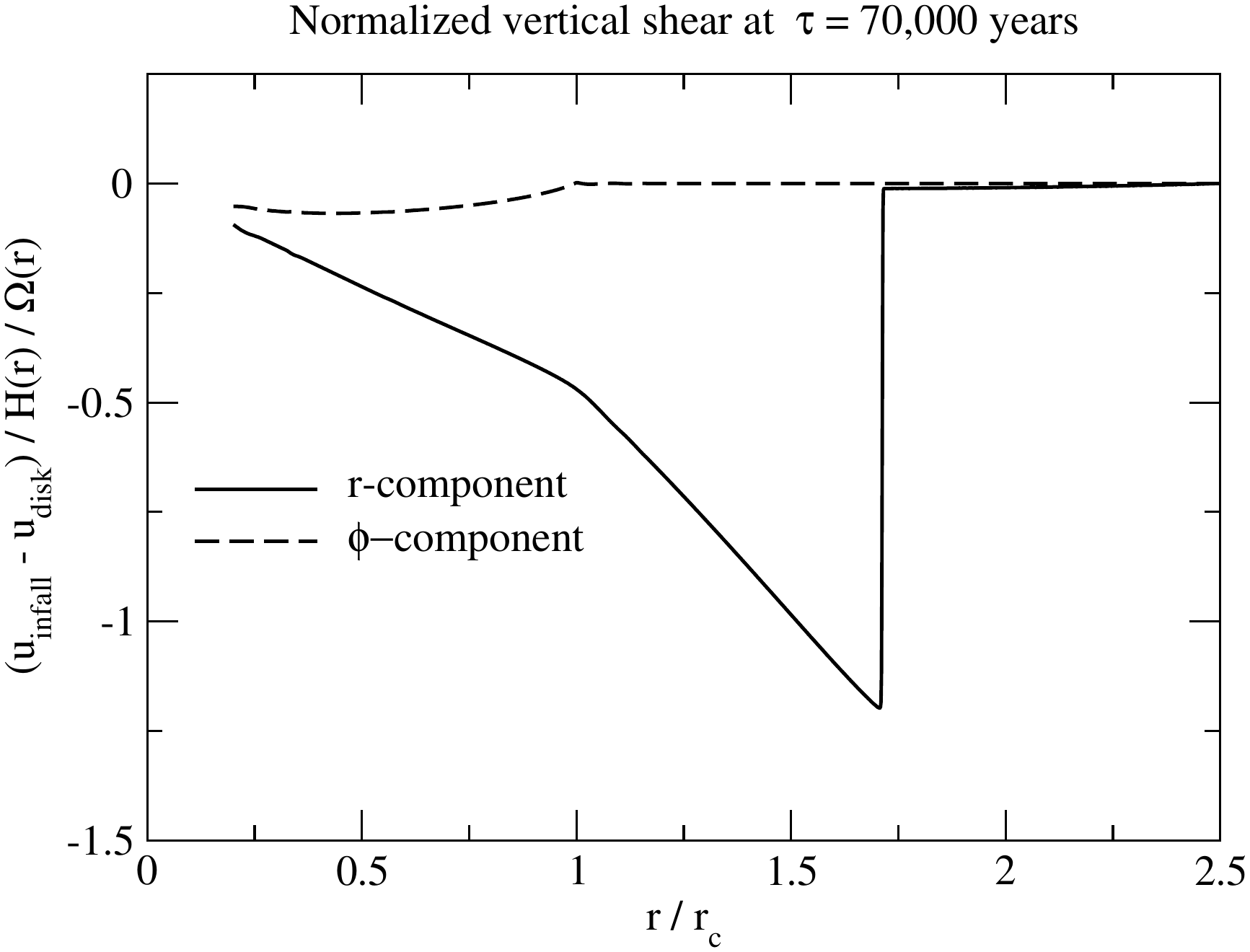}
\caption{Normalized vertical shear.}
\label{fig:v_shear}
\end{figure}
%%%%%%%%%%%%%%%%%%%%%%%%%%%%%%%%%%

Finally, in order to assess the possibility of shear driven turbulence, Figure~\ref{fig:v_shear} plots the average vertical shear 
\be
  \left(\frac{\p u}{\p z}\right)_\mathrm{ave} / \Omega(r) =
  \frac{u_\mathrm{infall} - u_\mathrm{disk}}{H(r) \Omega(r)}, \eql{v_shear}
\ee
normalized by the local orbital frequency $\Omega(r)$ for $u \to u_r$ and $u_\phi$.  Since these components of the infall velocity are continuous across the surface shock, equation \eqp{v_shear} represents the vertical shear within the disk interior.  One observes that the shear for the $r$-component is everywhere larger than for the $\phi$-component and its negative sign means that the negative $u_r$ velocity at the top of the disk is more negative than at the midplane.  Therefore $\p u_r / \p u_z$ shear driven turbulence remains possibility.

The negative sign of the $\phi$-component reflects the fact that $u_\phi$ is sub-Keplerian at the top of the disk and nearly Keplerian at the midplane.

%%%%%%%%%%%%%%%%%%%%%%%%%%%%
\section{Comparison to L1527 ALMA observations} 
\label{sec:comparison}

Except for the placement of the radial shock at the centrifugal radius in the observations of \citet[][henceforth Sa14a]{Sakai_etal_2014_Nature}, many of the simulation results are in qualitative agreement with observations of the disk around L1527 IRS.

\subsection{Comparison with Sakai etal. (2014a)}

Sa14a detect a region of high SO emission suggestive of a ring at $r = 100$ au; see Figure 1d in Sa17.  It is thought that a radial shock is present at this location that sublimates SO from icy grains.  Figure~\ref{fig:concluded}a shows that the temperature rises to $\approx 35$ K after the radial shock and quickly cools back down 30 K; this is insufficient for SO sublimation whose required temperature is 50--60 K (Sa17).  Using the \textsc{radex} code, Sa17 infer a post-shock temperature of $194 (^{+146}_{-60})$ K and an envelope temperature of $29(^{+26}_{-11})$ K (their pg. L80).
To account for SO sublimation, in \S\ref{sec:SO} we will follow \cite{Aota_etal_2015} and consider non-LTE effects in a thin post-shock layer.

In the simulation, after the shock, $T$ drops to the pre-shock temperature which is consistent with the freeze-out of SO inferred from the observations.  Further inward, the temperature rises as the star is approached which is consistent with the reappearance of SO emission in observations.

Finally, and crucially, Sa14a place the radial shock at the centrifugal barrier which is at half the centrifugal radius: $\rrshock = 0.5\rc$.  On the other hand, the simulation gives $\rrshock \approx 1.5\rc$.  This discrepancy is addressed in \S\S \ref{sec:argument} and \ref{sec:radial_infall_case}.

\subsection{Comparison with Ohashi etal. (2014)}

\begin{figure}
\centering
\includegraphics[width=3.5truein]{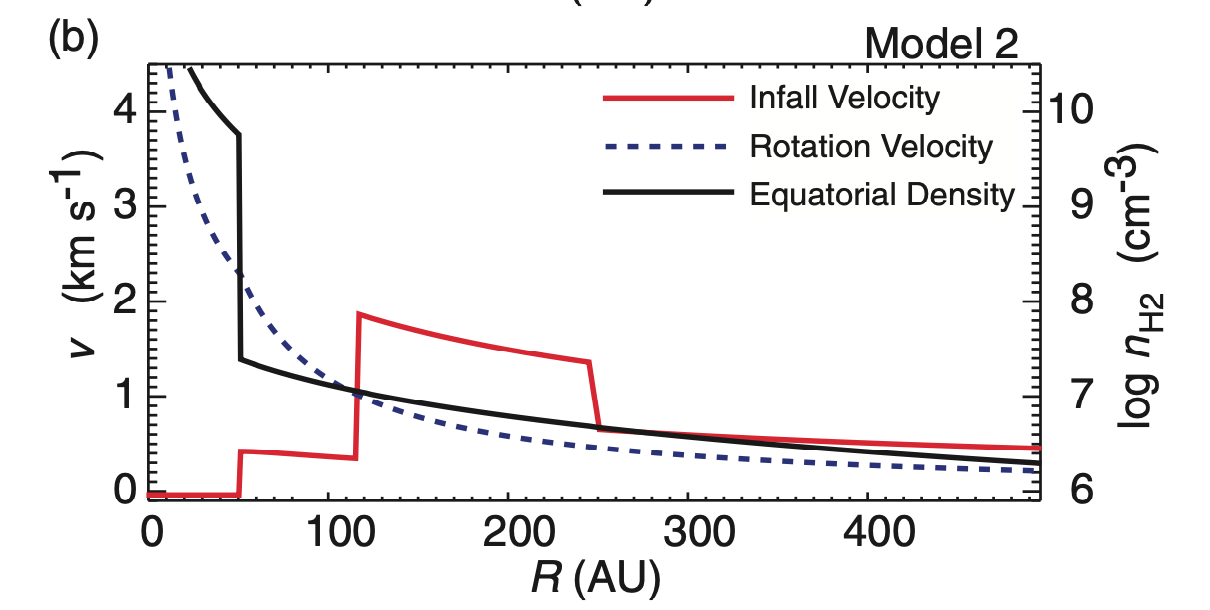}
\caption{Model 2 from \citet{Ohashi_etal_2014}.  This is their Figure 6b reproduced with permission.  In this figure, the radial shock is located at $r = 120$ au.}
\label{fig:Ohashi}
\end{figure}
\citet[][hereafter O14]{Ohashi_etal_2014} obtained a transition from an angular momentum preserving $u_\phi$ to a Keplerian $u_\phi$ at $r = 54$ au; following our previous reasoning, this is also the location of the centrifugal radius.
Figure~\ref{fig:Ohashi}, reproduced from O14, shows the result of their Model 2 constructed to fit their observations.  This model fits their line profiles much better than their Model 1.  Figure~\ref{fig:Ohashi} shows a radial shock located at $\rrshock = 120$ au, i.e., outward of the centrifugal radius, in agreement with our simulations.  Our pre-shock value of $u_r \approx -2$ km s$^{-1}$ is close to the pre-shock value in Figure~\ref{fig:Ohashi}.
One difference is that
in the O14 model, the midplane density jump does not coincide with the shock (as it should physically) but occurs at $r = \rc = 54$ au.  In the O14 model, $\log_{10}\nHtwo$ jumps from 7.4 to 9.8 across the shock while in the simulations it jumps at the shock from $\approx 6.5$ to 8 at $\tau = 40,000$ years.  The O14 post-jump value is actually closer to our peak value which occurs at $r = \rc$.

From an LVG (Large Velocity Gradient) analysis, O14 conclude that the gas temperature in the region of SO emission is $\approx 32$ K.  This is comparable to the post-shock temperature the simulation.  Since this is smaller than the SO sublimation temperature of 40--60 K, O14 suggested (following \citet{Aota_etal_2015}) that SO sublimation occurs in a thin layer behind the shock.

\subsection{Comparison with Aso etal. (2017)}

In ALMA cycle 1, \cite[][hereafter A17]{Aso_etal_2017} (see their Figure 5) identified a region of $u_\phi \sim r^{-1.22}$, i.e., approximately $u_\phi \sim r^{-1}$ which is $j_z$ preserving.  This transitions to a Keplerian $r^{-0.5}$ rotational velocity profile at $r \approx 74$ au (corrected for beam resolution).  As mentioned earlier, the location of Keplerian angular velocity in a $j_z$ preserving region gives the centrifugal radius.  Therefore, $\rc =74$ au. This is in line with our simulations which indicate a transition from $u_\phi \propto r^{-1}$ to $u_\phi \propto r^{-1/2}$ at $r = \rc$; see Figure~\ref{fig:case}d.  

Like O14, A14 obtain a density jump in their best fit model at a location that is close to the edge of the Keplerian disk.  Specifically, the edge of the Keplerian disk is at $r = 74$ au while the density jump is at $r = 84^{+16}_{-24}$ au.

\subsection{SO desorption and temperature rise across the radial shock}
\label{sec:SO}

The issue of SO desorption due to the putative radial shock of L1527 was taken up by \citet[][henceforth Ao15]{Aota_etal_2015}.  They performed 1D non-LTE calculations of shock structure accounting for gas cooling by collisions with dust, radiative emission from the dust, and SO desorption via thermal heating and sputtering.  They concluded that for our value of 2 km s$^{-1}$ for the pre-shock velocity, the required gas density for SO desorption is $\nH\gtrsim 10^9$ cm$^{-3}$; see their Figure 7.  On the other hand, the (midplane) pre-shock density in our case is $\nH = 6.3 \times 10^7$ cm$^{-3}$, a factor 16 smaller.  In the following discussion we shall assume that SO desorption nevertheless occurs with the parameters given by the simulation, and then go on to calculate the SO column density $N_\SO$ for comparison with the ALMA observations of S14a.
%
% Figure 8
\begin{figure}
\centering
\includegraphics[width=3.5truein]{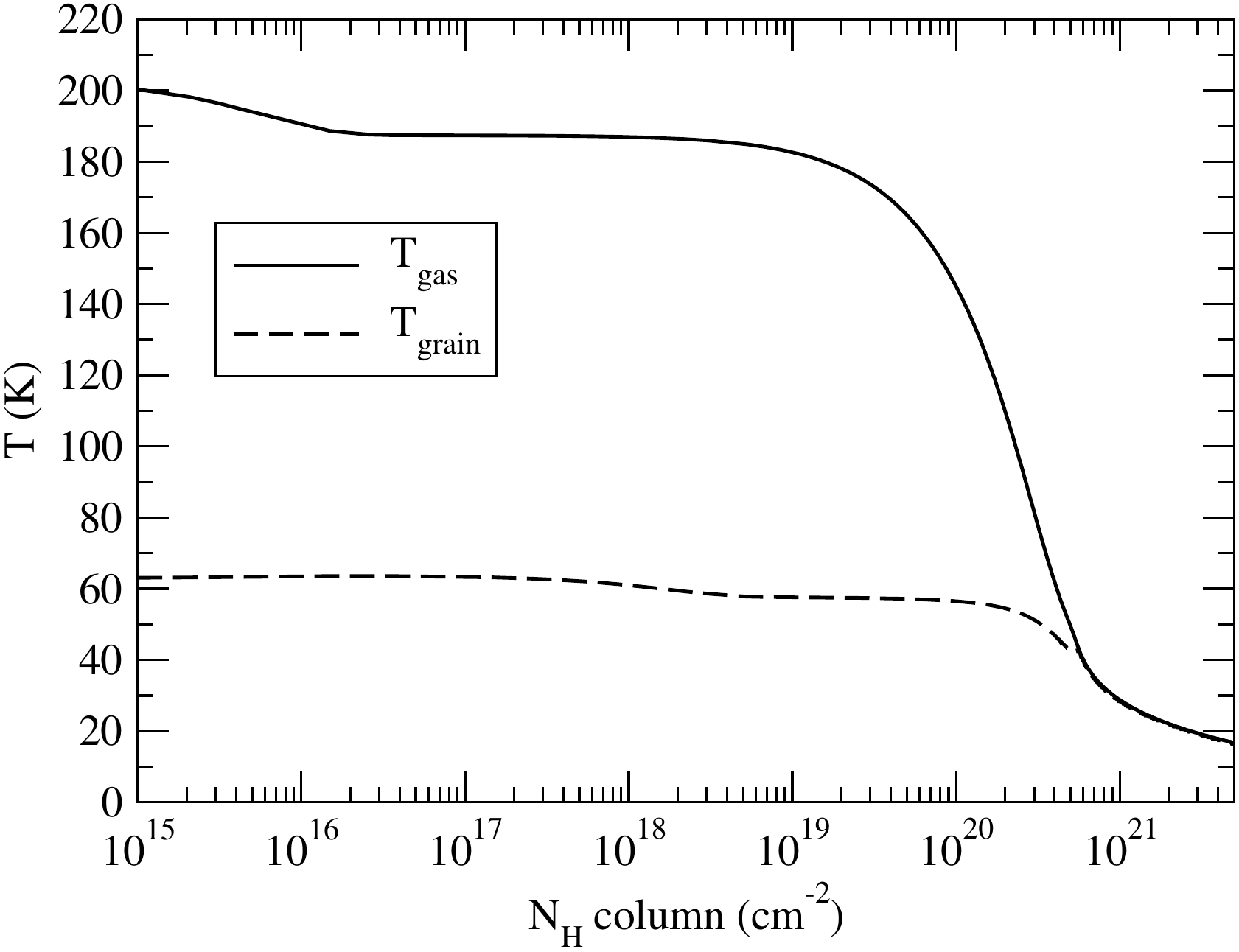}
\caption{Gas and grain temperature across the radial shock using the \citet{Neufeld_and_Hollenbach_1994} code.}
\label{fig:NH_radial_shock}
\end{figure}

The NH94 shock code does not predict desorption but otherwise can be used to calculate 1D shock structure.
We therefore ran the NH94 code to obtain the gas and dust temperature in a 1D shock with the pre-shock velocity and density given above, together with a pre-shock temperature of 25 K.  The result is shown in Figure \ref{fig:NH_radial_shock}.  
The abscissa is the $\NH$ column density measured from the shock front ($r_\mathrm{shock}$) and is defined as
\be
   \NH(r) = \int_r^{r_\mathrm{shock}} \nH(r^\prime)\, dr^\prime.
\ee
The column density $\NH(r)$ is a proxy for distance behind the shock.
One observes from Figure \ref{fig:NH_radial_shock} that
$T_\mathrm{gas} \geq100$ K for a column density of $\NH = 2.3 \times 10^{20}$ cm$^{-2}$ behind the shock and $T_\mathrm{dust} \geq 50$ K for a column density of $\NH = 3.1 \times 10^{20}$ cm$^{-2}$.
This dust temperature is in the middle of the range 40--60 K for SO sublimation \citep[e.g.][and the references therein]{Ohashi_etal_2014}.
The above values are comparable to those obtained in the non-LTE calculations of Ao15; see their Figures 3 and 4 for the gas and dust temperatures, respectively, and Figure 8 for the column density of warm gas.  Our post-shock density $\nH = 8 \times 10^8$ cm$^{-3}$ together with a warm dust column density of $\NH = 3.1 \times 10^{20}$ cm$^{-2}$ implies that the length $\ell_\warmdust = 0.026$ au.

Next, we use the time for SO to re-adsorb on to grains as given by Ao15 (their pg. 8, right column):
\be
   t_\adsorb = 10 \left(\frac{10^9}{n_\rmH}\right) \mathrm{yrs}.
\ee
From this we can obtain the distance over which re-adsorption occurs as
\be
   \ell_\adsorb = \urpost \times t_\adsorb,
\ee
where $\urpost = 0.05$ km s$^{-1}$ is the post-shock velocity and $n_\rmH = 8 \times 10^8$ cm$^{-3}$ which gives $\ell_\adsorb = 0.13$ au.
Hence the total radial length of the SO region is
\be
   \ell_\SO = \ell_\warmdust + \ell_\adsorb = 0.026\ \au + 0.13\ \au = 0.16\ \au. \eql{lSO}
\ee
Assuming an SO abundance of $10^{-7}$ following Ao15, \eqp{lSO} implies an SO column density of
\begin{align}
   N_\SO &= \left(10^{-7}\frac{\SO}{\rmH}\right) \left(8 \times 10^8 \frac{\rmH}{\cm^{3}}\right) \left(0.16\ 
   \au\right) \left(1.5 \times 10^{13} \frac{\cm}{\au}\right) \\
   &= 1.9 \times 10^{14} \cm^{-2}.
\end{align}
On the other hand, S14a state (last paragraph of their methods section) that the ``column density of SO is well constrained to be $4 \times 10^{14}$ cm$^{-2}$ for the central $1^{\prime\prime} \times 2^{\prime\prime}$ region'', which is comparable to our value.  Note that our column density is along the radial direction whereas the S14a value is along the line of sight which is purely radial only along the central line of sight.  Also note that if the observation measures both the region in front and behind the star, then our value should be multiplied by two for comparison with the observation.  This brings the two values into excellent agreement, which is fortuitous given the many uncertainties.
 
%%%%%%%%%%%%%%%%%%%%%%
\subsection{Sakai etal. (2014a) argument for equating the radial shock and centrifugal barrier locations}
\label{sec:argument}

Solving the midplane (ballistic) parabolic orbit equation \eqp{Em} for $u_r$ and determining its maximum value,   Sa14a (their Equation 6) obtain that
\be
   |u_r^\mathrm{max}| = \frac{1}{2} u_\phi^\mathrm{max}. \eql{ur_max}
\ee
We note for future use that $|u_r^\mathrm{max}|$ occurs at the centrifugal radius, $\rc$, and drops to zero at $r = \rcb = 0.5\rc$.

Equation \eqp{ur_max}, together with the fact that $u_\phi^\mathrm{max}$ occurs at the centrifugal barrier, allowed  Sa14a to determine the location of the centrifugal barrier from a position velocity (PV) plot of the envelope tracer cyclic-C$_3$H$_2$ (their Figure 2c).
Specifically, along the stellar position, the line of sight velocity is purely radial for the nearly edge-on L1527 disk.  From the PV plot, we have $u_r^\mathrm{max} = 1$ km s$^{-1}$ (line B on the red-shifted side).  On the other hand $u_\phi^\mathrm{max}$ in the envelope is seen to be $\approx 2$ km s$^{-1}$ at $r = 100$ AU.  Hence, relation \eqp{ur_max} holds and one concludes that $\rcb = 100$ au.  This coincides with the location of the radial shock (marked by SO emission) and justifies conflating the radial shock with a centrifugal barrier.
Furthermore, given the relation $\rc = 2\rcb$, Sa14a conclude that $\rc = 200$ au.  Finally, from this value they conclude that the stellar mass is $M = (0.18 \pm 0.05) \Msolar$, which is lower than the value $M = 0.45 \Msolar$ obtained by \cite{Aso_etal_2017} from the Keplerian velocity.

The Sa14a argument is indeed reasonable and cogent.  However, it should be noted that since the radial Mach number, $M_r$, upstream of the radial shock must be supersonic, the radial shock cannot exactly coincide with the centrifugal barrier where $u_r = 0$ by definition.  The values of $\rcb$ and $\rrshock$ can be close, however, provided that $M_r$ is large enough to produce the observed post-shock temperature.  

It should be pointed out that if we restrict ourselves to our envelope, i.e., the region outward of our radial shock (given that c-C$_3$H$_2$ is an envelope tracer), then our analog of \eqp{ur_max} is
\be
   |u_r^\mathrm{max}| = 1.55  u_\phi^\mathrm{max},
\ee
which is quite different from \eqp{ur_max}.  This is because our envelope is located so far outward that it does not access the higher $u_\phi$ region in the inner part of the disk.  The ratio $1.55$ was obtained from Figures~\ref{fig:case}a and (d) at $\tau = 50,000$ yr.  From panel (a), we have $|u_r^\mathrm{max}| = 2.12$ km s$^{-1}$at $r/\rc = 1.67$, just outward of the shock which is the end of the IRE.  From panel (d), from the same location we read-off that $u_\phi^\mathrm{max} = 1.37$ km s$^{-1}$.  Their ratio is 1.55.

The next sub-section shows that even when we perform a simulation with initial conditions set to the  Sa14a ballistic flow and allow only radial infall, the radial shock still positions itself at $r/\rc \sim 1.5$.

%%%%%%%%%%%%%%%%%
\subsection{Simulation with purely radial infall and  Sa14a ballistic initial velocities}
\label{sec:radial_infall_case}

\begin{figure*}
\vskip 0.2truecm
\centering
\includegraphics[width=3.1truein]{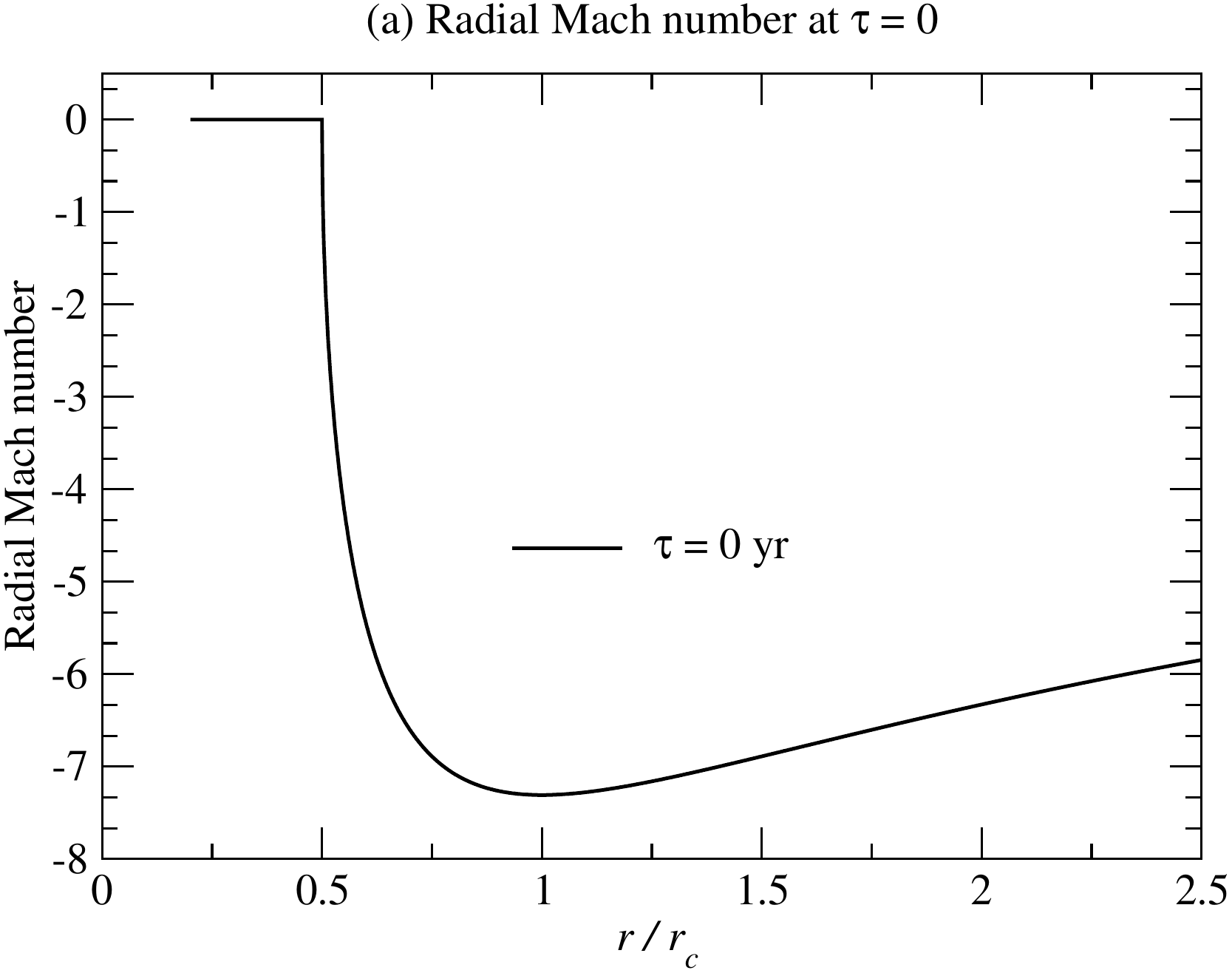}\hfill
\includegraphics[width=3.3truein]{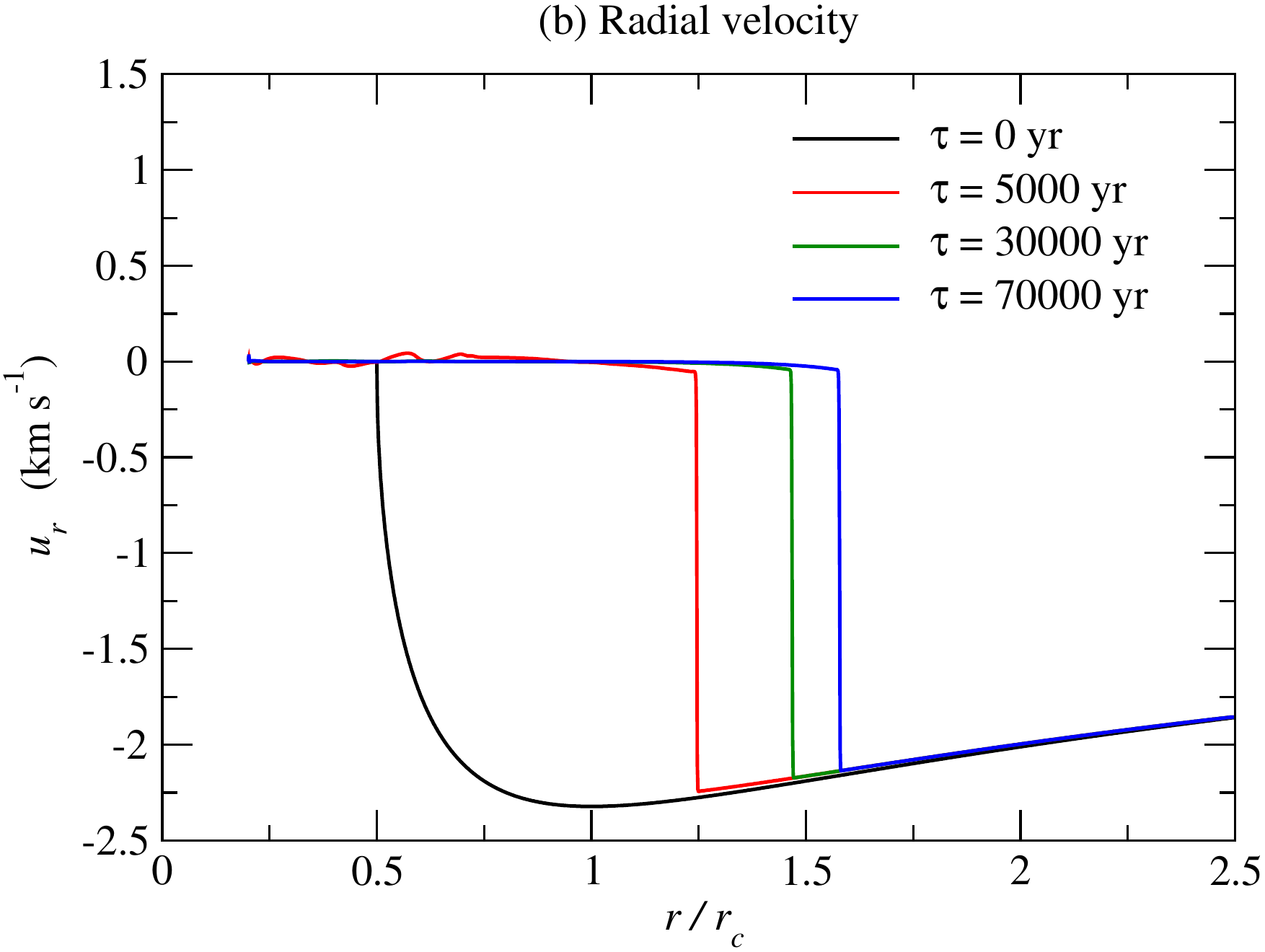}
\vskip 0.2truecm\centering
\includegraphics[width=3.3truein]{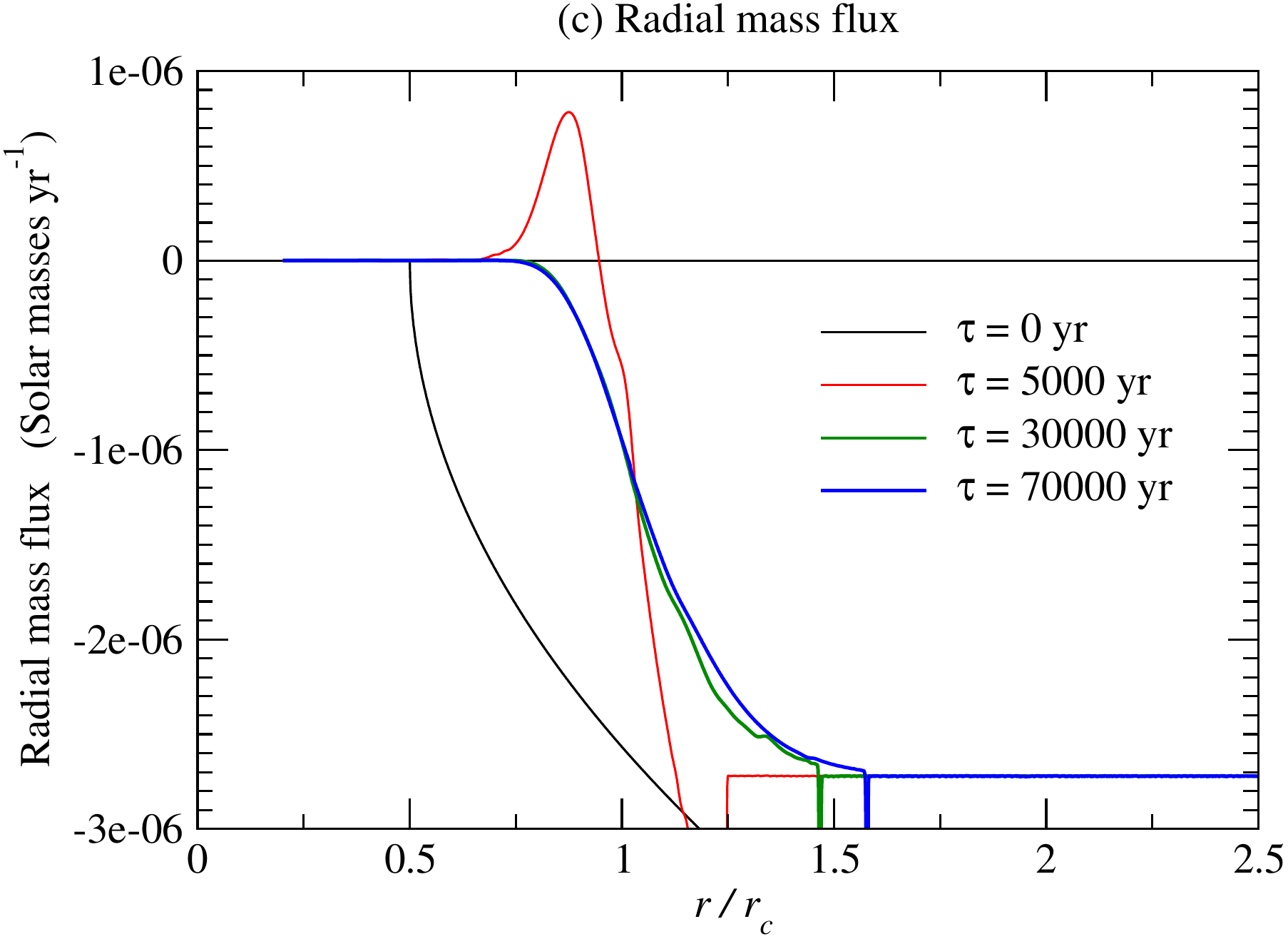}\hfill
\includegraphics[width=3.1truein]{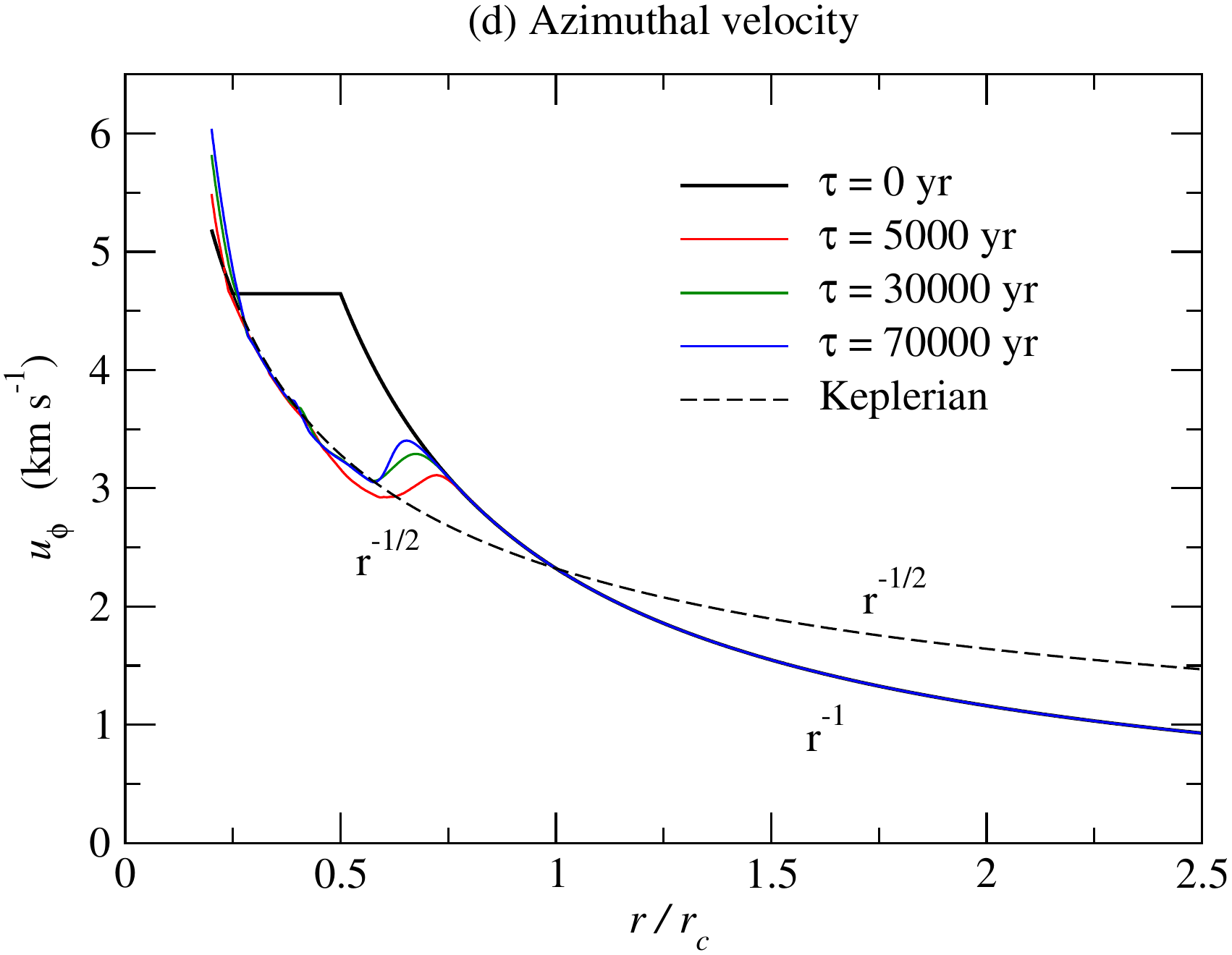}
\centering
\includegraphics[width=3.3truein]{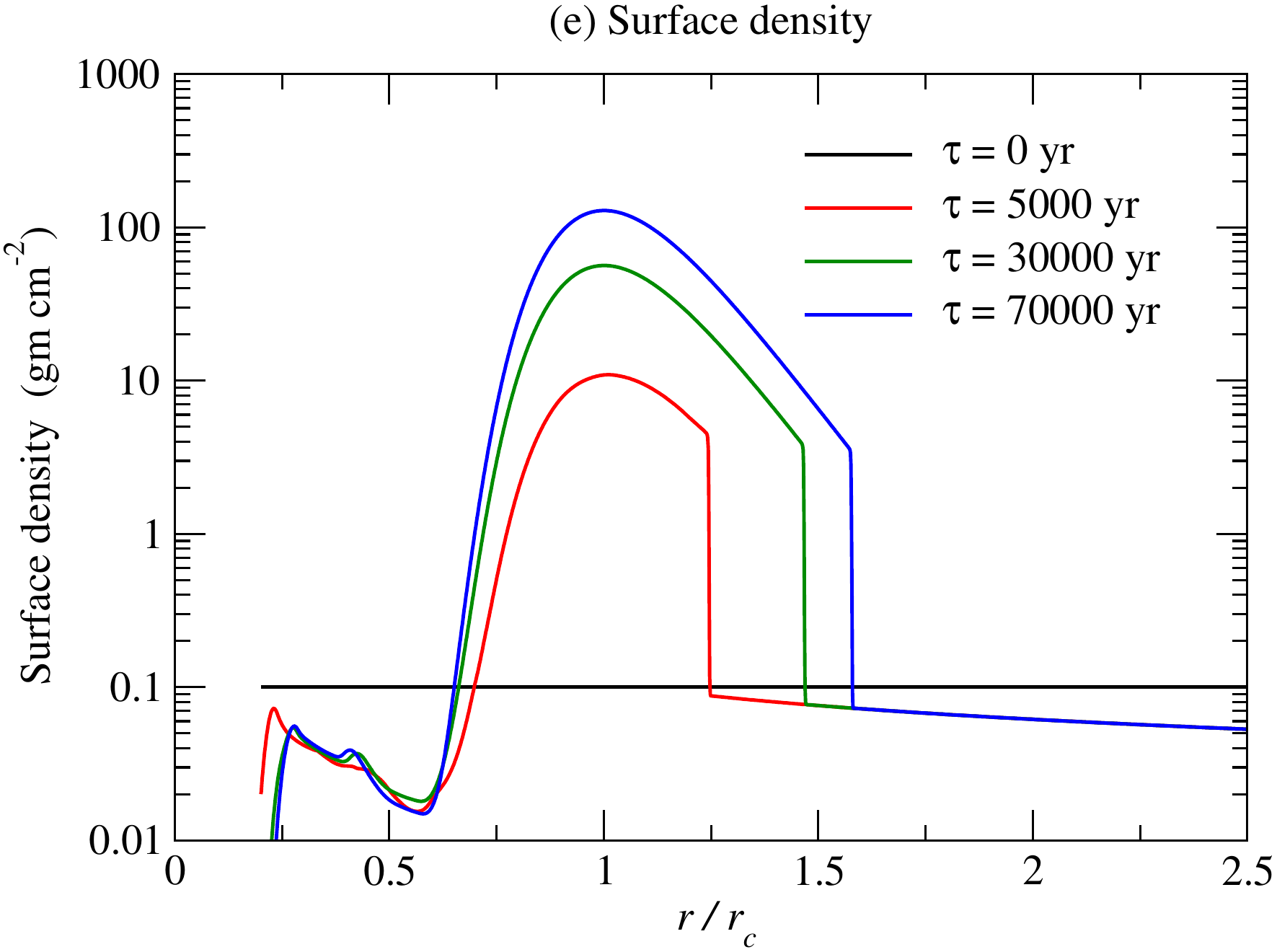}\hfill
\includegraphics[width=3.1truein]{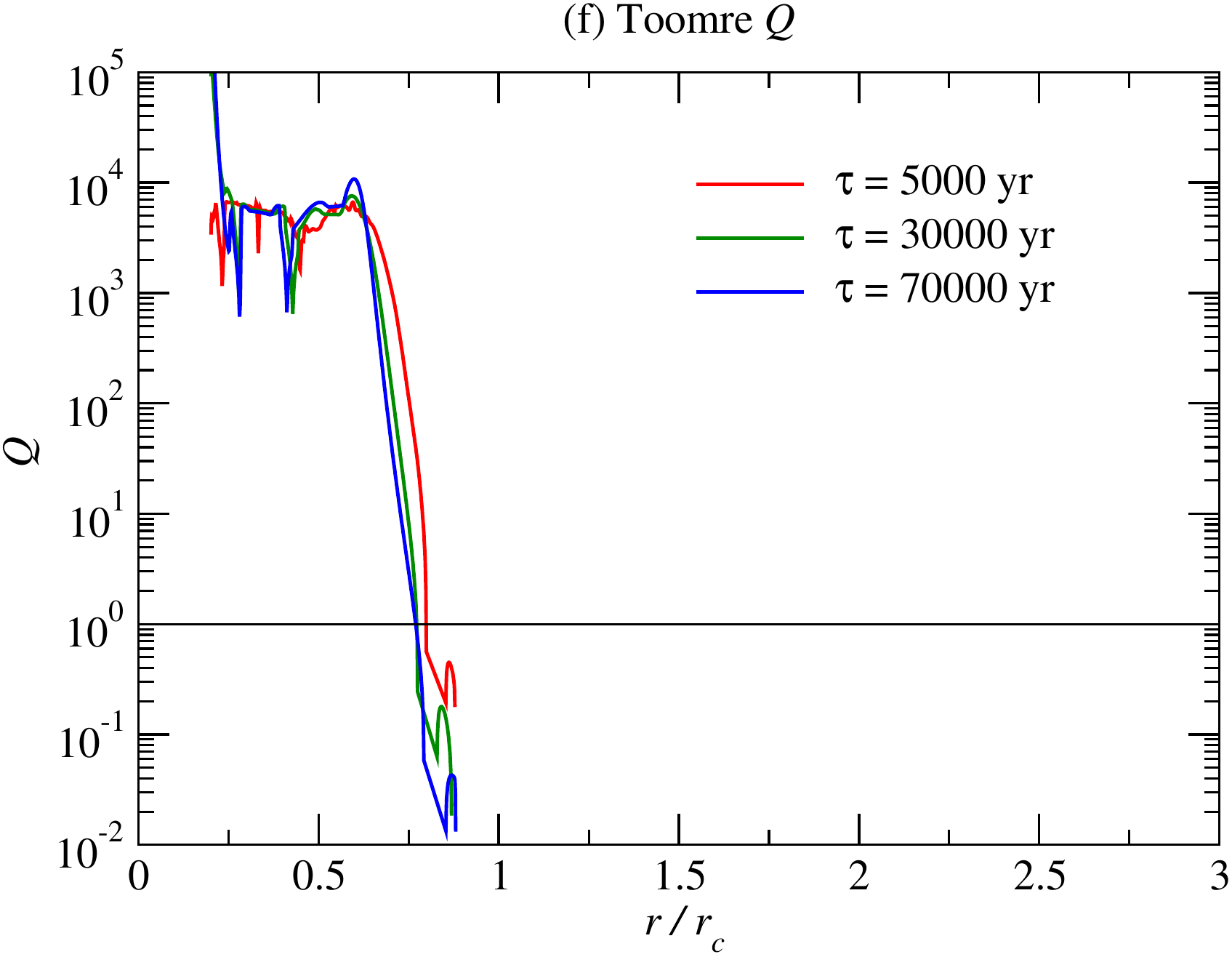}
\caption{Simulation with purely radial infall and Sa14a ballistic flow initial velocities for $r \geq \rc$.}
\label{fig:radial_infall_case}
\end{figure*}

Only radial infall is imposed and the initial velocity field is the Sa14a zero energy flow outside the centrifugal barrier ($r > \rc/2):$
\begin{align}
   u_r^\mathrm{outer}(r)     &= - \left(\frac{2GM}{r} - \frac{j_z^2}{r^2}\right)^{1/2}, \\
   u_\phi^\mathrm{outer}(r) &= j_z / r, \eql{uphi_Sakai} \\ 
   j_z & \equiv (GM \rc)^{1/2}.
\end{align}
For $r \leq \rc/2$ we would like to impose Keplerian angular velocity but it does not match \eqp{uphi_Sakai} at the interface.  To eliminate the discontinuity, a flat bridge in $u_\phi(r)$ is introduced to connect the two profiles:
\begin{align}
   u_r^\mathrm{inner}(r) &= 0, \\
   u_\phi^\mathrm{inner}(r) &=  
   \begin{cases}
      u_\phi^\mathrm{outer}(\rc/2), & \mathrm{\ if\ } u_\phi^\mathrm{Kepler}(r) < u_\phi^\mathrm{outer}(\rc/2),\\
      u_\phi^\mathrm{Kepler}(r),     & \mathrm{otherwise}.
   \end{cases}
\end{align}
Instead of this ungainly equation, it is better to consult the black curves in Figures \ref{fig:radial_infall_case}b and \ref{fig:radial_infall_case}d.  Initially $\Sigma = 0.1$ gm cm$^{-2}$, and $T = T_0 = 20.1175$ K.

We note that from their observations of the protostar Barnard 5 IRS1,
\citet{Velusamy_and_Langer_1998} suggest that an outflow can block infall from a biconical (or biparaboloidal) region such that most of the infall is radial.  This can also be observed in magnetic collapse simulations \citep[e.g.,][]{Zhao_etal_2016}.  The present simulation is also intended to mimic this situation, however, angular momentum transport by the magnetic field is not considered.

% Figure 9
%%%%%%%%%%%%%%%
%%%%%%%%%%%%%

Figure~\ref{fig:radial_infall_case}a shows the Mach number, $\MrB$, in the S14a ballistic flow which was used as the initial condition.  The interesting feature is that $\MrB$ is quite high close to the centrifugal barrier at $\rcb = 0.5\rc$, and therefore a strong shock could plausibly sit close to $\rcb$.  For instance, $\MrB = -6$ at $r/\rc = 0.64$.

Figure~\ref{fig:radial_infall_case}b shows the radial velocity.  Because in the initial condition, the inner disk has $u_r =0$, pressure waves propagate upstream in the subsonic region, and communicate to the oncoming flow the fact that $u_r$ has stagnated.  As a result, a shock is produced which greatly reduces $|u_r|$ behind it.  The shock propagates rapidly upstream up to $r/\rc \sim 1.5$, after which it propagates slowly.  This is similar to what we observed in the simulation with both vertical and radial infall.

A better view of the radial motion is obtained from the disk mass flow rate, $\Mdot$; see panel (c) and focus on times after the flow has settled down ($\tau \geq 30,000$ yr).  $\Mdot(r)$ is continuous across the shock as it should be and gradually reduces to zero (due to absence of Cassen-Moosman drag) rather than to a finite value as in the previous simulation.  As a result, the stellar mass accretion rate is zero.

Panel (d) shows that in the region of the initial $u_\phi$ bridge, the angular velocity changes to being nearly Keplerian The transition to $u_\phi \sim 1/r$ takes place at $r/\rc \approx 0.5$ rather than at $r/\rc \approx 1$ in the previous simulation.  We conclude that the larger region of Keplerian $u_\phi$ in the previous simulation is aided by the sub-Keplerian angular momentum of the vertical infall.

Panel (e) shows that the surface density peaks at $r/\rc = 1$.  This coincides with the location of inflection point in $\Mdot(r)$ as expected from the continuity equation
\be
   \frac{\p}{\p t} r\Sigma + \frac{\p\Mdot(r)}{\p r} = 0.
\ee
A local maximum with respect to $r$ in the first term implies that $\p_r^2 \Mdot(r) = 0$.  Due to the absence of both vertical infall and Cassen-Moosman drag, the inner disk is devoid of mass.  This would be quite different if magnetic and/or viscous torques were present.

We will skip showing the temperature profile since it is very similar to the previous simulation.

Finally, Toomre $Q$ in panel (f) shows that the region $\rc \gtrsim 0.75$ is susceptible to GI.  The Toomre criterion will be valid inward of the radial shock since there is very little radial infall in this region.

The main conclusion is that our attempt to mimic the Sa14a ballistic flow failed to place the radial shock anywhere near the centrifugal barrier, $\rcb = 0.5 \rc$.

Finally, we would like to raise the possibility that with the addition of an appropriate physical effect that we have not included, the radial shock can position itself on the left side of the maximum ballistic $|M_r|$ in Figure \ref{fig:radial_infall_case}, i.e, at $\rrshock < \rc$ as in the S14a observations.  We encourage investigators to consider various possibilities.

\section{Comparison with rotating core collapse simulations}\label{sec:Zhao}

\citet[][hereafter Z16]{Zhao_etal_2016} performed breakthrough simulations of magnetic cloud core collapse and found that removal of very small grains enhances ambipolar diffusion which weakens magnetic braking and allows the formation of a rotationally supported disk.  The presence of a radial shock is ubiquitous in their simulations with $\rrshock = 5$ to 40 AU depending on run parameters and time since the start of collapse.  

These authors also interpret the radial shock as a centrifugal barrier and state (their pg. 2065) that the ``presence of a centrifugal barrier naturally creates a shock by slowing down and piling up infalling materials.''  On the other hand, their radial velocity plots (e.g., their Figure 11, right hand panel) show quite the \textit{opposite}: the radial velocity speeds up as the shock is approached, just as in our simulations.  Since in the Sa14a ballistic model, the maximum radial velocity occurs at $\rc$, we conclude that the Z16 radial shock is outward of $\rc$, just as in our simulations.  This reasoning assumes that the flow into the Z16 radial shock is well described by the Sa14a ballistic model.

A shock does not form where $u_r= 0$.  Rather, a very weak shock will form where the characteristic speed $u_r + c_\rms = 0$ (for $u_r < 0$), i.e., where $|u_r|$ is sonic.  Stronger shocks form where the flow is supersonic.

A shock forms as a means of adjusting from supersonic flow upstream to given subsonic flow conditions downstream. For instance, supersonic flow approaching the nose of a bluff body forms a bow shock because no penetration conditions must be satisfied on the surface of the body.  A shock is a pile-up of acoustic waves, not because of slowing of the flow into the shock, but because upstream traveling acoustic waves from the subsonic region cannot travel into the (fast not slow) supersonic region. 

The azimuthal velocity is quite different between the Z16 and present simulations.  In Z16, $u_\phi$ is super-Keplerian in the region outward of the shock.  On the other hand, in our simulations $u_\phi$ is sub-Keplerian for $r > \rc$ which is simply a property of the UCM infall; see equation \eqp{uphi_rgtrc}.  This difference could simply be due to the fact that Z16 plot the Keplerian speed with respect to the central mass which is very small in their case.  It would be interesting to plot the Keplerian speed in Z16 taking into account the mass in the disk.

Earlier axisymmetric simulations of non-magnetic core collapse by \citet{Machida_etal_2010} also show a radial shock that is joined to a disk surface shock; for example, see their Figures 2f (density contours) and 3c (radial velocity profiles).  Again, the radial velocity increases approaching the shock and we arrive at the same conclusion as we did for the Z16 simulations.

In closing, if we have missed a physical effect that would place the shock closer to $\rcb=\rc/2$, then so have the collapse simulations.

%%%%%%%%%%%%%%%%%%%%%%%%%%%
\section{Review and comparison with previous 1D vertically integrated models} \label{sec:review}

\subsection{Models that assume Keplerian balance}
 
\citet[][CM81]{Cassen_and_Moosman_1981} pioneered the subject.  They adopt Ulrich's (\citeyear{Ulrich_1976}) infall flow field (described in Appendix \ref{sec:infall_field} and \ref{sec:infall_inward_of_rc}) and use Shu's (\citeyear{Shu_1977}) inside-out collapse solution to set the initial radius (in the parent cloud core) of infalling particles as well as the mass infall rate, $\dot{M}_0$.  CM81 apply the infall flow field evaluated at the midplane as a boundary condition to the vertically integrated thin disk equations.  The angular velocity is assumed to be Keplerian and the radial momentum equation then provides an explicit solution for $u_r$.  This is then substituted into the transport equation for surface density, leading to only one transport equation that needs to be solved.  With the exception of \cite{Stahler_etal_1994}, a similar procedure is applied in all the other works we discuss below.  Recall that in the present model, we do not assume that $u_\phi$ is Keplerian and  retain all relevant transport equations.  We do not include a turbulent viscosity because, as stated earlier, our goal is to first understand the basic state and the turbulence generating mechanisms it supports.

In CM81, solutions are obtained for both the inviscid and viscous cases with uniform turbulent viscosity.  The Keplerian assumption allows a local and analytical determination of the mass flow rate $\Mdot$ through the disk.  By local we mean that one does not need to solve a boundary value problem.  CM81 were the first to point out the existence of a star-ward accretion flow \textit{even in the absence of viscosity;} this arises because the sub-Keplerian angular momentum of the infall exerts a drag on the Keplerian flow in the disk.

\citet[][hereafter LP90]{Lin_and_Pringle_1990} have two transport equations.  The first is a diffusion equation for surface density as in accretion disk theory without infall \citep{Lynden-Bell_and_Pringle_1974}.  This diffusion equation is obtained by inserting the following expression for radial velocity:
\be
   u_r = \left(r \Sigma \Gamma^\prime\right) \frac{\p}{\p r}\left(r^3 \Sigma \nut \Omega^\prime\right), \eql{ur_LP}
\ee
into the mass conservation equation.  Here $\nut$ is the turbulent viscosity, $\Gamma \equiv u_\phi r$ is the specific angular momentum, and a prime denotes differentiation with respect to $r$.  Equation \eqp{ur_LP} is obtained from the angular momentum equation by (i) assuming that $\Gamma$ is time invariant or at least slowly varying compared to the time scale for viscous/turbulent diffusion; and (ii) neglecting Cassen-Moosman drag.  To specify $\Gamma(r)$, LP90 assume a modified Keplerian balance that accounts for self-gravity of the disk.  To specify $\nut$, LP90 include the treatment of \cite{Shakura_and_Sunyaev_1973} as well as self-gravity contributions.  The second transport equation in the LP90 model is the energy equation for temperature.

\citet[][hereafter HG05]{Hueso_and_Guillot_2005} adopt a model with a single transport equation, that for the surface density $\Sigma$.  For the radial velocity that this equation requires, they specialize \eqp{ur_LP} to Keplerian $u_\phi$:
\be
u_r = - \frac{3}{\Sigma r^{1/2}} \frac{\p}{\p r}\left(r^{1/2} \Sigma \nu_\mathrm{t}\right).
\ee
For the mass source term due to infall, HG05 do not use UCM infall flow field.  Instead, they develop a source term based on the assumption that cloud material with a given angular momentum ends up in the disk at the radius where the Keplerian disk has the same angular momentum.
\cite{Yang_and_Ciesla_2012}, whose interest is in the distribution of refractory materials, adopt a similar model for disk dynamics.

The models of \citet[][hereafter V09]{Visser_etal_2009} and \citet{Visser_and_Dullemond_2010}  are similar to HG05 but have some novel features.  (i) The mass source term from infall is evaluated at the disk surface rather than at the midplane as in all other works, including ours.  
V09 define the disk surface to be where the disk density (assumed to be in hydrostatic balance) matches the density of the infall field.  (ii) If the infall trajectory to a point on the boundary is obstructed by an outer part of the disk, V09 raise that point until it is no longer obstructed.

%%%%%%%%%%%%%%%%%%%%%%%%%
\subsection{Comparison with Stahler etal. (1994)}
\label{sec:Stahler}

The set-up of the \citet[][hereafter St94]{Stahler_etal_1994} analysis is the most similar to ours: it is also inviscid, vertically integrated, and prescribes the UCM infall boundary condition.  However, it does not include radial infall and assumes pressure-free steady flow.  For their outer disk, St94 obtain a singularity in surface density at $r/\rc = 0.345$ where $u_r \to 0$, which they interpret as a dense ring. 
They explain (their pg.~348, second column) the mechanism of its formation as being similar to that of a centrifugal barrier, namely, increase of centripetal acceleration leads to a stagnation of $u_r$.

Note that the St94 centrifugal barrier is further inward than the Sa14a barrier.  This is because vertical infall brings in inward radial momentum, and because Cassen-Moosman drag reduces $u_\phi$ which reduces the ability of a fluid element to oppose the pull of gravity.

This sub-section demonstrates that the singularity/dense ring is an artifact of their pressure-free assumption, which is equivalent to assuming that $u_r$ has infinite Mach number; this prevents upstream propagation of information.  
When finite pressure is allowed, $|u_r|$ will necessarily transition from being supersonic to subsonic before the stagnation point.  This permits information to travel upstream and invalidates the one-way pressure-free integration procedure in St94. 
Intuitively, when a subsonic flow region (in an otherwise supersonic flow) senses a stagnation of $u_r$, it sends pressure waves upstream to halt the flow via a shock rather than allow all the mass to pile-up at the stagnation point.  We show below that this is what happens when the pressure-free assumption is relaxed.  A shock is created that rapidly propagates radially outward, greatly reducing $|u_r|$ behind it.  The azimuthal velocity becomes nearly Keplerian behind the shock. 

Let us begin by showing that we reproduce St94 when their set-up is mimicked.  This is achieved by running isothermally with a uniform temperature of $T = 1$ K to approximate the pressure-free assumption, and placing the inner boundary of the computational domain at $\rmin = 0.355$, i.e., a little outward of the singularity where $|u_r|$ is still supersonic (the exit Mach number was found to be 17).  The rest of the parameters are the same as for the L1527 simulation.
%
%%% Figure 10
\begin{figure}
\centering
\includegraphics[width=3.0truein]{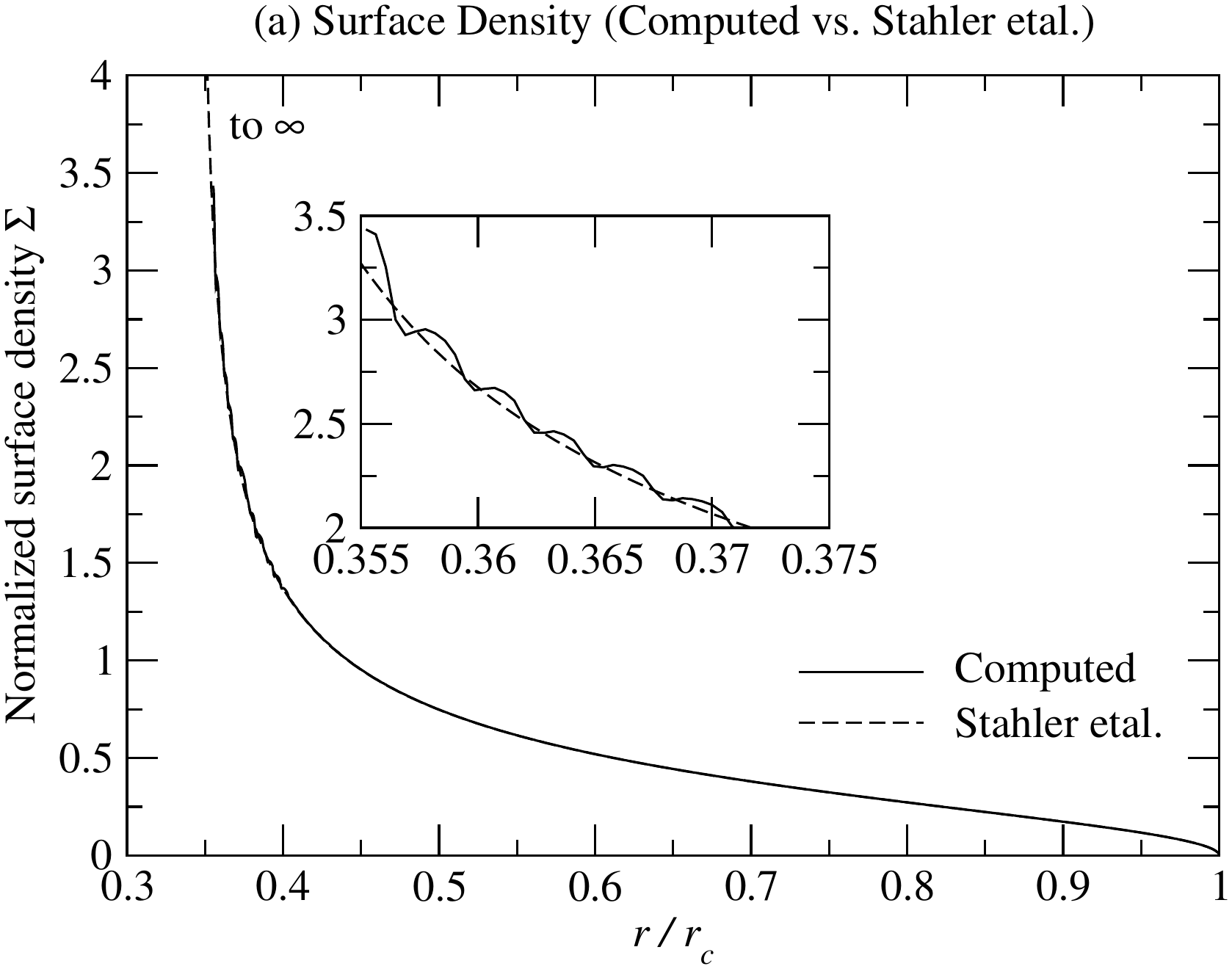}
\hfill
\includegraphics[width=3.0truein]{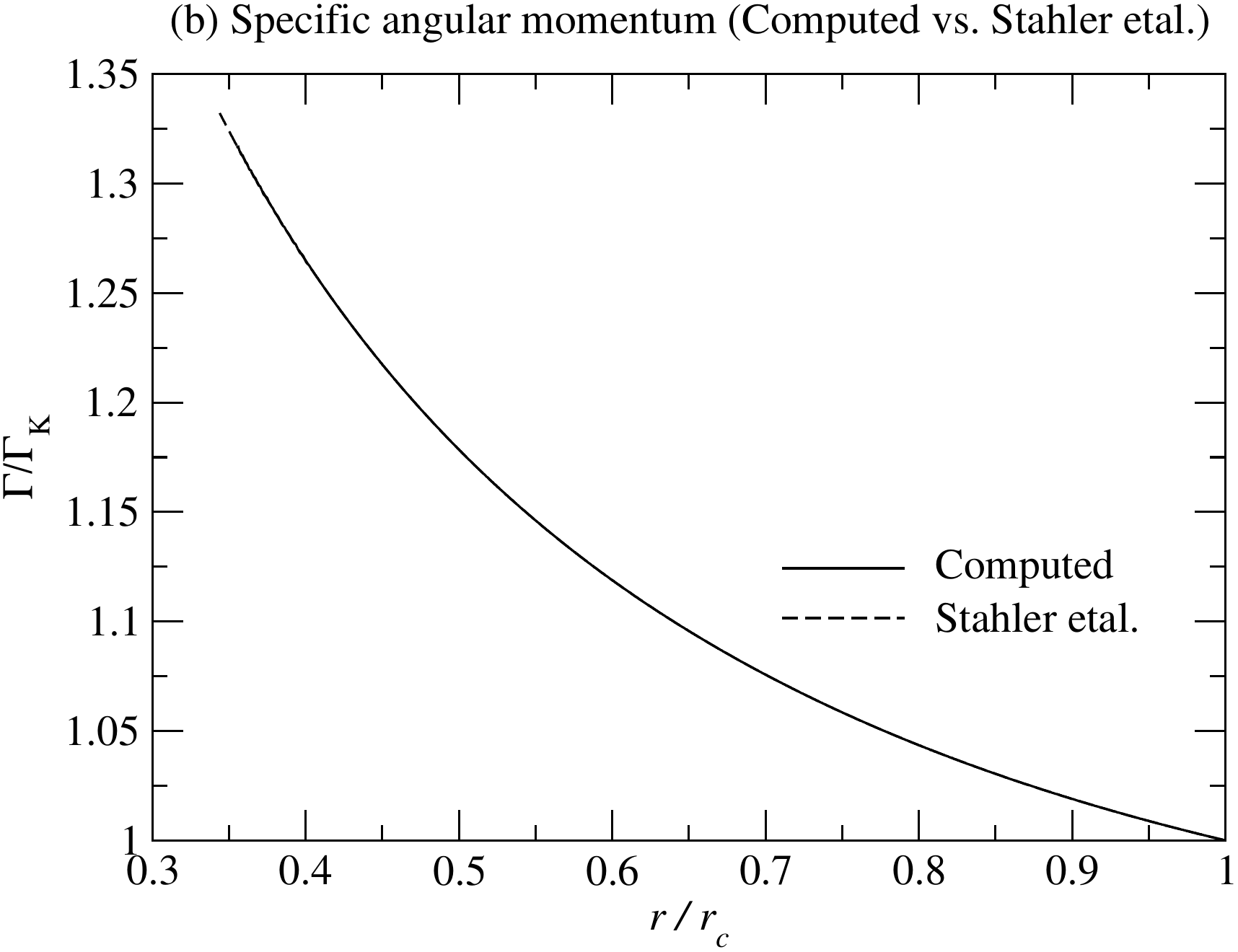}
\vskip 0.35truecm
\centering
\includegraphics[width=3.0truein]{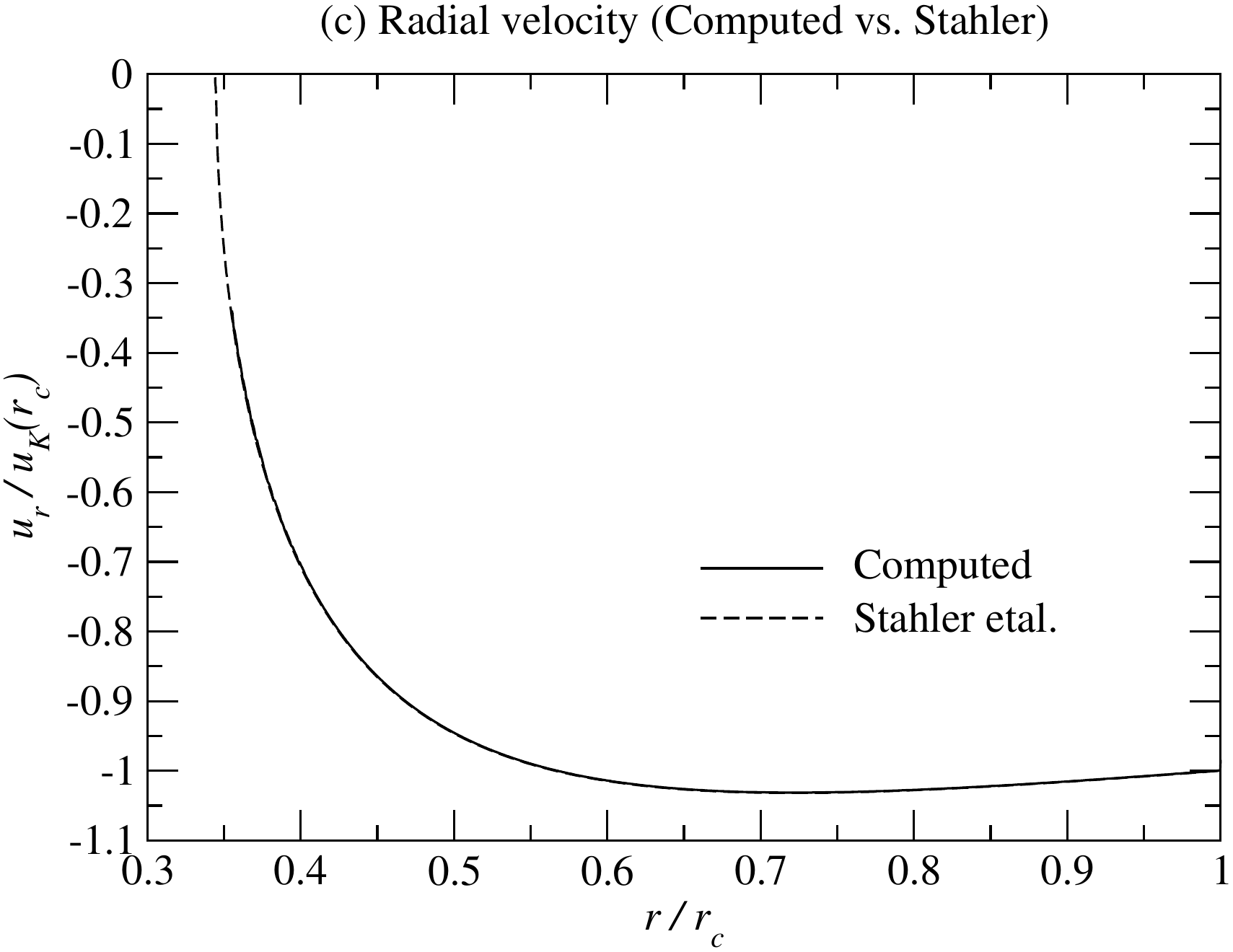}
\caption{Simulation to mimic the outer disk solution of \citet{Stahler_etal_1994}.  For this purpose the temperature is set at $1$ K and the inner boundary of the computational domain is placed at $r/\rc = 0.355$, i.e., a little outward of the St94 singularity.  See \eqp{Sigmat} for the normalization of surface density.}
\label{fig:Stahler}
\end{figure}

A steady state is reached quickly and Figure \ref{fig:Stahler} shows that the agreement between the exact and computed solution is excellent except for small numerical oscillations in $\Sigma$ where the gradient becomes large.  The ordinates are normalized in the same way as St94.  The normalized surface density is 
\be
\Sigmat \equiv \frac{\pi\rc\uK(\rc)}{\Mzdot} \Sigma, \eql{Sigmat}
\ee
where $u_\mathrm{K}(\rc)$ is the Keplerian speed at $r = \rc$.  The specific angular momentum $\Gamma \equiv r u_\phi$ is normalized by the local Keplerian value $\GammaK(r)$, and the radial velocity is normalized by $u_\mathrm{K}(\rc)$.
The angular momentum $\Gamma$ is super-Keplerian.
Note that the surface density $\Sigma \to \infty$ as $r/\rc \to 0.345$ where $u_r \to 0$.
% Figure 11
\begin{figure}
\centering
\includegraphics[width=3.0truein]{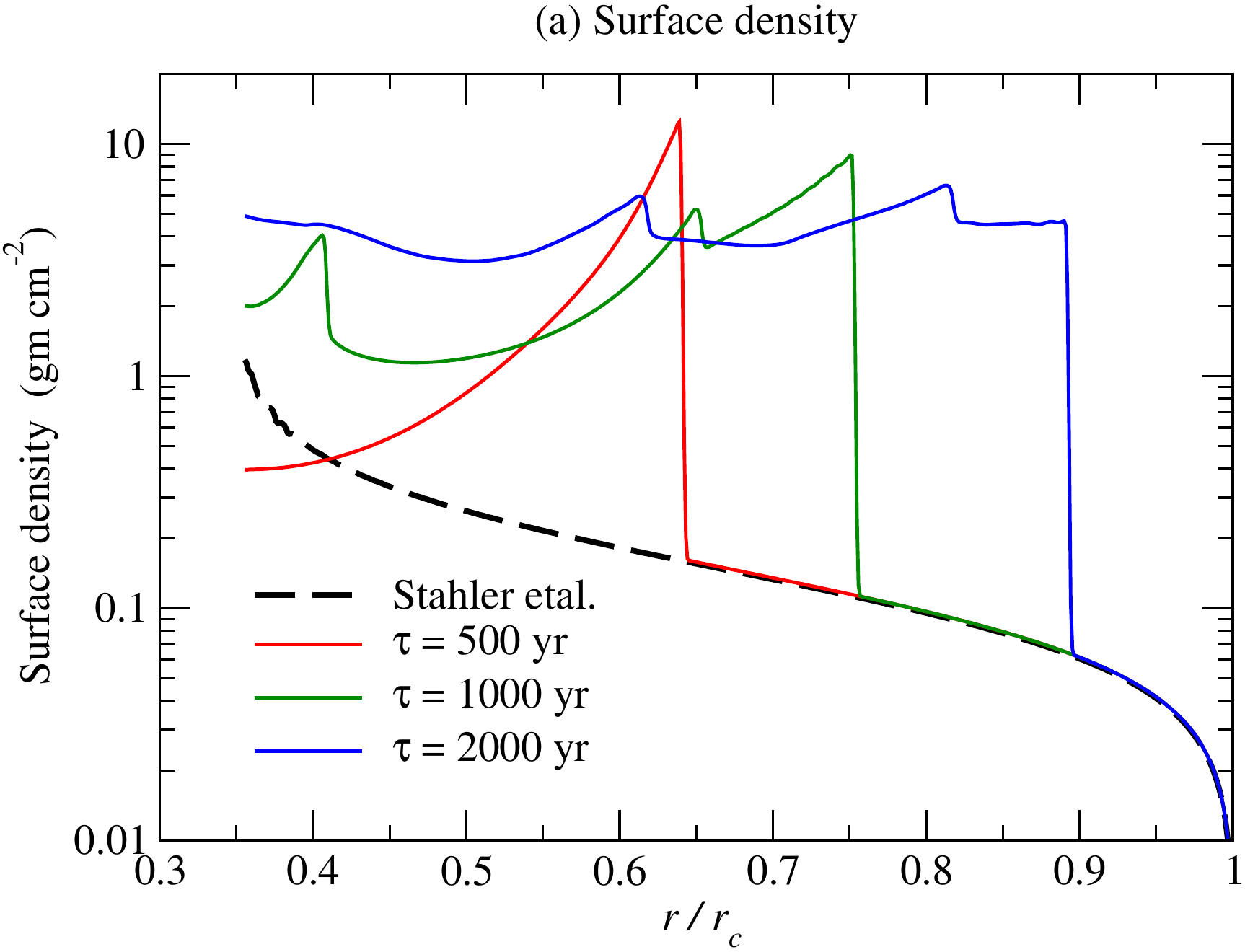}
\hfill
\includegraphics[width=3.0truein]{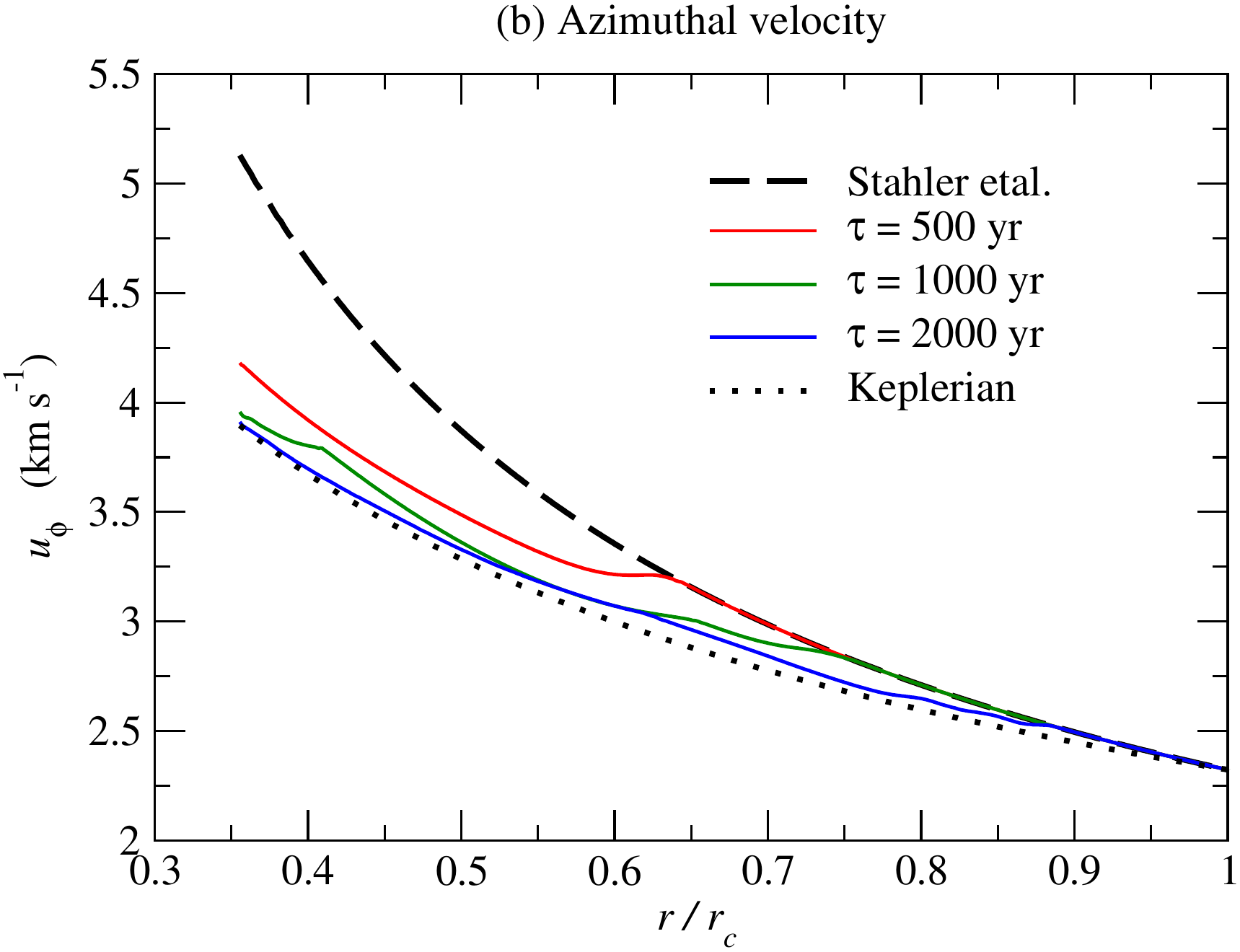}
\vskip 0.35truecm
\centering
\includegraphics[width=3.0truein]{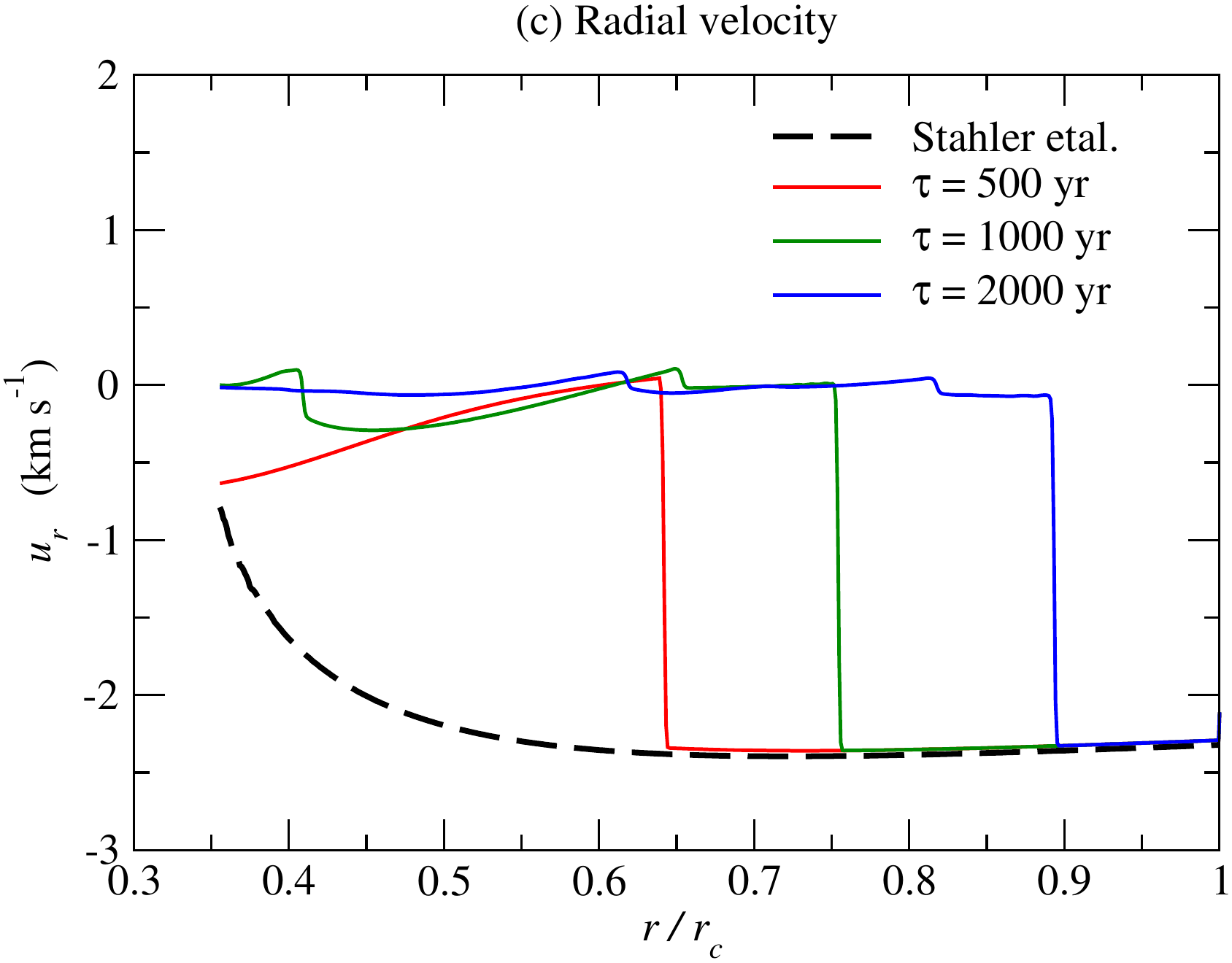}
\caption{Comparison with the theory of \citet{Stahler_etal_1994} when the pressure-free assumption is relaxed by setting $T = 20$ K.}
\label{fig:Stahler_20K}
\end{figure}

Next, the pressure-free assumption is relaxed by setting $T = 20$ K in which case $u_r$ becomes subsonic in the computational domain before the singularity in reached.  Figure \ref{fig:Stahler_20K} shows that a shock is created which propagates rapidly outward causing the solution to peel away from the St94 result.  Behind the shock, the radial velocity is greatly reduced and $u_\phi$ becomes nearly Keplerian.  If there were radial infall, the shock would come to rest where its propagation speed equals the radial infall speed.

Finally, for their inner disk ($r/\rc < 0.355$), St94 assume that the azimuthal velocity is Keplerian.  Therefore, in this region the results are very similar to the inviscid results of CM81.  In particular, there is an inward $u_r$ velocity induced by Cassen-Moosman drag as well as a contribution due to slow growth of stellar mass.  When this $u_r$ is inserted into the mass conservation equation, one obtains a surface density in the inner disk that is similar to the CM81 solution except that CM81 set the homogeneous solution to the differential equation equal to zero, while St94 do not in order to match fluxes across the dense ring.

%%%%%%%%%%%%%%%%%%%%%%%%%%%%%%%%%%%%%%%%%
\section{Closing remarks}
\label{sec:closing}

\subsection{Summary}

Simulations using a one-dimensional vertically integrated model of a protostellar disk with analytically prescribed infall \citep{Ulrich_1976, Cassen_and_Moosman_1981} were performed targeting L1527.  An implicit assumption of such models is that incoming flow quantities fully mix vertically with flow quantities in the disk at the radius of entry.  Viscous (turbulent), magnetic, and gravitational torques were not included.  One innovation compared to previous vertically integrated models is the inclusion of radial infall which is necessary for capturing the infalling rotating envelope (IRE) and a radial shock that separates the IRE from the main disk; radial infall is also a source of mass and momentum.
Another difference is our use of all the unsteady transport equations without neglect of pressure.  Most previous works assume Keplerian $u_\phi$ which eliminates many important effects such as pressure wave propagation, and reduces the problem to a single equation for surface density.

Except for a crucial disagreement, the simulation velocity and temperature profiles agree qualitatively with ALMA observations of L1527 \citep{Ohashi_etal_2014, Sakai_etal_2014_Nature, Sakai_etal_2014_ApJ, Sakai_etal_2017_MNRAS, Aso_etal_2017}.  
Specifically, the disk can be divided into three parts.  (i)  For $r > \rrshock$ (the IRE) there is radial infall with $u_\phi \propto r^{-1}$ which is angular momentum preserving.  Since the mechanical energy very closely satisfies $E = 0$ in this region,  the flow is ballistic with parabolic particle trajectories.
(ii) For $\rc < r < \rrshock$ ($\rc$ being the centrifugal radius), the radial velocity is greatly reduced by the radial shock  but the angular velocity remains $\propto r^{-1}$.  
A separate non-LTE 1D shock calculation showed that the grain temperature rises to 60 K across the shock and remains near this value in a thin region.  However, the pre-shock density was a factor of 16 smaller compared to the value necessary for SO desorption according to the study of \cite{Aota_etal_2015}.  Assuming that desorption nevertheless does take place gave a value of the SO column density close to that observed by \cite{Sakai_etal_2014_Nature}.  Given the assumptions, this agreement is very likely, fortuitous and further studies should be performed.
(iv) Inward of $\rc$, the azimuthal velocity is close to Keplerian with an accretion mass flow, $\Mdot(r)$, due to drag induced by sub-Keplerian vertical infall. $\Mdot(r)$ is nearly uniform over a significant range of $r$ and this implies subtle deviations from Keplerian $u_\phi$.

The crucial difference is that the ALMA observations of \cite{Sakai_etal_2014_Nature} are consistent with a ballistic maximum velocity relation \eqp{ur_max} and imply that the radial shock is located at $\rrshock = \rc/2$ and coincides with a ballistic centrifugal barrier, defined as the point of closest approach to the star of infalling parabolic orbits in the midplane.  On the other hand, simulations give $\rrshock \approx 1.5 \rc$.  This was the case even when we attempted to mimic the set-up of the \cite{Sakai_etal_2014_Nature} ballistic argument by including only radial infall and using the ballistic flow in the initial condition.

Since the flow just upstream of the radial shock must be sufficiently supersonic to account for the post-shock temperature inferred from observations, the radial shock cannot coincide with the centrifugal barrier where $u_r = 0$.  

Nevertheless, since the Mach number of the ballistic flow can be large close to the centrifugal barrier (Figure \ref{fig:radial_infall_case}a), it is possible that the radial shock is located close to the ballistic centrifugal barrier in reality.  There might be a physical effect lacking in our simulation that would displace the shock from the region $r > \rc$ to the region $r < \rc$ where the ballistic Mach number is comparably large.

It was pointed out that that Model 2 of \cite{Ohashi_etal_2014} fit to their observations gives a jump in radial velocity at $r = 120$ au, which is outward of their centrifugal radius at $r = 54$ au.  This is qualitatively in line with our simulations.

We noted (\S\ref{sec:Zhao}) that the radial shock in the magnetic core collapse simulations of \cite{Zhao_etal_2016} is also not in the vicinity of a centrifugal barrier and is in fact outward of $\rc$.
Therefore, if we are missing a shock-displacing physical effect, the same must be true of these simulations.

Some observers identify the shock location with a transition to Keplerian $u_\phi$.  In the simulation, however, angular momentum is preserved across the radial shock at $\rrshock \approx 1.5\rc$ and $u_\phi$ remains $\propto 1/r$ until $r = \rc$ where the transition to Keplerian flow occurs.  If the observers are correct, then there must be an angular momentum reduction process at the shock that the simulation does not include.  In favor of the simulations, it may be stated that according to \cite{Aso_etal_2017}, the transition to Keplerian flow occurs at 74 au while the radial shock is located at 100 au according to Sa14a.

Finally, the Toomre $Q$ parameter suggests gravitational instability in the outer disk where the radial velocity is small and therefore the Toomre analysis is valid.

\subsection{Suggestions for further investigation}

\begin{enumerate}
\item Is there a physical effect which neither we nor magnetic rotating-collapse simulations have included, that could move the radial shock to the left of the maximum ballistic Mach number in Figure~\ref{fig:radial_infall_case}a?
\item The vertical structure of the disk should be elucidated, first via $(r, z)$ axisymmetric simulations.  The vertically integrated model assumes instantaneous mixing of infall momentum with disk momentum.  In really, this may occur by turbulence which may also lead to angular momentum transport and disk growth.
As pointed out by \cite{Stahler_etal_1994} and shown in Figure~\ref{fig:v_shear}, there is vertical shear of the planar velocity components $u_r$ and $u_\phi$.  Axisymmetric and 3D simulations should be performed to study whether this shear leads to turbulence.  Such shear is absent when there is purely radial infall.  What is the structure of the flow and turbulence in this case?
\item Magnetic core collapse simulations have shown that outflows impinge upon and shut-off infall at some locations on the disk surface.  The physics of this process should be studied: for instance, what is the ram pressure of the outflow relative to the infall?  Is the impingement turbulent and does it affect grain growth and transport?  How does the lack of vertical infall together with an outflow affect disk structure and evolution?
\item How valid is the UCM infall model when compared with magnetic collapse simulations.  For instance, where vertical infall is present (i.e., not blocked by an outflow) is this infall sub-Keplerian as in the UCM model?  Is the flow in the IRE described by zero energy parabolic orbits?  A step toward such a comparison has been taken by \cite[][their \S5.5.3]{Lee_etal_2021} who compare the UCM mass source function with their simulations.
\item The simulations suggested gravitational instability in the outer disk and where the radial velocity is small, which renders Toomre $Q$ criterion valid.  This should be investigated even in the context of a planar $(r, \phi)$ treatment with self-gravity.

\end{enumerate}

\section*{Acknowledgements}

We thank the referee for his/her detailed and useful comments, N.~Ohashi for allowing reproduction of his figure and useful discussions, N.~Sakai for useful discussions and use of her table data, S.~Terebey for useful discussions, and D.~Hollenbach for helpful discussions and for giving us the Neufeld-Hollenbach code.
We thank P.~Estrada for catching equation typos and A.~Wray for suggesting helpful clarifications of the model in their respective internal reviews.  We thank J. Cuzzi for his encouragement and bringing to our attention several previous papers on vertically integrated models with infall.  The inspiration for this work was \cite{Stahler_etal_1994} and we thank S.  Stahler for useful discussions.

%%%%%%%%%%%%%%%%%%%%%%%%%%%%%%%%%%%%%%%%%%%%%%%%%%
\section*{Data Availability}

The data produced in this work is available upon reasonable request to the authors.


\bibliographystyle{mnras}
\bibliography{paper}

\begin{thebibliography}{}
\makeatletter
\relax
\def\mn@urlcharsother{\let\do\@makeother \do\$\do\&\do\#\do\^\do\_\do\%\do\~}
\def\mn@doi{\begingroup\mn@urlcharsother \@ifnextchar [ {\mn@doi@}
  {\mn@doi@[]}}
\def\mn@doi@[#1]#2{\def\@tempa{#1}\ifx\@tempa\@empty \href
  {http://dx.doi.org/#2} {doi:#2}\else \href {http://dx.doi.org/#2} {#1}\fi
  \endgroup}
\def\mn@eprint#1#2{\mn@eprint@#1:#2::\@nil}
\def\mn@eprint@arXiv#1{\href {http://arxiv.org/abs/#1} {{\tt arXiv:#1}}}
\def\mn@eprint@dblp#1{\href {http://dblp.uni-trier.de/rec/bibtex/#1.xml}
  {dblp:#1}}
\def\mn@eprint@#1:#2:#3:#4\@nil{\def\@tempa {#1}\def\@tempb {#2}\def\@tempc
  {#3}\ifx \@tempc \@empty \let \@tempc \@tempb \let \@tempb \@tempa \fi \ifx
  \@tempb \@empty \def\@tempb {arXiv}\fi \@ifundefined
  {mn@eprint@\@tempb}{\@tempb:\@tempc}{\expandafter \expandafter \csname
  mn@eprint@\@tempb\endcsname \expandafter{\@tempc}}}

\bibitem[\protect\citeauthoryear{{Aota}, {Inoue}  \& {Aikawa}}{{Aota}
  et~al.}{2015}]{Aota_etal_2015}
{Aota} T.,  {Inoue} T.,   {Aikawa} Y.,  2015, \mn@doi [\apj]
  {10.1088/0004-637X/799/2/141}, 799, 141

\bibitem[\protect\citeauthoryear{{Aso} et~al.,}{{Aso}
  et~al.}{2015}]{Aso_etal_2015}
{Aso} Y.,  et~al., 2015, \mn@doi [\apj] {10.1088/0004-637X/812/1/27}, 812, 27

\bibitem[\protect\citeauthoryear{{Aso} et~al.,}{{Aso}
  et~al.}{2017}]{Aso_etal_2017}
{Aso} Y.,  et~al., 2017, \mn@doi [\apj] {10.3847/1538-4357/aa8264}, \href
  {https://ui.adsabs.harvard.edu/abs/2017ApJ...849...56A} {849, 56}

\bibitem[\protect\citeauthoryear{Audard et~al.,}{Audard
  et~al.}{2014}]{Audard_etal_2014}
Audard M.,  et~al., 2014, \mn@doi [Protostars and Planets VI]
  {10.2458/azu\_uapress\_9780816531240-ch017}

\bibitem[\protect\citeauthoryear{{Cassen} \& {Moosman}}{{Cassen} \&
  {Moosman}}{1981}]{Cassen_and_Moosman_1981}
{Cassen} P.,  {Moosman} A.,  1981, \mn@doi [\icarus]
  {10.1016/0019-1035(81)90051-8}, 48, 353

\bibitem[\protect\citeauthoryear{{Chevalier}}{{Chevalier}}{1983}]{Chevalier_1983}
{Chevalier} R.~A.,  1983, \mn@doi [\apj] {10.1086/160997}, 268, 753

\bibitem[\protect\citeauthoryear{{Chick} \& {Cassen}}{{Chick} \&
  {Cassen}}{1997}]{Chick_and_Cassen_1997}
{Chick} K.~M.,  {Cassen} P.,  1997, \mn@doi [\apj] {10.1086/303700}, 477, 398

\bibitem[\protect\citeauthoryear{{Codella} et~al.,}{{Codella}
  et~al.}{2018}]{Codella_etal_2018}
{Codella} C.,  et~al., 2018, \mn@doi [\aap] {10.1051/0004-6361/201832765}, 617,
  A10

\bibitem[\protect\citeauthoryear{{Hueso} \& {Guillot}}{{Hueso} \&
  {Guillot}}{2005}]{Hueso_and_Guillot_2005}
{Hueso} R.,  {Guillot} T.,  2005, \mn@doi [\aap] {10.1051/0004-6361:20041905},
  442, 703

\bibitem[\protect\citeauthoryear{{Imai}, {Oya}, {Sakai}, {L{\'o}pez-Sepulcre},
  {Watanabe}  \& {Yamamoto}}{{Imai} et~al.}{2019}]{Imai_etal_2019}
{Imai} M.,  {Oya} Y.,  {Sakai} N.,  {L{\'o}pez-Sepulcre} A.,  {Watanabe} Y.,
  {Yamamoto} S.,  2019, \mn@doi [\apjl] {10.3847/2041-8213/ab0c20}, 873, L21

\bibitem[\protect\citeauthoryear{Kurganov \& Tadmor}{Kurganov \&
  Tadmor}{2000}]{Kurganov_and_Tadmor_2000}
Kurganov A.,  Tadmor E.,  2000, \mn@doi [J. Comp. Phys.]
  {https://doi.org/10.1006/jcph.2000.6459}, 160, 241

\bibitem[\protect\citeauthoryear{Lee, Li, Ho, Hirano, Zhang  \& Shang}{Lee
  et~al.}{2017}]{Lee_etal_2017}
Lee C.-F.,  Li Z.-Y.,  Ho P. T.~P.,  Hirano N.,  Zhang Q.,   Shang H.,  2017,
  \mn@doi [\apj] {10.3847/1538-4357/aa7757}, 843, 27

\bibitem[\protect\citeauthoryear{{Lee}, {Charnoz}  \& {Hennebelle}}{{Lee}
  et~al.}{2021}]{Lee_etal_2021}
{Lee} Y.-N.,  {Charnoz} S.,   {Hennebelle} P.,  2021, \mn@doi [\aap]
  {10.1051/0004-6361/202038105}, 648, A101

\bibitem[\protect\citeauthoryear{{Lin} \& {Pringle}}{{Lin} \&
  {Pringle}}{1990}]{Lin_and_Pringle_1990}
{Lin} D.~N.~C.,  {Pringle} J.~E.,  1990, \mn@doi [\apj] {10.1086/169004}, 358,
  515

\bibitem[\protect\citeauthoryear{{Lynden-Bell} \& {Pringle}}{{Lynden-Bell} \&
  {Pringle}}{1974}]{Lynden-Bell_and_Pringle_1974}
{Lynden-Bell} D.,  {Pringle} J.~E.,  1974, \mn@doi [\mnras]
  {10.1093/mnras/168.3.603}, \href
  {https://ui.adsabs.harvard.edu/abs/1974MNRAS.168..603L} {168, 603}

\bibitem[\protect\citeauthoryear{Machida, ichiro Inutsuka  \&
  Matsumoto}{Machida et~al.}{2010}]{Machida_etal_2010}
Machida M.~N.,  ichiro Inutsuka S.,   Matsumoto T.,  2010, \mn@doi [\apj]
  {10.1088/0004-637x/724/2/1006}, 724, 1006

\bibitem[\protect\citeauthoryear{{Mendoza}, {Tejeda}  \& {Nagel}}{{Mendoza}
  et~al.}{2009}]{Mendoza_etal_2009}
{Mendoza} S.,  {Tejeda} E.,   {Nagel} E.,  2009, \mn@doi [\mnras]
  {10.1111/j.1365-2966.2008.14210.x}, 393, 579

\bibitem[\protect\citeauthoryear{{Nakamoto} \& {Nakagawa}}{{Nakamoto} \&
  {Nakagawa}}{1994}]{Nakamoto_and_Nakagawa_1994}
{Nakamoto} T.,  {Nakagawa} Y.,  1994, \mn@doi [\apj] {10.1086/173678}, 421, 640

\bibitem[\protect\citeauthoryear{Nakatani, Liu, Ohashi, Zhang, Hanawa,
  Chandler, Oya  \& Sakai}{Nakatani et~al.}{2020}]{Nakatani_etal_2020}
Nakatani R.,  Liu H.~B.,  Ohashi S.,  Zhang Y.,  Hanawa T.,  Chandler C.,  Oya
  Y.,   Sakai N.,  2020, \mn@doi [\apj] {10.3847/2041-8213/ab8eaa}, 895, L2

\bibitem[\protect\citeauthoryear{{Neufeld} \& {Hollenbach}}{{Neufeld} \&
  {Hollenbach}}{1994}]{Neufeld_and_Hollenbach_1994}
{Neufeld} D.~A.,  {Hollenbach} D.~J.,  1994, \mn@doi [\apj] {10.1086/174230},
  428, 170

\bibitem[\protect\citeauthoryear{Ohashi et~al.,}{Ohashi
  et~al.}{2014}]{Ohashi_etal_2014}
Ohashi N.,  et~al., 2014, \mn@doi [\apj] {10.1088/0004-637x/796/2/131}, 796,
  131

\bibitem[\protect\citeauthoryear{{Okoda}, {Oya}, {Sakai}, {Watanabe},
  {J{\o}rgensen}, {Van Dishoeck}  \& {Yamamoto}}{{Okoda}
  et~al.}{2018}]{Okoda_etal_2018}
{Okoda} Y.,  {Oya} Y.,  {Sakai} N.,  {Watanabe} Y.,  {J{\o}rgensen} J.~K.,
  {Van Dishoeck} E.,   {Yamamoto} S.,  2018, \mn@doi [\apjl]
  {10.3847/2041-8213/aad8ba}, 864, L25

\bibitem[\protect\citeauthoryear{{Oya}, {Sakai}, {Lefloch},
  {L{\'o}pez-Sepulcre}, {Watanabe}, {Ceccarelli}  \& {Yamamoto}}{{Oya}
  et~al.}{2015}]{Oya_etal_2015}
{Oya} Y.,  {Sakai} N.,  {Lefloch} B.,  {L{\'o}pez-Sepulcre} A.,  {Watanabe} Y.,
   {Ceccarelli} C.,   {Yamamoto} S.,  2015, \mn@doi [\apj]
  {10.1088/0004-637X/812/1/59}, 812, 59

\bibitem[\protect\citeauthoryear{{Oya}, {Sakai}, {L{\'o}pez-Sepulcre},
  {Watanabe}, {Ceccarelli}, {Lefloch}, {Favre}  \& {Yamamoto}}{{Oya}
  et~al.}{2016}]{Oya_etal_2016}
{Oya} Y.,  {Sakai} N.,  {L{\'o}pez-Sepulcre} A.,  {Watanabe} Y.,  {Ceccarelli}
  C.,  {Lefloch} B.,  {Favre} C.,   {Yamamoto} S.,  2016, \mn@doi [\apj]
  {10.3847/0004-637X/824/2/88}, 824, 88

\bibitem[\protect\citeauthoryear{{Oya} et~al.,}{{Oya}
  et~al.}{2017}]{Oya_etal_2017}
{Oya} Y.,  et~al., 2017, \mn@doi [\apj] {10.3847/1538-4357/aa6300}, 837, 174

\bibitem[\protect\citeauthoryear{{Oya} et~al.,}{{Oya}
  et~al.}{2018}]{Oya_etal_2018}
{Oya} Y.,  et~al., 2018, \mn@doi [\apj] {10.3847/1538-4357/aaa6c7}, 854, 96

\bibitem[\protect\citeauthoryear{Pudritz \& Ray}{Pudritz \&
  Ray}{2019}]{Pudritz_and_Ray_2019}
Pudritz R.~E.,  Ray T.~P.,  2019, \mn@doi [Frontiers in Astronomy and Space
  Sciences] {10.3389/fspas.2019.00054}, 6, 54

\bibitem[\protect\citeauthoryear{{Sai} et~al.,}{{Sai}
  et~al.}{2020}]{Sai_etal_2020}
{Sai} J.,  et~al., 2020, \mn@doi [\apj] {10.3847/1538-4357/ab8065}, 893, 51

\bibitem[\protect\citeauthoryear{Sakai}{Sakai}{2019}]{Sakai_Slides}
Sakai N.,  2019, Chemical Diversity and Evolution toward Protoplanetary Disks,
  Presentation Slides for ALMA2019: Science Results and Cross-Facility
  Synergies, held 14-18 October, 2019 in Cagliari, Italy.,
  \mn@doi{10.5281/zenodo.3585280}, \url {https://zenodo.org/record/3585280}

\bibitem[\protect\citeauthoryear{{Sakai} et~al.,}{{Sakai}
  et~al.}{2014a}]{Sakai_etal_2014_Nature}
{Sakai} N.,  et~al., 2014a, \mn@doi [Nature] {10.1038/nature13000}, 507, 78

\bibitem[\protect\citeauthoryear{Sakai et~al.,}{Sakai
  et~al.}{2014b}]{Sakai_etal_2014_ApJ}
Sakai N.,  et~al., 2014b, \mn@doi [\apjl] {10.1088/2041-8205/791/2/l38}, 791,
  L38

\bibitem[\protect\citeauthoryear{{Sakai} et~al.,}{{Sakai}
  et~al.}{2016}]{Sakai_etal_2016_TMC1}
{Sakai} N.,  et~al., 2016, \mn@doi [\apjl] {10.3847/2041-8205/820/2/L34}, 820,
  L34

\bibitem[\protect\citeauthoryear{{Sakai} et~al.,}{{Sakai}
  et~al.}{2017}]{Sakai_etal_2017_MNRAS}
{Sakai} N.,  et~al., 2017, \mn@doi [\mnras] {10.1093/mnrasl/slx002}, 467, L76

\bibitem[\protect\citeauthoryear{{Semenov}, {Henning}, {Helling}, {Ilgner}  \&
  {Sedlmayr}}{{Semenov} et~al.}{2003}]{Semenov_etal_2003}
{Semenov} D.,  {Henning} T.,  {Helling} C.,  {Ilgner} M.,   {Sedlmayr} E.,
  2003, \mn@doi [\aap] {10.1051/0004-6361:20031279}, 410, 611

\bibitem[\protect\citeauthoryear{{Shakura} \& {Sunyaev}}{{Shakura} \&
  {Sunyaev}}{1973}]{Shakura_and_Sunyaev_1973}
{Shakura} N.~I.,  {Sunyaev} R.~A.,  1973, \aap, 24, 337

\bibitem[\protect\citeauthoryear{Sheehan, Tobin, Federman, Megeath  \&
  Looney}{Sheehan et~al.}{2020}]{Sheehan_etal_2020}
Sheehan P.~D.,  Tobin J.~J.,  Federman S.,  Megeath S.~T.,   Looney L.~W.,
  2020, \mn@doi [apj] {10.3847/1538-4357/abbad5}, 902, 141

\bibitem[\protect\citeauthoryear{{Shu}}{{Shu}}{1977}]{Shu_1977}
{Shu} F.~H.,  1977, \mn@doi [\apj] {10.1086/155274}, 214, 488

\bibitem[\protect\citeauthoryear{{Stahler}, {Korycansky}, {Brothers}  \&
  {Touma}}{{Stahler} et~al.}{1994}]{Stahler_etal_1994}
{Stahler} S.~W.,  {Korycansky} D.~G.,  {Brothers} M.~J.,   {Touma} J.,  1994,
  \mn@doi [\apj] {10.1086/174489}, 431, 341

\bibitem[\protect\citeauthoryear{{Terebey}, {Shu}  \& {Cassen}}{{Terebey}
  et~al.}{1984}]{Terebey_etal_1984}
{Terebey} S.,  {Shu} F.~H.,   {Cassen} P.,  1984, \mn@doi [\apj]
  {10.1086/162628}, 286, 529

\bibitem[\protect\citeauthoryear{{Tsukamoto}}{{Tsukamoto}}{2016}]{Tsukamoto_2016}
{Tsukamoto} Y.,  2016, \mn@doi [\pasa] {10.1017/pasa.2016.6}, 33, e010

\bibitem[\protect\citeauthoryear{{Ulrich}}{{Ulrich}}{1976}]{Ulrich_1976}
{Ulrich} R.~K.,  1976, \mn@doi [\apj] {10.1086/154840}, 210, 377

\bibitem[\protect\citeauthoryear{{Velusamy} \& {Langer}}{{Velusamy} \&
  {Langer}}{1998}]{Velusamy_and_Langer_1998}
{Velusamy} T.,  {Langer} W.~D.,  1998, \mn@doi [\nat] {10.1038/33624}, 392, 685

\bibitem[\protect\citeauthoryear{{Visser} \& {Dullemond}}{{Visser} \&
  {Dullemond}}{2010}]{Visser_and_Dullemond_2010}
{Visser} R.,  {Dullemond} C.~P.,  2010, \mn@doi [\aap]
  {10.1051/0004-6361/200913604}, 519, A28

\bibitem[\protect\citeauthoryear{{Visser}, {van Dishoeck}, {Doty}  \&
  {Dullemond}}{{Visser} et~al.}{2009}]{Visser_etal_2009}
{Visser} R.,  {van Dishoeck} E.~F.,  {Doty} S.~D.,   {Dullemond} C.~P.,  2009,
  \mn@doi [\aap] {10.1051/0004-6361/200810846}, 495, 881

\bibitem[\protect\citeauthoryear{{Yang} \& {Ciesla}}{{Yang} \&
  {Ciesla}}{2012}]{Yang_and_Ciesla_2012}
{Yang} L.,  {Ciesla} F.~J.,  2012, \mn@doi [Meteoritics and Planetary Science]
  {10.1111/j.1945-5100.2011.01315.x}, 47, 99

\bibitem[\protect\citeauthoryear{Yen et~al.,}{Yen et~al.}{2014}]{Yen_etal_2014}
Yen H.-W.,  et~al., 2014, \mn@doi [\apj] {10.1088/0004-637x/793/1/1}, 793, 1

\bibitem[\protect\citeauthoryear{{Zhao}, {Caselli}, {Li}, {Krasnopolsky},
  {Shang}  \& {Nakamura}}{{Zhao} et~al.}{2016}]{Zhao_etal_2016}
{Zhao} B.,  {Caselli} P.,  {Li} Z.-Y.,  {Krasnopolsky} R.,  {Shang} H.,
  {Nakamura} F.,  2016, \mn@doi [\mnras] {10.1093/mnras/stw1124}, 460, 2050

\bibitem[\protect\citeauthoryear{{Zhao} et~al.,}{{Zhao}
  et~al.}{2020}]{Zhao_etal_2020}
{Zhao} B.,  et~al., 2020, \mn@doi [\ssr] {10.1007/s11214-020-00664-z}, 216, 43

\bibitem[\protect\citeauthoryear{{Zhao}, {Caselli}, {Li}, {Krasnopolsky},
  {Shang}  \& {Lam}}{{Zhao} et~al.}{2021}]{Zhao_etal_2021}
{Zhao} B.,  {Caselli} P.,  {Li} Z.-Y.,  {Krasnopolsky} R.,  {Shang} H.,   {Lam}
  K.~H.,  2021, \mn@doi [\mnras] {10.1093/mnras/stab1295}, 505, 5142

\makeatother
\end{thebibliography}

%%%%%%%%%%%%%%%%%%%%%%%%%%%%%%%%%%%%%%%%%%%%%%%%%%

%%%%%%%%%%%%%%%%% APPENDICES %%%%%%%%%%%%%%%%%%%%%

\appendix

\section{Infall field}\label{sec:infall_field}

\begin{figure}
\centering
\includegraphics[width=2.8truein]{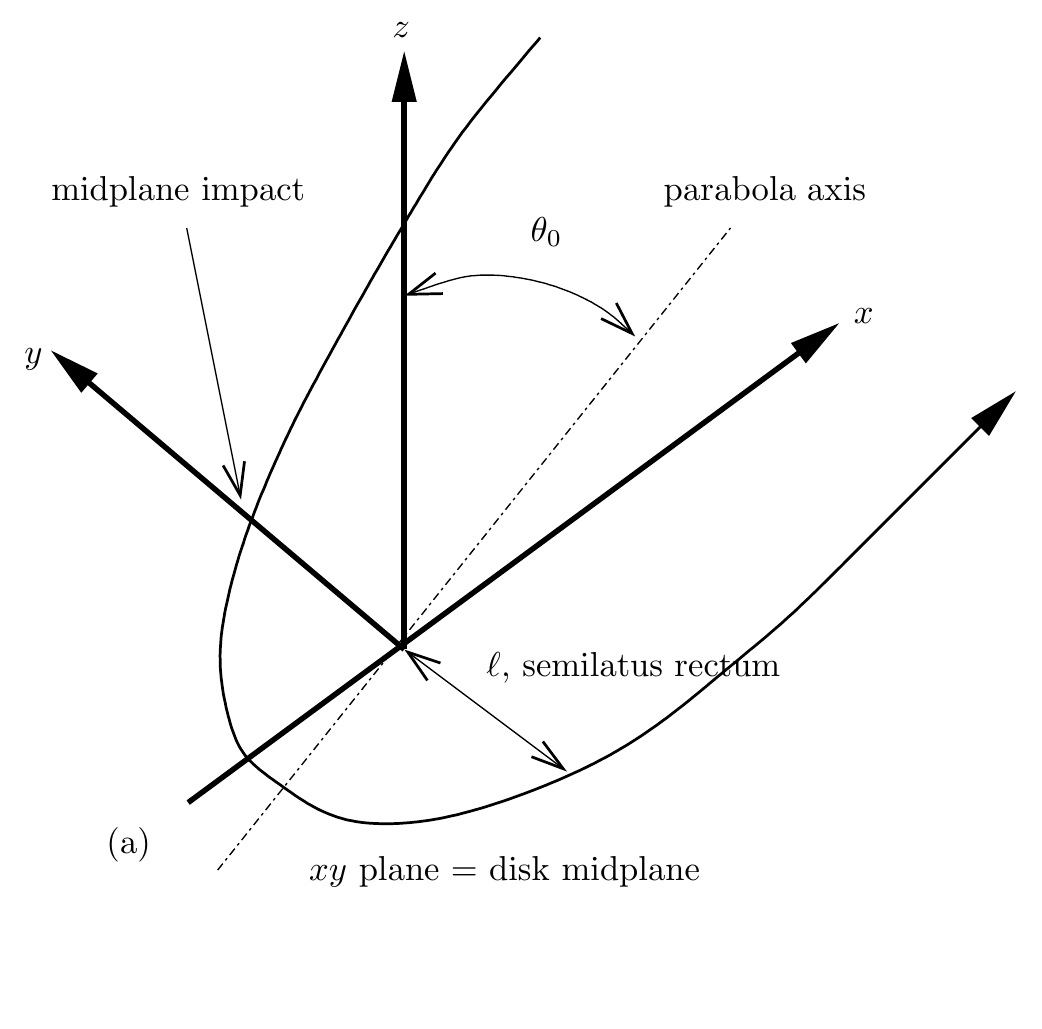}
\includegraphics[width=2.8truein]{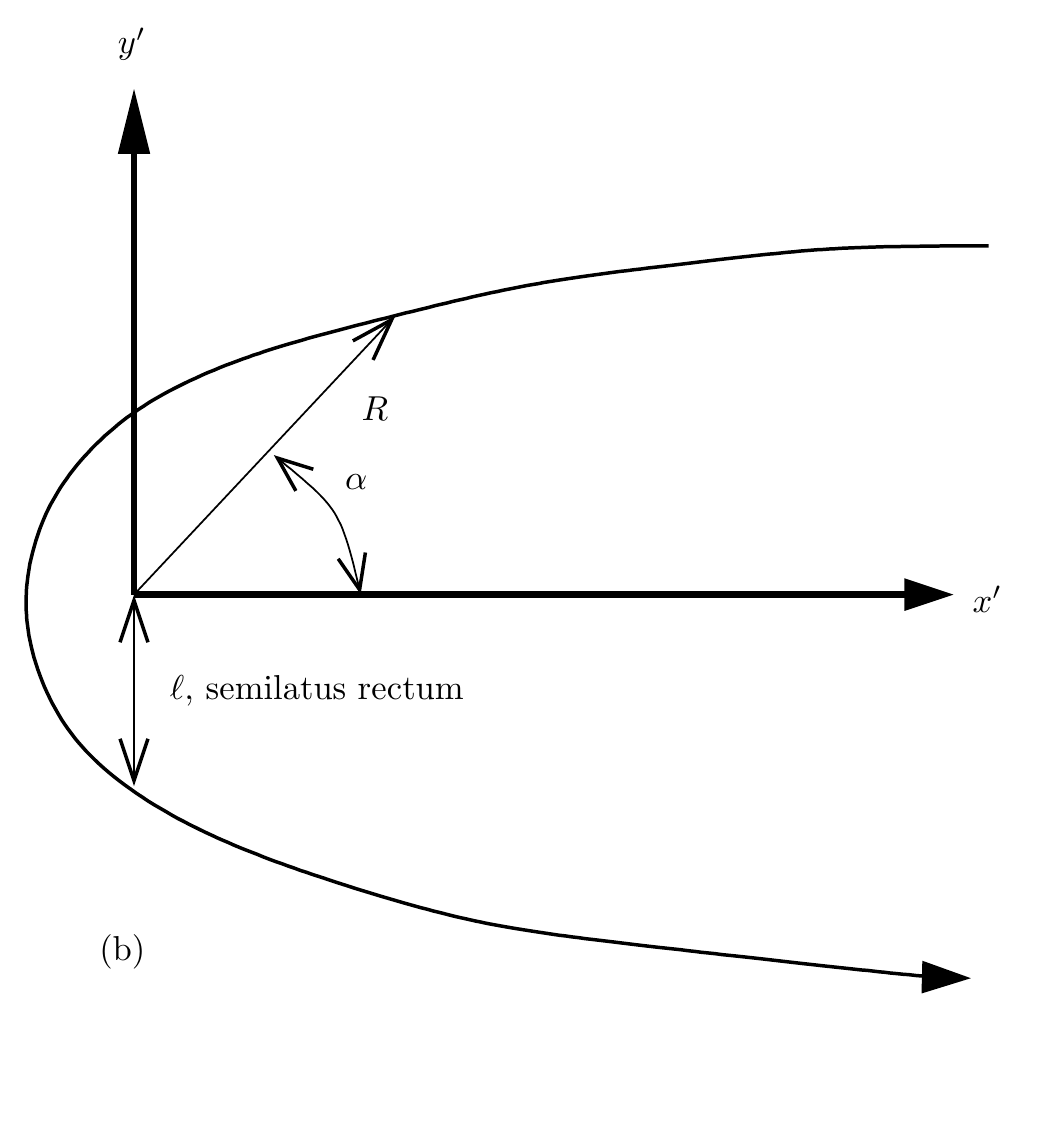}
\centering
\includegraphics[width=2.8truein]{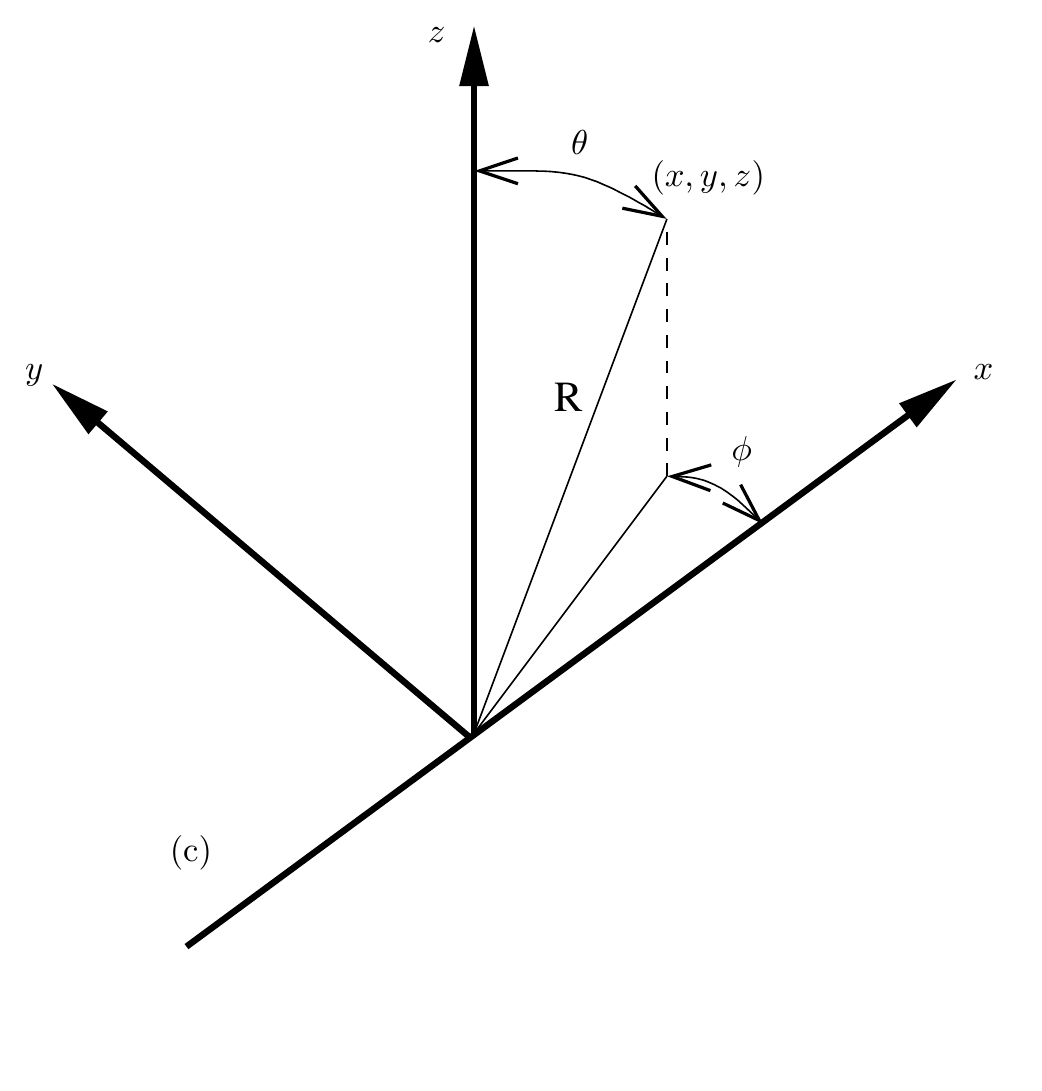}
\caption{(a) The inclination of the parabola defined as a rotation by angle $\pi/2 - \theta_0$ about the $y$-axis.
(b) Parabola in its parent plane ($x^\prime, y^\prime$).  (c) Spherical coordinates.}
\label{fig:parabola}
\end{figure}

To facilitate future modification or comparison with rotating collapse simulations (which we encourage), the UCM infall model is presented in some detail.  Spherical and cylindrical radii are $R$ and $r$, respectively.  Note that $R = r$ at the midplane.  When starting their infall, particles are assumed to be so far from the star that their initial potential energy is negligible.  Similarly, their initial kinetic energy due to cloud rotation is assumed to be negligible.  Both quantities are being compared to their values near the disk.  This puts particles on zero energy orbits, i.e., parabolas (see Figure~\ref{fig:parabola}a).  Note in passing: \cite{Mendoza_etal_2009} have extended Ulrich's model by assuming that particles start with a non-zero radial velocity at a finite radius.

The specific angular momentum of an infalling particle is defined as
\be
   \vec{j} \equiv \vec{R} \times \vec{u}.
\ee
Let $(\Rinit, \theta_\rmi, \phi_\rmi)$ denote the initial position of the particle in spherical coordinates.
Initially, due to cloud rotation, $\vec{u} = u_\phi \vec{e}_\phi$ and $\vec{R} = \Rinit \vec{e}_R$.  Since
\be
   \vec{e}_R \times \vec{e}_\phi = - \vec{e}_\theta
\ee
we have
\be
   \vec{j} = - u_\phi \Rinit \et = -\Omega_0 R_\mathrm{i}^2 \sin\theta_\rmi \et, \eql{j}
\ee
where $\Omega_0$ is the cloud rotation rate (assumed to be uniform) about the $z$-axis.
Equation \eqp{j} says that the angular momentum vector, and therefore the normal, $\vec{n}$, to the parabola is in the $-\et$ direction.
We therefore get the situation shown in Figure~\ref{fig:parabola}a in which the parabola has some inclination or ``pitch'' angle $\pi/2 - \theta_0$ about the $y$-axis but \textit{no} ``roll'' angle.  

Figure~\ref{fig:parabola}b shows the parabola in its parent plane $(x^\prime, y^\prime)$ in which its parametric equation is
\be
   R(\alpha) = \ell \left(1 - \cos\alpha\right)^{-1}, \eql{polar}
\ee
where the length $\ell$ of the semi-latus rectum is related to the angular momentum $j_n$ normal to its plane as
\be
   \ell = j_n^2 / GM.
\ee
From Figure \ref{fig:parabola}a observe that the latus rectum $\ell$ equals the radius where the parabolic orbit lands on the midplane.
If we assume that $\Rinit \gg \ell$ then $\theta_\rmi \approx \theta_0$.  Then
from equation \eqp{j} we have that
\be 
   j_n = \Omega_0 \Rinit^2 \sin\theta_0. \eql{jn}
\ee

The parametric equation of the parabola in Cartesian coordinates in its parent plane is
\be
   x^\prime = R(\alpha)\cos\alpha, \hskip 0.5truecm y^\prime = R(\alpha)\sin\alpha, \hskip 0.5truecm z^\prime = 0.
\ee
Then, applying a rotation by $\pi/2 - \theta_0$ about the $y$-axis we get
\begin{align}
\left(\begin{array}{c} x\\y\\z\end{array}\right) &=
\left(
\begin{array}{ccc}
\cos(\pi/2 - \theta_0)  & \,0\,   &   -\sin(\pi/2 - \theta_0) \\
  0                          & \,1\,   &  0 \\
\sin(\pi/2 - \theta_0)& \,0\,  &   \cos(\pi/2 - \theta_0) 
\end{array}
\right)
\left(\begin{array}{c} R\cos\alpha\\R\sin\alpha\\0\end{array}\right), \\
&= \left(
\begin{array}{ccc}
\sin\theta_0  & \,0\,   &   -\cos\theta_0 \\
  0                          & \,1\,   &  0 \\
\cos\theta_0& \,0\,  &   \sin\theta_0
\end{array}
\right)
\left(\begin{array}{c} R\cos\alpha\\R\sin\alpha\\0\end{array}\right). \eql{parabola}
\end{align}
Equating \eqp{parabola} to the representation of $(x, y, z)$ in spherical coordinates $(R, \theta, \phi)$, we get (see Figure~\ref{fig:parabola}c)
\begin{align}
R \cos\alpha\sin\theta_0 &= R\sin\theta\cos\phi, \\
R\sin\alpha &= R\sin\theta\sin\phi, \\
R\cos\alpha\cos\theta_0 &= R\cos\theta. \eql{third}
\end{align}
From \eqp{third} we have $\cos\alpha = \cos\theta/\cos\theta_0$ which when substituted into \eqp{polar} gives
\be
   R = \frac{\Omega_0^2 \Rinit^4}{GM} \sin^2\theta_0 \left(1 - \frac{\cos\theta}{\cos\theta_0}\right)^{-1}, \eql{parab_spherical}
\ee
for the equation of the parabola in spherical coordinates.  Let us define the centrifugal radius
\be
   \rc \equiv \frac{\Omega_0^2 \Rinit^4}{GM}, \eql{rc}
\ee
whose interpretation as the furthest radius where particles land on the midplane with non-zero $u_z$ will become clear later.  Note the strong dependence of $\rc$ on the cloud rotation rate $\Omega_0$ and an even stronger dependence on the initial radius $\Rinit$.

If one defines $\zeta \equiv \rc/R$ \citep[][henceforth TSC]{Terebey_etal_1984}, then  \eqp{parab_spherical} becomes
\be
   \zeta = \frac{\cos\theta_0 - \cos\theta}{\sin^2\theta_0\cos\theta_0}, \eql{zeta}
\ee
which is eq. (84) in TSC.
Differentiating \eqp{zeta} with respect to time gives
\be
  u_\theta = \frac{\cos\theta - \cos\theta_0}{\sin\theta} u_r, \eql{ut}
\ee
after noting that $u_\theta = R\dot{\theta}$ and $u_R = \dot{R}$.  Conservation of angular momentum gives
\be
   u_\phi = \frac{j_z}{R \sin\theta}. \eql{uphi}
\ee
Substituting \eqp{jn} and \eqp{polar} into \eqp{uphi} gives $u_\phi$.  Then, use of \eqp{ut} and the zero energy relation
\be
   \frac{1}{2}\left(u_\phi^2 + u_\theta^2 + u_R^2 \right) - GM/R = 0
\ee
gives the velocity field as eqs. (88)--(90) in TSC:
\begin{align}
u_R          &= -\GMRfac \left(1 + \frac{\cos\theta}{\cos\theta_0}\right)^{1/2}, \eql{uR_TSC} \\
u_\theta &= \GMRfac \frac{\left(\cos\theta_0 - \cos\theta\right)}{\sin\theta} 
\left(1 + \frac{\cos\theta}{\cos\theta_0}\right)^{1/2}, \eql{ut_TSC}\\
u_\phi    &= \GMRfac \frac{\sin\theta_0}{\sin\theta}\left(1 - \frac{\cos\theta}{\cos\theta_0}\right)^{1/2}. \eql{up_TSC}
\end{align}
It should be noted that the expression for $u_\theta$ given by eq.~(8) in \citet{Ulrich_1976} has a slight error as does eq.~(6) in \citet{Chevalier_1983}.

The density field is derived from the mass conservation equation and given by eq. (92) in TSC as
\be
   \rho = - \frac{\Mzdot}{4\pi R^2 u_R} \left[1 + 2 \zeta P_2(\cos\theta_0)\right]^{-1}, \eql{rho_TSC}
\ee
where $P_2$ is the Legendre polynomial of order 2.

Finally, one needs to specify the initial radius $\Rinit$ of a particle arriving at the disk at $t_0$ since the beginning of collapse.  \citet[][their pg. 357]{Cassen_and_Moosman_1981} use the expansion wave collapse solution of \cite{Shu_1977} and state, without proof, that
\be
   \Rinit = \frac{1}{2} m_0 c_0 t_0, \eql{Ri}
\ee
where $c_0$ is the isothermal sound speed in the cloud and $m_0$ is a parameter that results from Shu's collapse solution and depends on the amplitude $A$ of the initial density profile (see eq.~(16) and Table 1 in Shu, \citeyear{Shu_1977}).  The value $A = 2$ corresponds to the marginally gravitationally unstable case and gives $m_0 = 0.975$; this is the value we use here.  Values $A > 2$ give $m_0 > 0.975$ and correspond to initial conditions that are not in hydrostatic balance and therefore collapse spontaneously.

The expansion wave collapse solution gives the mass flux from the cloud as
\be
   \Mzdot = m_0 c_0^3 / G, \eql{Mzdot}
\ee
and the stellar mass is obtained as $M = \Mzdot t_0$, which assumes that a relatively smaller mass goes into building the disk.
Substituting \eqp{Ri} and \eqp{Mzdot} into \eqp{rc} gives the centrifugal radius as
\be
   \rc = \frac{1}{16} m_0^3 \Omega_0^2 c_0 t_0^3. \eql{rc2}
\ee

The procedure for obtaining the velocity and density field at any point $(R, \theta)$ is to solve the
cubic equation \eqp{zeta} for $\cos\theta_0$ (the orbital inclination parameter) and then use \eqp{uR_TSC}--\eqp{rho_TSC}.
This is how Figure \ref{fig:CM} in the body of the paper was prepared.

%%%%%%%%%%%%%%%%%%%%%%%%%%%%%%
\section{Infall at the midplane inward of the centrifugal radius}\label{sec:infall_inward_of_rc}

Here we provide pre-shock infall source terms (those with a `1' superscript) in the conservation equations \eqp{mass}--\eqp{amom} and \eqp{e}.
These are evaluated at the midplane where $\theta = \pi/2$ and $R = r$.
With $\xi \equiv \cos\theta$ and $\xi_0 \equiv \cos\theta_0$, \eqp{zeta} is
\be
   \zeta = \frac{\xi_0 - \xi}{(1 - \xi_0^2) \xi_0}. \eql{xi_eq}
\ee
Given any evaluation point $(R, \theta)$, \eqp{xi_eq} is a cubic equation for the orbital inclination parameter $\xi_0$.
Let us evaluate \eqp{xi_eq} at the midplane $\xi = 0$.  Consider orbits that are not parallel to the midplane, i.e., $\xi_0 \neq 0$.  Then we may cancel $\xi_0$ in the numerator and denominator to get
\be
   \zeta = \frac{1}{1 - \xi_0^2}. \eql{zeta1}
\ee
Now $0 \leq |\xi_0| \leq 1$ $\implies$ $1 \leq \zeta < \infty$ or $0 < r \leq \rc$.  In other words, those orbits that are not parallel to the midplane impact the midplane at radii $r \in [0, \rc]$.  Solving \eqp{zeta1} for $\xi_0$ we get
\be
   \xi_0 = \pm (1 - r/\rc)^{1/2}.
\ee
With this substitution \eqp{uR_TSC}--\eqp{rho_TSC} can be evaluated at the midplane as
\begin{align}
   u_r &= -\GMrfac, \eql{ur_mp} \\
   u_\theta = -u_z &= \GMrfac \left(1 - \eta\right)^{1/2}, \eql{uz_mp} \\
   u_\phi &= \GMrfac \eta^{1/2},  \eql{uphi_mp} \\
   \rho &= \frac{\Mzdot}{8\pi r^2}\left(\frac{GM}{r}\right)^{-1/2} \frac{\eta}{1 - \eta}, \eql{rho_mp}
\end{align}
where $\eta \equiv r/\rc$ and we used the fact that $R = r$ and $\theta = \pi/2$ at the midplane.  These are the same as eqs. (4a)--(4d) in \cite{Stahler_etal_1994}.  After some algebra, the infall source terms are
\begin{align}
   S_\mathrm{mass} &\equiv (2 \rho u_z r)_1 = - \frac{\Mzdot}{4\pi \rc} \left(1 - \eta\right)^{-1/2}, \\
   S_\mathrm{r-mom} &\equiv (2 \rho u_z r u_r)_1 
   = \frac{\Mzdot}{4\pi\rc}\left(\frac{GM}{\rc}\right)^{1/2}\left(1-\eta\right)^{-1/2} \eta^{-1/2}, \\
   S_\mathrm{\phi-mom} &\equiv (2 \rho u_z r u_\phi r)_1 = -\frac{\Mzdot}{4\pi}\left(\frac{GM}{\rc}\right)^{1/2} (1 - \eta)^{-1/2} \eta, \\
   S_\mathrm{energy} &\equiv (2 u_z e r)_1 = (2 \rho u_z r c_v \Tpcl)_1 = S_\mathrm{mass} c_v \Tpcl,
\end{align}
where we recall that $\Tpcl$ is the post cooling-layer temperature.

Since a finite-volume numerical method is employed, the above source terms are averaged over a cell interval $[r_i, r_{i+1}]$.  The integrals of the various functions of $\eta$ needed for the averages can be easily obtained using Mathematica so we will not display them here.  To second-order, the average equals the midpoint value and so using averages could be dispensed with for second-order schemes.  However, the source terms have an integrable singularity at $\eta = 1$ and this was the real motivation for using cell averages.

%%%%%%%%%%%%%%%%%%%%%%%%
\section{Radial infall at the midplane outward of the centrifugal radius} \label{sec:radial_infall}

Unlike previous vertically integrated models, we also impose radial infall at the outer boundary of the computational domain.
We saw that non-horizontal parabolas ($\xi_0 \neq 0$) impact the midplane at $r < \rc$.  Therefore parabolas pertinent to $r > \rc$ must be horizontal ($\xi_0 = 0$).  In other words, the infall velocity at the midplane is purely radial for $r > \rc$.
To linear order in $\xi_0$, \eqp{xi_eq} gives
\be
   \xi_0 = \xi (1 - \zeta)^{-1}. \eql{res}
\ee
When $\zeta \equiv \rc/r$ is not close to unity (and $< 1$), equation \eqp{res} says that the parabolas pertinent for flow field evaluation near the midplane ($\xi$ small) are nearly horizontal ($\xi_0$ is small).
Substituting \eqp{res} into the general flowfield expressions \eqp{uR_TSC}--\eqp{rho_TSC} gives after some algebra:
\begin{align}
   (u_r)_\rmmp         &= - \GMrfac \left(\frac{2\eta -1}{\eta}\right)^{1/2}, \\
   (u_z)_\rmmp        &=  0, \\
   (u_\phi)_\rmmp    &= \GMrfac \eta^{-1/2}, \eql{uphi_rgtrc} \\
   \rho_\rmmp       &= - \frac{\dot{M}_0}{4 \pi r^2 u_r} \frac{\eta}{\eta - 1},
\end{align}
where the subscript `mp' denotes the midplane.
One can verify that the above expressions conserve angular momentum and mechanical energy per unit mass.
Radial infall is imposed in the code via boundary conditions at $r_\mathrm{max}$.  The surface density is
\be
   \Sigma(\rmax) = (2\pi)^{1/2} \rho_\rmmp(\rmax) \Hrmax,
\ee
where the $(2\pi)^{1/2}$ factor is the integral of a Gaussian (hydrostatic) density profile with unit amplitude and
the scale height
\be
   \Hrmax = \frac{c_0}{\Omega(\rmax)} = \frac{c_0}{(u_\phi)_\rmmp(\rmax) / \rmax}.  
\ee
Given $\Sigma(\rmax)$, $(u_r)_\rmmp$, $(u_\phi)_\rmmp$ and the cloud temperature $T_0$, all of the variables in the transport equations can be specified at $\rmax$.

\section{Plots of UCM infall quantities at the midplane}

\begin{figure}
\centering
\includegraphics[width=2.8truein]{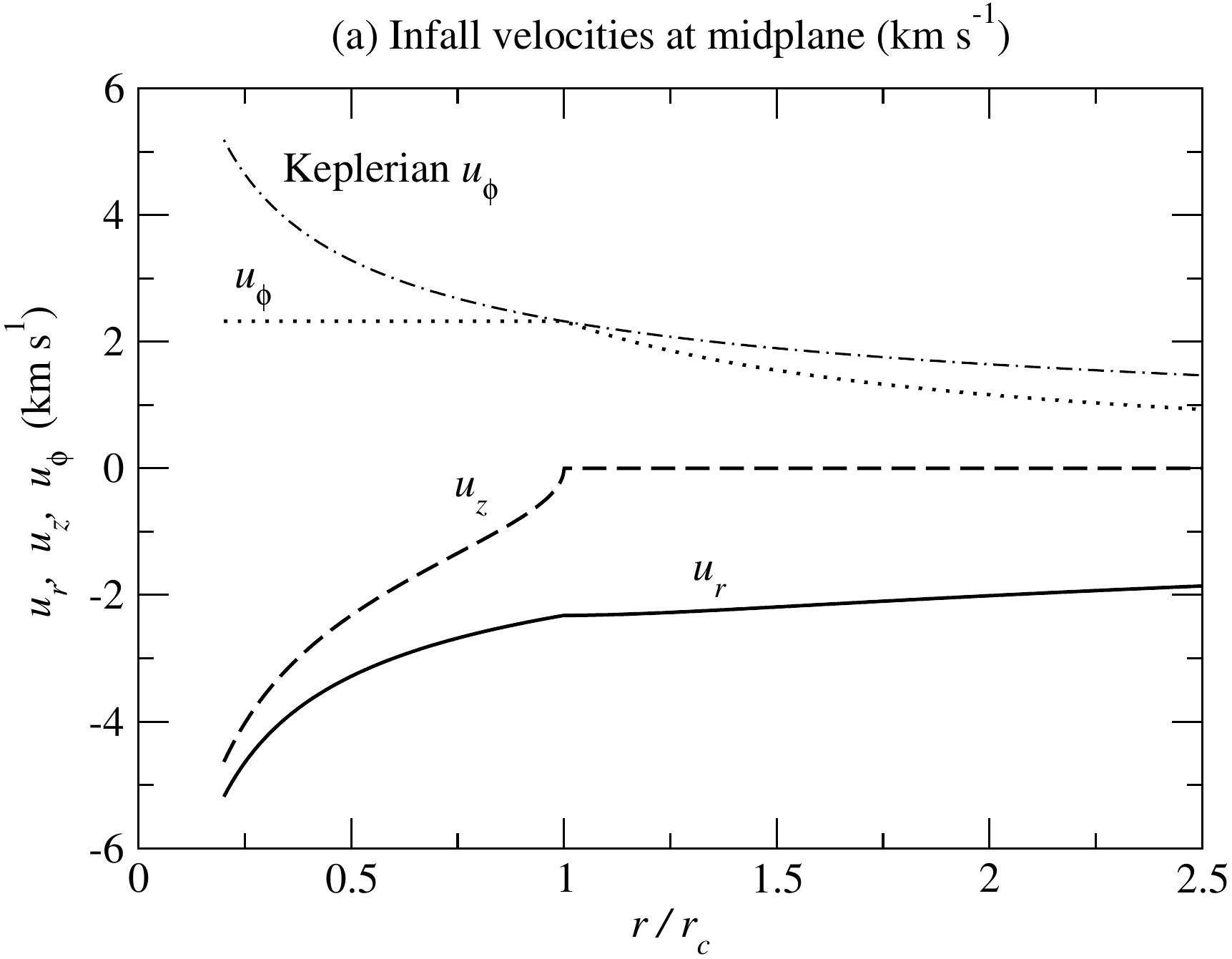}
\includegraphics[width=2.8truein]{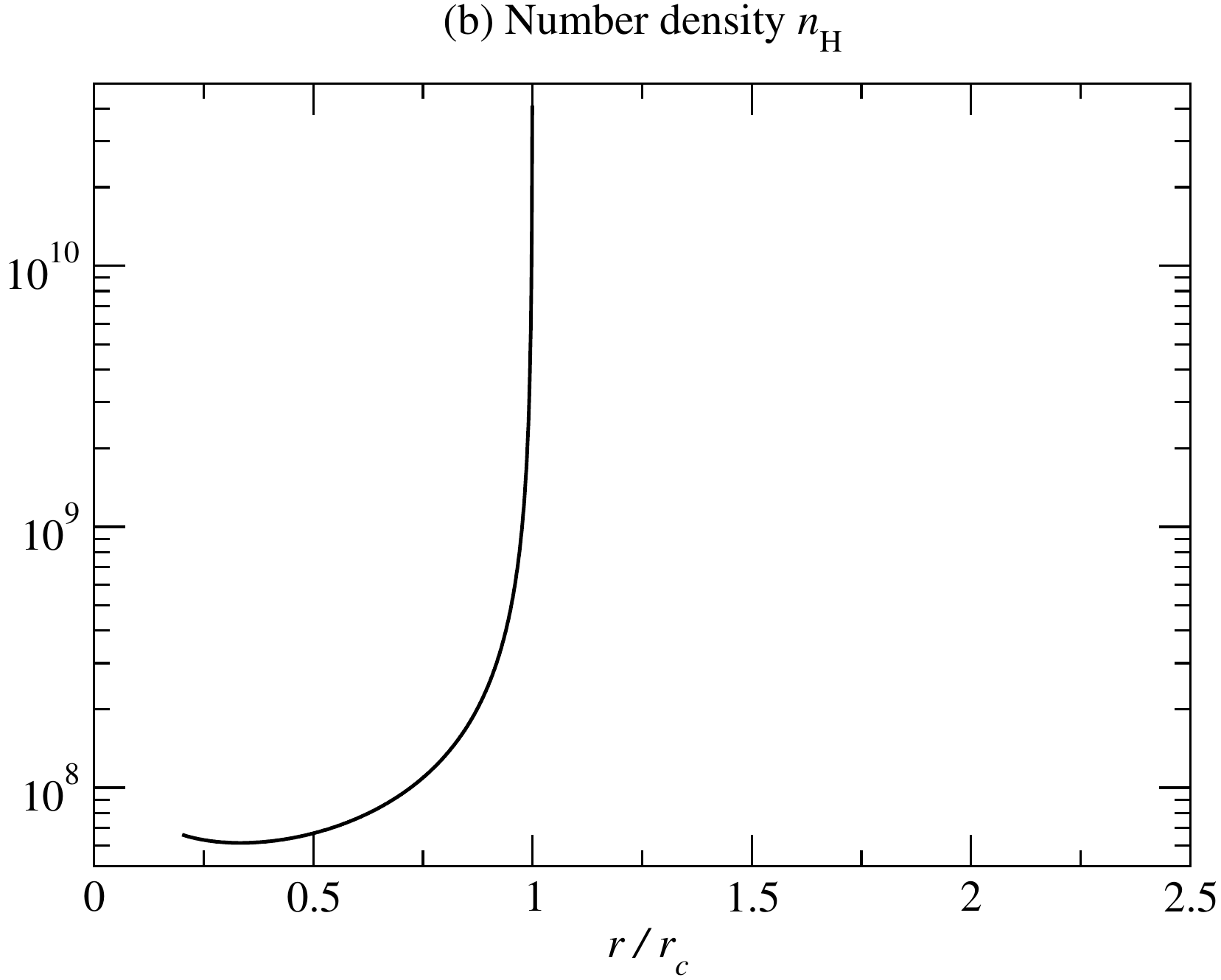}
\includegraphics[width=2.8truein]{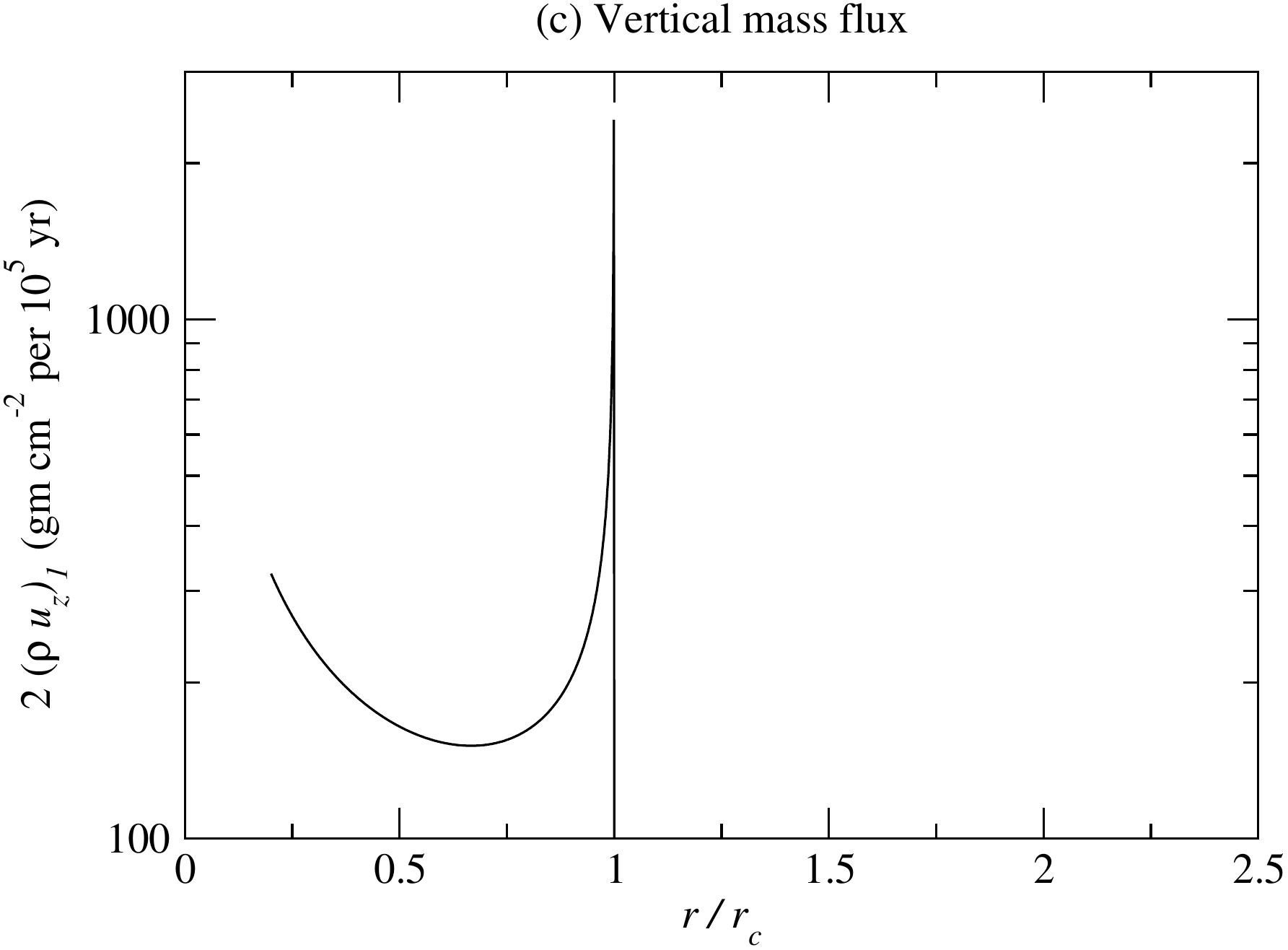}
\caption{Properties of the UCM vertical infall evaluated at the midplane for the L1527 simulation performed in the body of the paper.}
\label{fig:infall_quants}
\end{figure}
At the suggestion of the referee, Figure~\ref{fig:infall_quants} plots properties of the UCM infall evaluated at the midplane.  The parameters used were those of the L1527 simulation in the paper.  In panel (a) is to be noted that $u_\phi$ is sub-Keplerian, in fact constant for $r \leq \rc$, and exactly Keplerian for $r = \rc$.  There is no vertical impact ($u_z = 0$) for $r \geq \rc$.  Overall, the radial velocity is the dominant component.

Panel (b) plots the number density $n_\mathrm{H}$ which, in addition $u_z$, is an input parameter to the \citet{Neufeld_and_Hollenbach_1994} shock and cooling-layer code.

Panel (c) plots $2 \rho |u_z|$, the mass flux via vertical infall.  It has an integrable singularity at $r = \rc$, i.e., the mass entering a finite area that covers $r = \rc$ is finite.  As anticipated by the referee, the curve correlates well with the surface density observed in the simulations (Figure \ref{fig:case}e).  This is because the region $r < \rc$ receives very little mass from the region $r > \rc$ as noted earlier, and because there is very little radial transport for $r < \rc$.

%%%%%%%%%%%%%
\section{Vertical integration procedure and assumptions}\label{sec:vert_int}

To illustrate the vertical integration procedure, consider the mass conservation equation for axisymmetric flow
\be
   \frac{\p\rho}{\p t} + \frac{\p}{\p z}\left(\rho u_z\right) + \frac{1}{r} \frac{\p}{\p r}\left(r\rho u_r\right) = 0.
\ee
For possible future developments, we note the need for care: since the disk surface $z = Z(r, t)$ is a function of $r$ and $t$, vertical integration does not commute with
$\p/\p t$ and $\p/\p r$. This is unlike standard accretion disk theory \citep{Lynden-Bell_and_Pringle_1974} where the limits of integration are $z = \pm\infty$.  Instead, one must use Leibniz' rule, e.g., for the $r$ derivative
\be
   \int_0^{Z(r)} \frac{\p F(r,z)}{\p r}\, dz  = \frac{\p}{\p r} \int_0^{Z(r)} F(r, z)\, dz - F(r, Z(r)) \frac{\p Z}{\p r}, \eql{Leibniz}
\ee
and similarly for the $t$ derivative.  The extra term is the last one in \eqp{Leibniz} and represents the flux of mass via the radial velocity when the disk surface is inclined.
In the present work, where a razor thin disk is assumed, $\p Z/\p r = 0$ and this term disappears.  Similarly, vertical integration of time derivatives produces an extra term with a factor of $\p Z/\p t$ which is also zero in the limit of a razor-thin disk.  Finally, as in standard accretion disk theory, within the disk one allows for vertical variation in $\rho$ only and treats velocities as being uniform in $z$.   In other words, vertical variations in velocity are neglected.  In disks without infall, $u_\phi(r)$ depends weakly on $z$ due to a radial temperature gradient.  In disks with infall, velocity gradients with respect to $z$ are expected to be much larger due to the difference in velocity between the infalling gas and gas in the disk.

%%%%%%%%%%%%%%%%%%%%%%%%%%%%%%%%
\section{Extended ballistic argument}\label{sec:extended_ballistic}

\begin{figure*}
\vskip 0.5truecm
\centering
\includegraphics[width=3.4truein]{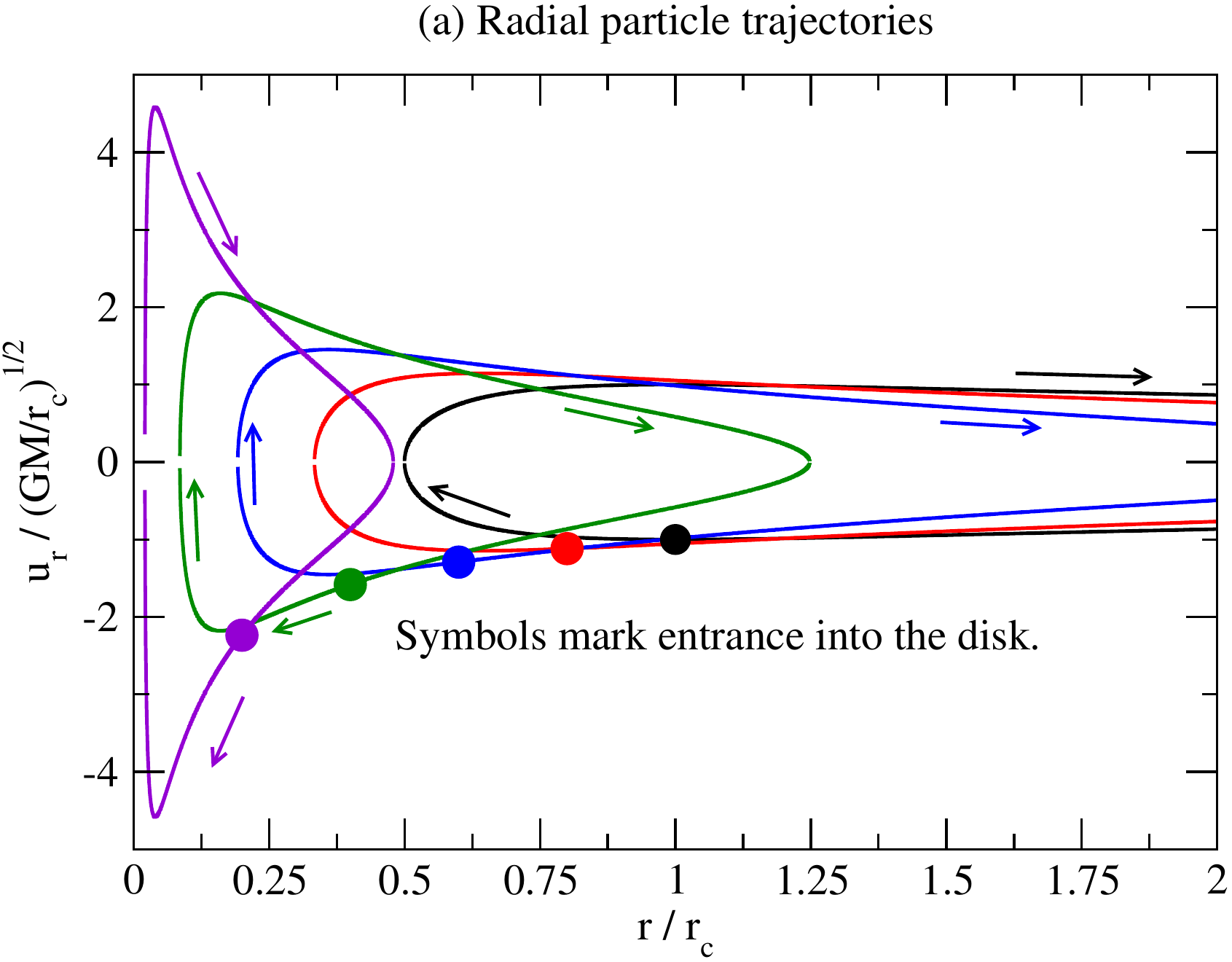}
\hfill
\includegraphics[width=3.4truein]{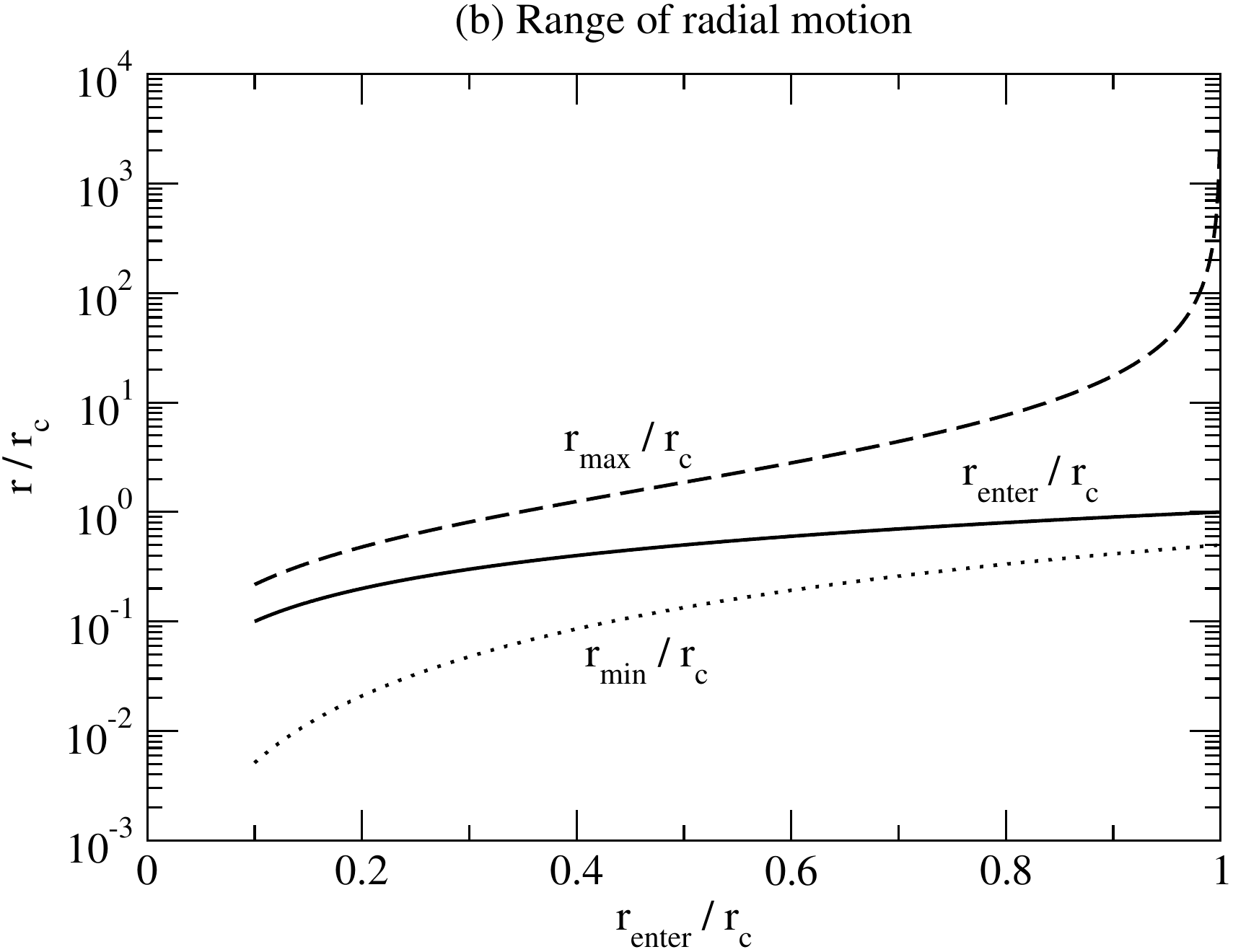}
\vskip 0.5truecm
\caption{Ballistic particle trajectories in the disk.  (a) Radial velocity versus radial position for particles entering at different radii.  (b) The range of radial motion for particles entering at different radii.}
\label{fig:reversal}
\end{figure*}

\citet{Sakai_etal_2014_Nature} show that the periastron of ballistic particles radially infalling along the midplane is at $r = \rc/2$, referred to as a centrifugal barrier.
This section considers ballistic trajectories of particles entering the disk via vertical infall at different $\renter \leq \rc$ and shows that particles initially head toward the star, stagnate at different locations and then reverse course.  Their orbits are ellipses of varying eccentricity.  This implies collisions of inwardly and outwardly moving fluid elements.  When collisions occur, the ballistic treatment is no longer valid, and therefore, as put by the referee, the outward moving portion of each ellipse is ``fictive.''  Even if one restricts attention to the inwardly moving portion of each orbit, one observes (Figure \ref{fig:reversal}a) that $u_r$ is not unique at a given $r$.  This is not a problem because orbits for different $\renter$ lie on different streamsurfaces having a different heights $z$ at a given $r$.

The ballistic flow is also invalid when the flow becomes subsonic and pressure effects must be taken into account.  One must consider not only the region where the original ballistic flow is subsonic, but also be mindful of the fact that the passage of a shock can render the flow subsonic over a larger region than in the original ballistic flow.

It is assumed that the disk-surface shock is flat so that $u_\phi$ and $u_r$ of a particle does not change across it and that, inside the disk, $u_z$ is much smaller in comparison.
Then, the mechanical energy of a particle inside the disk is
\be
  E = \frac{1}{2}\left(u_r^2 + u_\phi^2\right) - GM/r, \eql{Edisk}
\ee
where we have set $u_z = 0$.  Conservation of specific angular momentum implies that
\be
   u_\phi r = (u_\phi r)_\enter = \left(\frac{GM}{\rc}\right)^{1/2} \renter, \eql{sam}
\ee
where the subscript ``enter'' refers to quantities evaluated where the particle enters the disk, and \eqp{uphi_mp} was used for $(u_\phi)_\enter$.  Substituting \eqp{sam} into \eqp{Edisk} gives
\be
   E  = \frac{1}{2}\left(\frac{GM}{\rc}\frac{\renter^2}{r^2} + u_r^2\right) - \frac{GM}{r}.
\ee

The mechanical energy of the particle just after it crosses the accretion shock at a radius $r = \renter$ is:
\be
   E = \frac{1}{2}\left(u_r^2 + u_\phi^2\right)_\mathrm{enter} - GM/\renter.
\ee
Using equations \eqp{ur_mp} and \eqp{uphi_mp} for $u_r$ and $u_\phi$ at $\renter$ we get
\be
   E = \frac{GM}{2}\left(\frac{1}{\rc} - \frac{1}{\renter}\right). \eql{E0}
\ee 
Since $\renter \leq \rc$ for particles entering at the disk surface, Equation \eqp{E0} implies that $E_0 \leq 0$, where equality holds for $\renter = \rc$.  Negative and zero energy orbits are ellipses and parabolas respectively.

Setting $E = E_0$ gives (after some algebra) the radial velocity as
\be
   \frac{u_r^2}{(GM/\rc)} = 1 - \frac{1}{\eta_\enter} + \frac{2}{\eta} - \frac{\eta_\enter^2}{\eta^2}, \eql{ur}
\ee
where
\be
   \eta_\enter \equiv \renter/\rc.
\ee

The range of radial motion, i.e., the periastron and apastron radii $\eta_\mathrm{min}$ and $\eta_\mathrm{max}$, can be obtained by setting $u_r = 0$ in \eqp{ur}.  This leads to a quadratic equation for $1/\eta$ whose solutions are:
\be
   \eta_\mathrm{min, max} = \frac{\etaenter^2}{1 \pm \left(1 + \etaenter^2 -\etaenter\right)^{1/2}} \eql{roots}
\ee
%%%%%%%%%%%%%%%%%%%%%%%%%%%%
%%%%%%%%%%%%%%%%%%%%%%%%%%%%%%%%

Figure~\ref{fig:reversal}a uses Equation \eqp{ur} to plot particle trajectories in the normalized $(r, u_r)$ plane for particles entering the disk at different radii.  
The black particle enters at $r  = \rc$, reverses course at $r = 0.5\rc$ and heads out to $r = \infty$ because it is on a zero energy parabolic orbit.  
This is the same as the trajectory of all other particles infalling along the midplane as considered by \cite{Sakai_etal_2014_Nature} .
The green particle enters at $r = 0.275\rc$, reverses course at $r \approx 0.125$ and then heads out to its apastron at $r = 1.25\rc$.  Figure~\ref{fig:reversal}b plots the range of radial motion using equation \eqp{roots}.

Figure~\ref{fig:reversal} illustrates that ballistic particles entering the disk at $\renter > 0$ can never fall into the star.  For stellar accretion, angular momentum removal is required.

%%%%%%%%%%%%%%%%%%%%%%%%%%%%%%%%%%%%%%%%%%%%%%%%%%%%%%%%%%%
\section{Cassen-Moosman inviscid formula for the mass flow-rate through the disk} \label{sec:Mdot_CM}

In the body of the paper we plotted the mass flow rate in the disk due the drag induced by sub-Keplerian vertical infall and showed that it agreed with the simulation result.  Here we derive the formula used for that diagnostic by following \citet[][p.~360]{Cassen_and_Moosman_1981} and removing various effects (such as the slow increase in centrifugal radius with epoch $t_0$) that are not pertinent to the present work.  The main assumption is that $u_\phi(r)$ is steady.

The mass flow-rate in the disk is
\be
   \Mdot(r) \equiv 2\pi r \Sigma u_r. \eql{Mdot_def}
\ee
The equation for mass conservation \eqp{mass} is rewritten as
\be
   \frac{\p}{\p t}\left(\Sigma r\right) = - \frac{1}{2\pi} \frac{\p \Mdot}{\p r} - g(r),\eql{mass1}
\ee
where
\be
g(r) = 
\begin{cases}
2r (\rho u_z)_1, & r \le \rc; \\
0, & r > \rc;
\end{cases}
\ee
is the source term from infall.  Likewise, the angular momentum equation \eqp{amom} is rewritten as
\be
   \frac{\p}{\p t}\left(\Sigma r \Gamma\right) + \frac{1}{2\pi}\frac{\p}{\p r}\left(\Mdot(r)\Gamma\right) + g(r) \Gamma_1 = 0, \eql{amom2}
\ee
where $\Gamma \equiv u_\phi r$ is the specific angular momentum in the disk and $\Gamma_1 \equiv (u_\phi r)_1$ is the specific angular momentum of the infalling material.  Expanding derivatives in \eqp{amom2} and using \eqp{mass1}, one obtains
\be
   \frac{\p\Gamma}{\p t} + u_r \frac{\p\Gamma}{\p r} = \frac{g}{\Sigma r} (\Gamma - \Gamma_1). \eql{Gamma}
\ee
Equation \eqp{Gamma} is a Kelvin circulation theorem and says that circulation is preserved following a circular material line in the absence of infall ($g = 0$) while the right-hand-side of \eqp{Gamma} is a torque due to infall.
Solving \eqp{Gamma} for $u_r$, assuming that $\Gamma(r)$ has reached a steady-state, and substituting the result into \eqp{Mdot_def} gives
\be
   \Mdot(r) = 2\pi g(r) (\Gamma - \Gamma_1) \left(\p\Gamma/\p r\right)^{-1}, \eql{Mdot_CM_general}
\ee
which applies for \textit{general} but steady $\Gamma(r)$.  We inserted the simulation $\Gamma(r)$ into \eqp{Mdot_CM_general} to obtain the dotted line in Figure \ref{fig:case}c, indicated in the legend as ``CM81 formula using simulation $u_\phi$''.

From \eqp{uz_mp} and \eqp{rho_mp}  we have that
\be
g(r) = \begin{cases} -\frac{\Mzdot}{4\pi\rc} (1-\eta)^{-1/2}, & \eta \le 1;\\
0, & \eta > 1,
\end{cases} \eql{g}
\ee
where $\eta \equiv r/\rc$.
Note that $g(r)$ has an integrable singularity at $\eta = 1$ meaning that the mass flux is finite for any interval $\eta \in [1 - \epsilon, 1]$.
From \eqp{uphi_mp} 
\be
\Gamma_1 = 
\begin{cases}
\Gamma_\mathrm{K} \eta^{1/2}, & \eta \le 1;\\
0, & \eta > 1;
\end{cases}
\ee
where $\Gamma_\mathrm{K} \equiv (G M r)^{1/2}$ is the Keplerian value, i.e., the infall angular momentum is sub-Keplerian.  

In the special case where the disk is Keplerian, we have
\be
  g(\Gamma - \GammaK) = g \GammaK (1 - \eta^{1/2}) < 0, \eql{drag}
\ee
Thus, sub-Keplerian infall exerts a drag when mixed with the disk angular momentum; this drives an accretion mass flow which can be obtained from
\eqp{Mdot_CM_general} as
\be
   \Mdot(r) = -\Mzdot\eta(1 - \eta^{1/2})(1 - \eta)^{-1/2},
   \hskip 0.5truecm \eta \le 1.\eql{Mdot_CM}
\ee
Equation \eqp{Mdot_CM} was used to plot the dashed line in Figure \ref{fig:case}b, labeled in the legend as ``Cassen-Moosman Keplerian formula.''
Equation \eqp{Mdot_CM} agrees with the equation after (13) in CM81 who note that $\Mdot \to 0$ as $\eta \to 0$ and 1.  The former limit means that infall drag is insufficient for stellar accretion if $\Gamma(r)$ is Keplerian.

%%%%%%%%%%%%%%%%%%%%%%%%%%%%%%%%%%%%%%%%%

\section{Calculation of shock height}\label{sec:height}

Here we show how the disk-surface shock height $\Hshock(r, t)$ was calculated as a diagnostic and to help future development of a more sophisticated model that accounts for the actual height and shape of the disk surface.  Since the cooling layer is thin, we assume that the shock height is the same as height of the end of the cooling later.

As in \S\ref{sec:mass_and_mom} let subscript 1 denote pre-shock quantities and let subscript `2' denote quantities evaluated at the end of the cooling layer.  The velocity $u$ is the component normal to the shock, e.g., $u_z$ in the present razor thin disk model.
Conservation of mass and momentum across the shock and cooling layer give
\ba
\rho_1 u_1 &=& \rho_2 u_2,  \eql{mass_as}\\
p_1 + \rho_1 u_1^2 &=& p_2 + \rho_2 u_2^2. \eql{mom_as}
\ea
Using the ideal gas law for pressures, solving \eqp{mass_as} for $u_2$, substituting the result into \eqp{mom_as}, dividing through by $c_1 \equiv (\Rgas T_1)^{1/2}$, the pre-shock isothermal sound speed, one obtains the following quadratic equation for the density ratio $a \equiv \rho_2/\rho_1$:
\be
  \left(\frac{T_2}{T_1}\right) a^2 - a (1 + M_1^2) + M_1^2 = 0, \eql{quad}
\ee 
where $M_1 \equiv u_1 / c_1$ is the shock Mach number.  Recall that $T_2 = \Tpcl$ is known and given by \eqp{Tpcl}.
The roots of \eqp{quad} are
\be
   a = \frac{1}{2}\frac{T_1}{T_2} \left\{
   (1 + M_1^2) \pm \left[(1 + M_1^2)^2 - 4 (T_2/T_1) M_1^2\right]^{1/2}
   \right\}. \eql{qroots}
\ee
The `+' root in \eqp{qroots} is the physical one since it gives $a = \rho_2/\rho_1 = 1$ when $M_1 = 0$.  Since $u_z$ is small within the disk compared to $u_r$ and $u_\phi$, we can assume vertical hydrostatic balance so that
\be
   \rho_2(r) = \rhomp \exp(-\Hshock^2/(2 H^2)), \eql{rho2}
\ee
To determine the midplane density, $\rhomp$, we use the definition of the surface density:
\be
   \Sigma = 2 \rhomp \int_0^{\Hshock} \exp(-z^2/(2 H^2)) \, dz.
\ee
Performing the integral and solving for $\rhomp$ gives
\be
   \rhomp = \Sigma / \Htilde, \eql{rhomp}
\ee
where
\be
   \Htilde \equiv (2 \pi)^{1/2} H \erf(\Hshock/(\sqrt{2} H)).
\ee
Substituting \eqp{rhomp} into \eqp{rho2} gives the transcendental equation
\be
   \beta = \left[\ln\left(\frac{\Sigma}{\rho_2 \Htilde(\beta)}\right)\right]^{1/2}, \eql{trans}
\ee
for the ratio $\beta \equiv \Hshock / (\sqrt{2} H)$.  Equation \eqp{trans} is iterated to convergence.

%%%%%%%%%%%%%%%%%%%%%%%%%%%%%%%%%%%%%%%%%

% Don't change these lines
\bsp	% typesetting comment
\label{lastpage}
\end{document}